\documentclass[12pt]{article}
\usepackage{amsfonts,amsmath}
\usepackage{theorem}
\usepackage{rotating}
\newtheorem{definition}{Definition}[section]

\newtheorem{theorem}[definition]{Theorem}

\theorembodyfont{\rmfamily}
\newtheorem{remark}{Remark}[section]

\def\cG{{\cal G}}          \def\cH{{\cal H}}          \def\cI{{\cal I}}

\newcommand{\RR}{{\mathbb R}}
\newcommand{\ZZ}{{\mathbb Z}}
\newcommand{\eps}{{\varepsilon}}

\newcommand{\sfrac}[2]{{\textstyle{\frac{#1}{#2}}}}
\newcommand{\half}{{\sfrac{1}{2}}}

\newcommand{\evn}{{\overline{0}}}
\newcommand{\odd}{{\overline{1}}}
\newcommand{\bkl}{{[\hspace{-2pt}[}}
\newcommand{\bkr}{{]\hspace{-2pt}]}}
\newcommand{\ifrac}[2]{{\genfrac{}{}{0pt}{1}{#1}{#2}}}

\def\ad{\mathop{\rm ad}\nolimits}

\numberwithin{equation}{section}

\newcommand{\greydot}{\begin{picture}(20,20)
  \put(0,0){\circle{14}}
  \put(-5,-5){\line(1,1){10}}
  \put(-5,5){\line(1,-1){10}}
  \end{picture}}
\newcommand{\vertexbn}{\begin{picture}(20,20)
  \put(6,-3){\line(1,0){30}}\put(6,3){\line(1,0){30}}
  \put(27,0){\line(-1,1){10}}\put(27,0){\line(-1,-1){10}}
  \end{picture}}
\newcommand{\vertexcn}{\begin{picture}(20,20)
  \put(6,-3){\line(1,0){30}}\put(6,3){\line(1,0){30}}
  \put(17,0){\line(1,-1){10}}\put(17,0){\line(1,1){10}}
  \end{picture}}
\newcommand{\vertexdn}{\begin{picture}(20,40)
  \put(5,5){\line(3,2){19}} \put(5,-5){\line(3,-2){19}}
  \put(31,20){\circle{14}}\put(31,-20){\circle{14}}
  \end{picture}}
\newcommand{\vertexvn}{\begin{picture}(20,40)
  \put(6,18){\line(3,-2){19}} \put(6,-18){\line(3,2){19}}
  \put(-1,20){\circle{14}}\put(-1,-20){\circle{14}}
  \end{picture}}
\newcommand{\vertexgreydn}{\begin{picture}(20,40)
  \put(5,5){\line(2,1){20}}\put(5,-5){\line(2,-1){20}}
  \put(31,20){\circle{14}}\put(31,-20){\circle{14}}
  \put(26,15){\line(1,1){10}}\put(26,25){\line(1,-1){10}}
  \put(26,-25){\line(1,1){10}}\put(26,-15){\line(1,-1){10}}
  \put(28,-14){\line(0,1){28}}\put(34,-14){\line(0,1){28}}
  \end{picture}}
\newcommand{\vertexgcn}{\begin{picture}(20,20)
  \put(6,-4){\line(1,0){30}}\put(6,4){\line(1,0){30}}
  \put(17,0){\line(1,1){10}}\put(17,0){\line(1,-1){10}}
  \put(7,0){\line(1,0){28}}
  \end{picture}}
\newcommand{\vertexgbn}{\begin{picture}(20,20)
  \put(6,-4){\line(1,0){30}}\put(6,4){\line(1,0){30}}
  \put(27,0){\line(-1,1){10}}\put(27,0){\line(-1,-1){10}}
  \put(7,0){\line(1,0){28}}
  \end{picture}}
\newcommand{\vertexccn}{\begin{picture}(20,20)
  \put(6,-2){\line(1,0){29}}\put(6,2){\line(1,0){29}}
  \put(3,-6){\line(1,0){35}}\put(3,6){\line(1,0){35}}
  \put(17,0){\line(1,1){10}}\put(17,0){\line(1,-1){10}}
  \end{picture}}
\newcommand{\vertexbbn}{\begin{picture}(20,20)
  \put(6,-2){\line(1,0){29}}\put(6,2){\line(1,0){29}}
  \put(3,-6){\line(1,0){35}}\put(3,6){\line(1,0){35}}
  \put(27,0){\line(-1,1){10}}\put(27,0){\line(-1,-1){10}}
  \end{picture}}
\newcommand{\vertexdowndbl}{\begin{picture}(20,20)
  \put(3,-6){\line(0,-1){30}}\put(-3,-6){\line(0,-1){30}}
  \put(0,-27){\line(-1,1){10}}\put(0,-27){\line(1,1){10}}
  \end{picture}}
\newcommand{\vertexupdbl}{\begin{picture}(20,20)
  \put(3,6){\line(0,1){30}}\put(-3,6){\line(0,1){30}}
  \put(0,27){\line(-1,-1){10}}\put(0,27){\line(1,-1){10}}
  \end{picture}}
\newcommand{\arrowse}{\begin{rotate}{-30}\begin{picture}(10,10)%
  \put(0,0){\line(-1,1){10}}\put(0,0){\line(-1,-1){10}}%
  \end{picture}\end{rotate}}
\newcommand{\arrowne}{\begin{rotate}{30}\begin{picture}(10,10)%
  \put(0,0){\line(-1,1){10}}\put(0,0){\line(-1,-1){10}}%
  \end{picture}\end{rotate}}
\newcommand{\arrowno}{\begin{rotate}{-30}\begin{picture}(10,10)%
  \put(0,0){\line(1,-1){10}}\put(0,0){\line(1,1){10}}%
  \end{picture}\end{rotate}}
\newcommand{\arrowso}{\begin{rotate}{30}\begin{picture}(10,10)%
  \put(0,0){\line(1,-1){10}}\put(0,0){\line(1,1){10}}%
  \end{picture}\end{rotate}}

\begin{document}

\pagestyle{empty} 

\begin{center}

{\Large \textbf{Hyperbolic Kac--Moody superalgebras}}

\vspace{10mm}

{\large L. Frappat$^a$ and A. Sciarrino$^{b}$}

\vspace{10mm}

\emph{$^a$ Laboratoire d'Annecy-le-Vieux de Physique Th{\'e}orique LAPTH}

\emph{CNRS, UMR 5108, associ{\'e}e {\`a} l'Universit{\'e} de Savoie}

\emph{BP 110, F-74941 Annecy-le-Vieux Cedex, France}

\emph{Member of Institut Universitaire de France}

\vspace{7mm}

\emph{$^b$ Dipartimento di Scienze Fisiche, Universit{\`a} di Napoli 
``Federico II''}

\emph{and I.N.F.N., Sezione di Napoli}

\emph{Complesso Universitario di Monte S. Angelo, Via Cintia, I-80126 
Napoli, Italy}

\end{center}

\vspace{12mm}

\begin{abstract}
We present a classification of the hyperbolic Kac--Moody (HKM)
superalgebras. The HKM superalgebras of rank $r \ge 3$ are finite in number
(213) and limited in rank (6). The Dynkin--Kac diagrams and the corresponding
simple root systems are determined. We also discuss a class of singular
sub(super)algebras obtained by a folding procedure.
\end{abstract}

\vfill
\vfill

\rightline{LAPTH-1068/04}
\rightline{DSF-TH-28/04}
\rightline{math-ph/0409041}
\rightline{September 2004}

\clearpage
\pagestyle{plain}

\section{Introduction}
  
Affine Kac--Moody are presently well established tools of theoretical
physics. The indefinite Kac--Moody (KM) algebras \cite{Kacbook} form a so
general set of algebras that they defy any general classification. A
subclass of these KM algebras, called hyperbolic, which are defined by the
property that the diagrams (generally disconnected) obtained taking away a
dot from their defining diagrams define a direct sum of finite and/or
affine KM algebras have been classified in \cite{Sac89,Wan88}. It has been
found that these algebras are finite in number (238 in which 142 have a
symmetric or symmetrizable Cartan matrix) and bounded in rank (10). These
algebras have in the last decade attracted the attention of physicists, as
they appear in a variety of physical models in two-dimensional field
theories (supergravity, string theory, cosmological billiards)
\cite{GNW95,HM96,DHN02,KN04}. Therefore it seems natural to study the
corresponding partners in the realm of Kac--Moody superalgebras
\cite{Kac77a,Kac78b,VdL85}. From these motivations, the authors of Ref.
\cite{TDP03} have recently classified hyperbolic Kac--Moody (HKM)
superalgebras by a procedure quite close to that followed to classify the
hyperbolic Kac--Moody ones, showing that they are limited in rank, now the
maximum rank being 6, and that they are finite in number (for rank $>$ 2).
However, as we remarked that many diagrams are missing, while some of the
proposed Dynkin--Kac diagrams correspond in fact to diagrams of untwisted
or twisted affine Lie superalgebras (sometimes in the not distinguished
basis), we present here a, hopefully exhaustive, classification of HKM
superalgebras, together with a corresponding simple roots basis and we
discuss a class of singular subalgebras.

\medskip

The article is organized as follows: in section 2 we recall the definition of
a superalgebra, the relation between Dynkin--Kac diagrams and generalized
Cartan matrices, the action of the (super)Weyl o generalized Weyl
transformations on the simple roots systems and the structure of the
supplementary or non Serre relations. Although most of the material is not
new, we believe it is worthwhile to report it in some details for several
reasons: (i) the standard rules of translating matrices in diagrams have to
be slightly and suitably defined to include the case of indefinite, in
particular hyperbolic, KM superalgebras; (ii) the deformation of the
Dynkin--Kac diagrams has to be carefully handled, otherwise one is lead to
naively include diagrams for HKM superalgebras, which really correspond to
more general indefinite KM superalgebras; (iii) the action of the
generalized Weyl transformations, which provides also in the case of HKM
superalgebras all the not equivalent simple roots systems, allows one not
to be worried about the appearance of new non Serre relations. In section 3
and the related appendices we present the diagrams corresponding to HKM
superalgebras, together with their (not unique) system of simple roots and
with their maximal regular subalgebras; in section 4 we list the singular
subalgebras, obtained by the procedure of folding.
  
\section{Kac--Moody superalgebras}

\subsection{Generalized Cartan matrices and Dynkin diagrams}

Let $A$ be a $r \times r$ matrix and $\{i_{1},\ldots,i_{p}\}$ be a subset
of indices of $I = \{1,\ldots,r\}$. The principal
$\{i_{1},\ldots,i_{p}\}$-submatrix of $A$, of order $r-p$, is a matrix
obtained from $A$ by deleting the rows and columns labelled by
$i_{1},\ldots,i_{p}$. A principal submatrix of order $r-1$ is called
leading.

We start by defining the notion of generalized Cartan matrix. In the case 
of $\ZZ_{2}$-graded algebras, it is convenient to deal with a recursive
definition.
\begin{definition}
  \label{def:GCM}
  A $r \times r$ matrix $A$ with integral entries $a_{ij}$ is called a
  \emph{generalized Cartan matrix} if for each $i \in \{1,\ldots,r\}$, the
  leading principal $\{i\}$-submatrix of $A$ is a generalized Cartan matrix
  (which may be of block diagonal form). \\
  The Cartan matrices of the simple Lie algebras -- $A_{n}$, $B_{n}$,
  $C_{n}$, $D_{n}$, $E_{6,7,8}$, $F_{4}$ and $G_{2}$ -- and of the basic
  Lie superalgebras -- $A(m,n)$, $B(m,n)$, $C(n+1)$, $D(m,n)$, $F(4)$,
  $G(3)$ and $D(2,1,\alpha)$ -- are generalized Cartan matrices. \\
  The matrix $A$ is called \emph{symmetrizable} if it exists an invertible
  diagonal matrix $D$ such that $DA$ is a symmetric matrix. The matrix $A$
  is called \emph{indecomposable} if it cannot be reduced to a block
  diagonal form by reordering rows and columns. 
\end{definition}
We will only consider generalized Cartan matrices which are indecomposable
and symmetrizable. Moreover, we assume that the generalized Cartan matrices
are properly normalized, i.e. $a_{ii}=2$ or $a_{ii}=0$ for each $i$. If one
defines the matrix $D_{ij} = d_i \, \delta_{ij}$ where the rational
coefficients $d_i$ satisfy $d_i \, a_{ij} = d_j \, a_{ji}$, the symmetric
Cartan matrix $A'$ is given from the generalized Cartan matrix $A$ by $A' =
DA$. Note that, due to fact that off-diagonal entries of a row of a Cartan
matrix corresponding to $a_{ii} = 0$ may have different signs, the diagonal
entries of the symmetric Cartan matrices are not necessarily positive.
\begin{remark}
  It follows from the definition that the Cartan matrices of the affine
  (untwisted or twisted) Kac--Moody algebras and superalgebras are
  generalized Cartan matrices.
\end{remark}

\begin{definition}
  \label{def:KMSA}
  Let $\tau$ be a subset of $I = \{ 1,\dots,r \}$. To a given generalized
  Cartan matrix $A$ and subset $\tau$, we associate a complex
  contragredient Lie superalgebra $\cG(A,\tau)$ -- called Kac--Moody
  superalgebra -- with $3r$ generators $h_i$, $e_i^\pm$ and
  $\ZZ_2$-gradation defined by $\deg e_i^\pm = \evn$ if $i \notin \tau$,
  $\deg e_i^\pm = \odd$ if $i \in \tau$ and $\deg h_i = \evn$ for all $i$.
  The generators $h_i$ and $e_i^\pm$ are subject the following set of
  relations:
  \begin{eqnarray}
    && [ h_i,h_j ] = 0 \label{eq:serre1} \\
    && [ h_i,e_j^{\pm} ] = \pm a_{ij} e_j^{\pm} \label{eq:serre2} \\
    && \bkl e_i^{+},e_j^{-} \bkr = \delta_{ij} h_i \label{eq:serre3} \\
    && \bkl e_i^{\pm},e_i^{\pm} \bkr = 0 \quad \mbox{if $a_{ii} = 0$}
    \label{eq:serre4}
  \end{eqnarray}
  and
  \begin{equation}
    \label{eq:serrerel}
    \left(\ad e_i^{\pm}\right)^{1-\widetilde{a}_{ij}} \, e_j^{\pm} = 0
  \end{equation}
  where the matrix $\widetilde{A} = (\widetilde{a}_{ij})$ is deduced from
  the Cartan matrix $A = (a_{ij})$ of $\cG(A,\tau)$ by replacing all its
  positive off-diagonal entries by $-1$. Here $\;\ad\;$ denotes the adjoint
  action:
  \begin{equation}
    (\ad X) \, Y = \bkl X,Y \bkr = XY - (-1)^{\deg X.\deg Y} YX
  \end{equation}
\end{definition}
We denote by $\cG_{\evn}$ and $\cG_{\odd}$ the even and odd parts of the
Kac--Moody superalgebra $\cG(A,\tau)$. Let $\cH \subset \cG_{\evn}$ be the
subalgebra of $\cG$ generated by the $h_{i}$ (Cartan subalgebra). The
superalgebra $\cG(A,\tau)$ can be decomposed as $\cG = \bigoplus_{\alpha}
\cG_{\alpha}$ where $\cG_{\alpha} = \{ x \in \cG \,\vert\, [ h,x ] =
\alpha(h) \, x,\, h \in \cH \}$. By definition, the root system of $\cG$ is
the set $\Delta = \{ \alpha \in \cH^* \,\vert\, \cG_{\alpha} \ne 0 \}$. A
root $\alpha$ is called even (resp. odd) if $\cG_\alpha \cap \cG_\evn \ne
\emptyset$ (resp. $\cG_\alpha \cap \cG_\odd \ne \emptyset$). The set of
even (resp. odd) roots is denoted by $\Delta_\evn$ (resp. $\Delta_\odd$).
Since $\cG(A,\tau)$ clearly admits a Borel decomposition, one defines as
usual the notion of simple root system \cite{Kac77a,FSS}.

\medskip

To each superalgebra $\cG(A,\tau)$ can be associated a Dynkin diagram
according to the following rules~\cite{FSS}. We will always assume that 
$i \in \tau$ if $a_{ii} = 0$.
\begin{enumerate}
  \item
  Using the generalized Cartan matrix $A$:
  \begin{enumerate}
    \item
    One associates to each $i$ such that $a_{ii}=2$ and $i \notin \tau$ a
    white dot, to each $i$ such that $a_{ii}=2$ and $i \in \tau$ a black
    dot, to each $i$ such that $a_{ii}=0$ and $i \in \tau$ a grey dot.
    \begin{center}
      \begin{picture}(220,20) \thicklines 
        \put(7,7){\circle{14}}
        \put(7,22){\makebox(0,0){white dot}}
        \put(107,7){\circle*{14}}
        \put(107,22){\makebox(0,0){black dot}}
        \put(207,7){\circle{14}}
        \put(202,2){\line(1,1){10}}\put(202,12){\line(1,-1){10}}
        \put(207,22){\makebox(0,0){grey dot}}
      \end{picture}
    \end{center}
    \item
    The $i$-th and $j$-th dots will be joined by $\eta_{ij}$ lines where
    \begin{displaymath}
      \begin{array}{ll}
        \bigg. \eta_{ij} = \max\Big( |a_{ij}|,|a_{ji}| \Big) & \mbox{if
        $a_{ii} \ne 0$ or/and $a_{jj} \ne 0$ and $|a_{ij} \, a_{ji}| \le
        4$} \\
        \bigg. \eta_{ij} = |a_{ij}| = |a_{ij}| & \mbox{if $a_{ii} = a_{jj}
        = 0$ and $|a_{ij}|, |a_{ji}| \le 4$}
      \end{array}
    \end{displaymath}
    Otherwise, the $i$-th and $j$-th dots will be joined by a boldface line
    equipped with an ordered pair of integers $(|a_{ij}|,|a_{ji}|)$. Note
    that this latter case does not appear for finite or affine Kac--Moody
    superalgebras.
    \item
    We add an arrow on the lines connecting the $i$-th and $j$-th dots 
    when $\eta_{ij} > 1$ and $|a_{ij}| \ne |a_{ji}|$, pointing from $j$ 
    to $i$ if $|a_{ij}| > 1$.
    \item
    For $D(2,1;\alpha)$, $\eta_{ij} = 1$ if $a_{ij} \ne 0$ and 
    $\eta_{ij} = 0$ if $a_{ij} = 0$.  No arrow is put on the Dynkin 
    diagram.
  \end{enumerate}
  \item
  Using the symmetric Cartan matrix $A'$:
  \begin{enumerate}
    \item
    One associates to each $i$ such that $a'_{ii} \ne 0$ and $i \notin
    \tau$ a white dot, to each $i$ such that $a'_{ii} \ne 0$ and $i \in
    \tau$ a black dot, to each $i$ such that $a'_{ii}=0$ and $i \in \tau$ a
    grey dot (see pictures above).
    \item
    The $i$-th and $j$-th dots will be joined by $\eta_{ij}$ lines 
    where
    \begin{displaymath}
      \begin{array}{ll}
        \big.  \displaystyle{ \eta_{ij} = \frac{2|a'_{ij}|}{\min 
        (|a'_{ii}|,|a'_{jj}|)} } & \mbox{if $a'_{ii}.a'_{jj} \ne 
        0$ and ${a'_{ij}}^{2} \le |a'_{ii}.a'_{jj}|$} \\
        \bigg.  \displaystyle{ \eta_{ij} = 
        \frac{2|a'_{ij}|}{\min(|a'_{ii}|,2)} } & \mbox{if $a'_{ii} 
        \ne 0, \; a'_{jj} = 0$ and $\eta_{ij} \le 4$} \\
        \bigg.  \displaystyle{ \eta_{ij} = |a'_{ij}| } & \mbox{if 
        $a'_{ii} = a'_{jj} = 0$ and $|a'_{ij}| \le 4$}
      \end{array}
    \end{displaymath}
    Otherwise, the $i$-th and $j$-th dots will be joined by a boldface line
    equipped with an ordered pair of integers $(|a_{ij}|,|a_{ji}|)$.
    \item
    We add an arrow on the lines connecting the $i$-th and $j$-th dots when 
    $\eta_{ij} > 1$, pointing from $i$ to $j$ if $a'_{ii}.a'_{jj} \ne 0$ 
    and $|a'_{ii}| > |a'_{jj}|$ or if $a'_{ii} = 0$, $a'_{jj} \ne 0$, 
    $|a'_{jj}| < 2$, and pointing from $j$ to $i$ if $a'_{ii} = 0$, 
    $a'_{jj} \ne 0$, $|a'_{jj}| > 2$.
    \item
    For $D(2,1;\alpha)$, $\eta_{ij} = 1$ if $a'_{ij} \ne 0$ and 
    $\eta_{ij} = 0$ if $a'_{ij} = 0$.  No arrow is put on the Dynkin 
    diagram.
  \end{enumerate}
\end{enumerate}
Although the rules seem more complicated when using the symmetric Cartan
matrix $A'$, the computation of the Cartan matrix $A$ is often more
involved than the symmetric Cartan matrix $A'$.

\begin{remark}
  The entries of the symmetric Cartan matrices $A'$ can be obtained as the
  scalar products of the simple roots, i.e. $a'_{ij} =
  (\alpha_{i},\alpha_{j})$ (up to a multiplication by a suitable factor in
  order to get integer entries).
\end{remark}

\begin{remark}
  The above rules imply that two white/black dots of square length $L$ and
  scalar product $S$ are connected by $|2S/L|$ lines. With this convention,
  the Dynkin diagram of the affine Kac--Moody algebra $A_{1}^{(1)}$ is
  simply given by two white dots connected by two lines without any arrow.
\end{remark}

\medskip

Note that for superalgebras Dynkin--Kac diagrams with the same ''topology''
may be different. For example the diagrams drawn below represent
respectively the superalgebras $sl(2|2)$, $osp(4|2)$, $osp(4|4)$ and a
hyperbolic Kac--Moody superalgebra of rank four (see definition
\ref{def:HKMSA}). The root systems are described in terms of the orthogonal
vectors $\eps_{i}$, $\delta_{i}$ and $K^{\pm}$ (see Appendix B for
conventions).

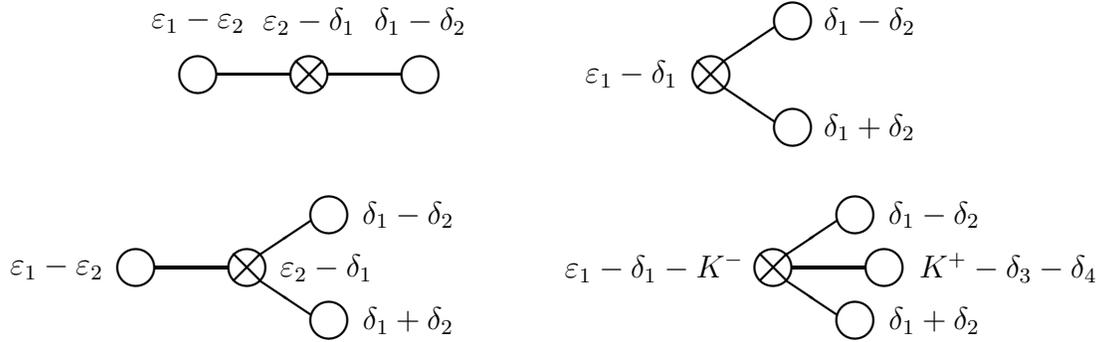
\begin{displaymath}
  \begin{array}{ccc}
    \begin{picture}(80,20) \thicklines 
      \put(0,0){\circle{14}} \put(42,0){\greydot} \put(84,0){\circle{14}} 
      \put(7,0){\line(1,0){28}} \put(49,0){\line(1,0){28}} 
      \put(0,20){\makebox(0,0){$\eps_{1}-\eps_{2}$}}
      \put(42,20){\makebox(0,0){$\eps_{2}-\delta_{1}$}}
      \put(84,20){\makebox(0,0){$\delta_{1}-\delta_{2}$}}
    \end{picture}
    &\qquad\qquad\qquad\qquad& 
    \begin{picture}(60,20) \thicklines 
      \put(0,0){\greydot} \put(0,0){\vertexdn} 
      \put(-30,0){\makebox(0,0){$\eps_{1}-\delta_{1}$}} 
      \put(60,20){\makebox(0,0){$\delta_{1}-\delta_{2}$}} 
      \put(60,-20){\makebox(0,0){$\delta_{1}+\delta_{2}$}} 
    \end{picture}
  \end{array}
\end{displaymath}
\vspace*{36pt}
\begin{displaymath}
  \begin{array}{ccc}
    \begin{picture}(80,20) \thicklines 
      \put(0,0){\circle{14}} \put(42,0){\greydot} \put(42,0){\vertexdn} 
      \put(7,0){\line(1,0){28}} 
      \put(-30,0){\makebox(0,0){$\eps_{1}-\eps_{2}$}} 
      \put(72,0){\makebox(0,0){$\eps_{2}-\delta_{1}$}} 
      \put(103,20){\makebox(0,0){$\delta_{1}-\delta_{2}$}} 
      \put(103,-20){\makebox(0,0){$\delta_{1}+\delta_{2}$}} 
    \end{picture}
    &\qquad\qquad\qquad\qquad\qquad\qquad& 
    \begin{picture}(60,20) \thicklines 
      \put(0,0){\greydot} \put(0,0){\vertexdn} \put(42,0){\circle{14}}
      \put(7,0){\line(1,0){28}} 
      \put(-45,0){\makebox(0,0){$\eps_{1}-\delta_{1}-K^{-}$}} 
      \put(89,0){\makebox(0,0){$K^{+}-\delta_{3}-\delta_{4}$}} 
      \put(61,20){\makebox(0,0){$\delta_{1}-\delta_{2}$}} 
      \put(61,-20){\makebox(0,0){$\delta_{1}+\delta_{2}$}} 
    \end{picture}
  \end{array}
\end{displaymath}

\vspace*{36pt}

\subsection{Non Serre type relations and generalized Weyl transformations}

In the case of finite and affine Kac--Moody superalgebras, it is known that
the description given by the Serre relations (\ref{eq:serre1}) to
(\ref{eq:serre4}) may lead to superalgebras with non trivial ideals
\cite{LSS90, Sch93}. In order to obtain a simple superalgebra, it is
necessary to write supplementary relations involving more than two
generators, in order to quotient the bigger superalgebra. These
supplementary non Serre type conditions appear when one deals with
isotropic odd roots (that is $a_{ii} = 0$).

\medskip

The supplementary conditions depend on the different kinds of vertices
which appear in the Dynkin diagrams. The vertices for finite and affine
superalgebras can be of the following type:

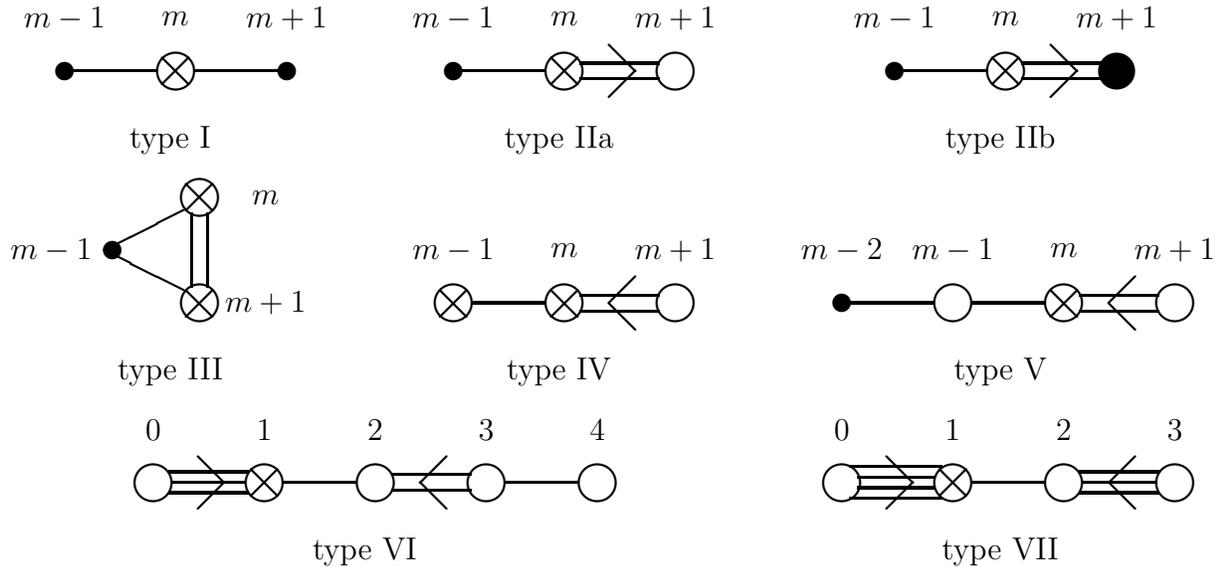
\begin{displaymath}
  \begin{array}{ccccc}
    \begin{picture}(80,20) \thicklines 
      \put(0,0){\circle*{7}} \put(42,0){\greydot} \put(84,0){\circle*{7}} 
      \put(3,0){\line(1,0){32}} \put(49,0){\line(1,0){32}}
      \put(0,20){\makebox(0,0){$m-1$}} 
      \put(42,20){\makebox(0,0){$m$}} 
      \put(84,20){\makebox(0,0){$m+1$}} 
    \end{picture}
    &\qquad\qquad& 
    \begin{picture}(80,20) \thicklines 
      \put(0,0){\circle*{7}} \put(42,0){\greydot} \put(84,0){\circle{14}} 
      \put(3,0){\line(1,0){32}} \put(42,0){\vertexbn} 
      \put(0,20){\makebox(0,0){$m-1$}} 
      \put(42,20){\makebox(0,0){$m$}} 
      \put(84,20){\makebox(0,0){$m+1$}} 
    \end{picture}
    &\qquad\qquad&
    \begin{picture}(80,20) \thicklines 
      \put(0,0){\circle*{7}} \put(42,0){\greydot} \put(84,0){\circle*{14}} 
      \put(3,0){\line(1,0){32}} \put(42,0){\vertexbn} 
      \put(0,20){\makebox(0,0){$m-1$}} 
      \put(42,20){\makebox(0,0){$m$}} 
      \put(84,20){\makebox(0,0){$m+1$}} 
    \end{picture}
    \\
    &&\\
    \mbox{type I} && \mbox{type IIa} && \mbox{type IIb} \\
    &&\\
    \begin{picture}(40,40) \thicklines 
      \put(-2,20){\circle*{7}} \put(0,20){\vertexgreydn}
      \put(5,25){\line(-2,-1){10}}\put(5,15){\line(-2,1){10}}
      \put(-25,20){\makebox(0,0){$m-1$}} 
      \put(56,40){\makebox(0,0){$m$}} 
      \put(56,0){\makebox(0,0){$m+1$}} 
    \end{picture}
    &\qquad\qquad& 
    \begin{picture}(80,20) \thicklines 
      \put(0,0){\greydot} \put(42,0){\greydot} \put(84,0){\circle{14}} 
      \put(7,0){\line(1,0){28}} 
      \put(42,0){\vertexcn} 
      \put(0,20){\makebox(0,0){$m-1$}} 
      \put(42,20){\makebox(0,0){$m$}} 
      \put(84,20){\makebox(0,0){$m+1$}} 
    \end{picture}
    &\qquad\qquad& 
    \begin{picture}(120,20) \thicklines 
      \put(0,0){\circle*{7}} \put(42,0){\circle{14}} 
      \put(84,0){\greydot} \put(126,0){\circle{14}} 
      \put(3,0){\line(1,0){32}} \put(49,0){\line(1,0){28}} 
      \put(84,0){\vertexcn} 
      \put(0,20){\makebox(0,0){$m-2$}} 
      \put(42,20){\makebox(0,0){$m-1$}} 
      \put(84,20){\makebox(0,0){$m$}} 
      \put(126,20){\makebox(0,0){$m+1$}} 
    \end{picture}
    \\
    &&\\
    \mbox{type III} && \mbox{type IV} && \mbox{type V} \\
    &&\\
    \multicolumn{3}{c}{
    \begin{picture}(160,20) \thicklines 
      \put(0,0){\circle{14}} \put(42,0){\greydot} 
      \put(84,0){\circle{14}} \put(126,0){\circle{14}} 
      \put(168,0){\circle{14}}
      \put(0,0){\vertexgbn} \put(49,0){\line(1,0){28}} 
      \put(84,0){\vertexcn} \put(133,0){\line(1,0){28}} 
      \put(0,20){\makebox(0,0){0}} 
      \put(42,20){\makebox(0,0){1}} 
      \put(84,20){\makebox(0,0){2}} 
      \put(126,20){\makebox(0,0){3}} 
      \put(168,20){\makebox(0,0){4}} 
    \end{picture}}
    &\qquad\qquad& 
    \begin{picture}(120,20) \thicklines 
      \put(0,0){\circle{14}} \put(42,0){\greydot} 
      \put(84,0){\circle{14}} \put(126,0){\circle{14}} 
      \put(49,0){\line(1,0){28}} 
      \put(0,0){\vertexbbn} \put(84,0){\vertexgcn}
      \put(0,20){\makebox(0,0){0}} 
      \put(42,20){\makebox(0,0){1}} 
      \put(84,20){\makebox(0,0){2}} 
      \put(126,20){\makebox(0,0){3}} 
    \end{picture}
    \\
    &&\\
    \multicolumn{3}{c}{\mbox{type VI}} && \mbox{type VII} \\
\end{array}
\end{displaymath}
where the small black dots represent either white dots associated to even 
roots or grey dots associated to isotropic odd roots. Hyperbolic 
superalgebras exhibit also more complicated vertices.

\medskip

The supplementary conditions take the following form \cite{LSS90, Sch93,
Yam94}:
\begin{eqnarray*}
  && \bigg. \mbox{- type I, IIa and IIb vertices:} \quad \bkl e_{m}^{\pm} ,
  \bkl e_{m+1}^{\pm} , \bkl e_{m}^{\pm} , e_{m-1}^{\pm} \bkr \bkr \bkr = 0
  \\
  && \bigg. \mbox{- type III vertex:} \quad \bkl e_{m}^{\pm} , \bkl
  e_{m+1}^{\pm} , e_{m-1}^{\pm} \bkr \bkr - \bkl e_{m+1}^{\pm} , \bkl
  e_{m}^{\pm} , e_{m-1}^{\pm} \bkr \bkr = 0 \\
  && \bigg. \mbox{- type IV vertex:} \quad \bkl e_{m}^{\pm} , \bkl \bkl
  e_{m+1}^{\pm} , \bkl e_{m}^{\pm} , e_{m-1}^{\pm} \bkr \bkr , \bkl
  e_{m}^{\pm} , e_{m-1}^{\pm} \bkr \bkr \bkr = 0 \\
  && \bigg. \mbox{- type V vertex:} \quad \bkl e_{m}^{\pm} , \bkl
  e_{m-1}^{\pm} , \bkl e_{m}^{\pm} , \bkl e_{m+1}^{\pm} , \bkl e_{m}^{\pm}
  , \bkl e_{m-1}^{\pm} , e_{m-2}^{\pm} \bkr \bkr \bkr \bkr \bkr \bkr = 0 \\
  && \bigg. \mbox{- type VI vertex:} \quad \bkl e_{2}^{\pm} , \bkl
  e_{1}^{\pm} , \bkl e_{3}^{\pm} , \bkl e_{2}^{\pm} , \bkl e_{1}^{\pm} ,
  e_{0}^{\pm} \bkr \bkr \bkr \bkr \bkr - 2 \; \bkl e_{1}^{\pm} , \bkl
  e_{2}^{\pm} , \bkl e_{3}^{\pm} , \bkl e_{2}^{\pm} , \bkl e_{1}^{\pm} ,
  e_{0}^{\pm} \bkr \bkr \bkr \bkr \bkr = 0 \\
  && \bigg. \mbox{- type VII vertex:} \quad 2 \; \bkl e_{2}^{\pm} , \bkl
  e_{1}^{\pm} , \bkl e_{3}^{\pm} , \bkl e_{2}^{\pm} , \bkl e_{1}^{\pm} ,
  e_{0}^{\pm} \bkr \bkr \bkr \bkr \bkr - 3 \; \bkl e_{1}^{\pm} , \bkl
  e_{2}^{\pm} , \bkl e_{3}^{\pm} , \bkl e_{2}^{\pm} , \bkl e_{1}^{\pm} ,
  e_{0}^{\pm} \bkr \bkr \bkr \bkr \bkr = 0
\end{eqnarray*}
For Kac--Moody superalgebras, there are in general many inequivalent simple
root systems (when they contain isotropic odd roots), up to a
transformation of the Weyl group $W(\cG)$ of $\cG$. Following \cite{LSS86},
the Weyl group $W(\cG)$ is extended by adding the following transformations
(called generalized Weyl transformations) associated to the isotropic odd
roots of $\cG$. For $\alpha \in \Delta_\odd$, one defines:
\begin{equation}
  \begin{array}{lll}
    \big. \displaystyle{ w_{\alpha}(\beta) = \beta - 2 \;
    \frac{(\alpha,\beta)}{(\alpha,\alpha)} \; \alpha } &\quad& \mbox{if}
    \;\; (\alpha,\alpha) \ne 0 \\
    \bigg.  w_{\alpha}(\beta) = \beta + \alpha &\quad& \mbox{if} \;\; 
    (\alpha,\alpha) = 0 \;\;\mbox{and}\;\; (\alpha,\beta) \ne 0 
    \\
    \bigg. w_{\alpha}(\beta) = \beta &\quad& \mbox{if} \;\; (\alpha,\alpha)
    = 0 \;\;\mbox{and}\;\; (\alpha,\beta) = 0 \\
    \bigg.  w_{\alpha}(\alpha) = - \alpha &&
  \end{array}
\end{equation}
The transformation associated to an isotropic odd root $\alpha$ cannot be
lifted to an automorphism of the superalgebra since $w_{\alpha}$ transforms
even roots into odd ones, and vice versa, and the $\ZZ_2$-gradation would
not be respected. \\
Let $\Delta^{0}$ be a simple root system of $\cG$ and $\alpha$ an isotropic
odd root. Then one has for any root $\gamma \ne n_{\alpha} \alpha$:
\begin{equation}
  \gamma = \sum_{\ifrac{\beta\ne\alpha\in\Delta^{0}}{a_{\alpha\beta} \ne
  0}} n_{\beta} \,\beta +
  \sum_{\ifrac{\beta\ne\alpha\in\Delta^{0}}{a_{\alpha\beta} = 0}} n_{\beta}
  \,\beta + n_{\alpha} \,\alpha = \sum_{\beta\ne\alpha\in\Delta^{0}}
  n_{\beta} \,w_{\alpha}(\beta) + s_{\gamma} \,w_{\alpha}(\alpha)
\end{equation}
where the coefficients $n_{\alpha}, n_{\beta} \in \ZZ_{\ge 0}$ and 
$s_{\gamma}$ is given by
\begin{equation}
 s_{\gamma} = \sum_{\ifrac{\beta\ne\alpha\in\Delta^{0}}{a_{\alpha\beta} \ne
 0}} n_{\beta} - n_{\alpha}
\end{equation}
Then, by induction on the height of the root $\gamma$, one can prove that
$s_{\gamma}$ is a non negative number, which shows that the transformed
simple root system $w_{\alpha}(\Delta^{0})$ is again a simple root system
\cite{LSS86}. The generalization of the Weyl group gives a method for
constructing all the simple root systems of $\cG$ and hence all the
inequivalent Dynkin diagrams. A simple root system $\Delta^{0}$ being
given, from any isotropic odd root $\alpha \in \Delta^{0}$, one constructs
the simple root system $w_{\alpha}(\Delta^{0})$ where $w_{\alpha}$ is the
generalized Weyl reflection with respect to $\alpha$ and one repeats the
procedure on the obtained system until no new basis arises.

\medskip

Note that this procedure is in fact very general and apply for any
Kac--Moody superalgebra whose simple root systems contain isotropic odd
roots. However, for Kac--Moody superalgebras which are neither of finite
type nor of affine one, one may obtain in certain cases simple root systems
containing even or odd root(s) of very large negative length. We will
comment this point in the next section in the peculiar case of hyperbolic
KM superalgebras.

\section{Hyperbolic Kac--Moody superalgebras}

\subsection{Definition}

Let $\cG(A,\tau)$ be a Kac--Moody superalgebra with generalized Cartan
matrix $A$ and $\ZZ_{2}$-gradation $\tau$. By convention, it will be called
indefinite Kac--Moody superalgebra if it is neither of finite nor of affine
type. Of course, when the $\ZZ_{2}$-gradation $\tau$ is trivial, one
recovers the usual classification of the Kac--Moody algebras.
  
\begin{definition}
  \label{def:HKMSA}
  Let $\cG(A,\tau)$ be an indefinite Kac--Moody superalgebra with
  generalized Cartan matrix $A$ and non trivial $\ZZ_{2}$-gradation $\tau$
  corresponding to a connected Dynkin--Kac diagram. $\cG(A,\tau)$ is called
  a \emph{hyperbolic Kac--Moody} (HKM) superalgebra if every leading
  principal submatrix of $A$ decomposes into constituents of finite or
  affine type, or equivalently, if deleting a vertex of the Dynkin diagram,
  one gets Dynkin diagrams of finite or affine type.
\end{definition}

\medskip

The hyperbolic superalgebras are divided into the following classes:
\begin{enumerate}
  \item
  strictly hyperbolic if every leading principal submatrix of $A$ 
  decomposes into constituents of finite type,
  \item
  purely hyperbolic if every leading principal submatrix of $A$ 
  decomposes into constituents of affine type,
  \item
  hyperfinite if at least one leading principal submatrix of $A$ 
  decomposes into constituents of finite type,
  \item
  hyperaffine if at least one leading principal submatrix of $A$ 
  decomposes into constituents of affine type.
\end{enumerate}

\begin{theorem}
  The hyperbolic Kac--Moody superalgebras of rank $r \ge 3$ are finite in
  number (213) and limited in rank (6). They are listed in Appendix A.
\end{theorem}

\begin{remark}
  As in the algebraic case, the HKM superalgebras are \emph{not} of finite
  growth \cite{Kac68,VdL85}. Let us remind the notion of growth: let
  $\cG(A,\tau)$ be a Kac--Moody superalgebra and $\cI \subset \cG$ a finite
  subset of $\cG$. The growth of $\cG$ is by definition the number
  \begin{equation}
    r(\cG) = \sup_{\cI} \overline{\lim_{n \to \infty}} (\ln d(\cI,n)/\ln n)
    \label{eq:growth}
  \end{equation}
  where $\cI$ runs over all finite subsets of $\cG$ and $d(\cI,n)$ is the
  dimension of the linear span of the commutators of length at most $n$ of
  elements of $\cI$. The superalgebra $\cG$ is of finite growth if $r(\cG)
  < \infty$.
\end{remark}

\medskip

By applying the definition \ref{def:HKMSA}, one gets a Dynkin diagram for a
given HKM superalgebra. The other Dynkin diagrams are obtained by means of
generalized Weyl transformations. Generally, the transformed Dynkin
diagrams do not satisfy the definition \ref{def:HKMSA} (see example below).
We conjecture that it always exists only one Dynkin diagram with the
minimal number of odd roots satisfying definition \ref{def:HKMSA}. Such a
Dynkin diagram will be called distinguished.

\subsection{Example}

As an illustration of the generalized Weyl transformations procedure, we
give below the different inequivalent simple root systems with the
corresponding Dynkin diagrams and symmetric Cartan matrices of a HKM
superalgebra of rank 6. Let
$(\eps_+=K^{+},\eps_-=K^{-},\eps_1,\eps_2,\eps_3,\eps_4,\eps_5=\delta)$ be
a basis of $\RR^{(5,2)}$ with metric $g_{ij} = (\eps_i,\eps_j)$ such that
$g_{+-} = g_{-+} = g_{11} = g_{22} = g_{33} = g_{44} = -g_{55} = 1$ and all
other values are zero. \\
$\bullet$ Simple root system $\Delta^0 = \{ \alpha_1 = \delta-\eps_1-K^{-},
\alpha_2 = \eps_1-\eps_2, \alpha_3 = \eps_2-\eps_3, \alpha_4 =
\eps_3-\eps_4, \alpha_5 = \eps_3+\eps_4, \alpha_6 = K^{+}-\eps_1-\eps_2 \}$
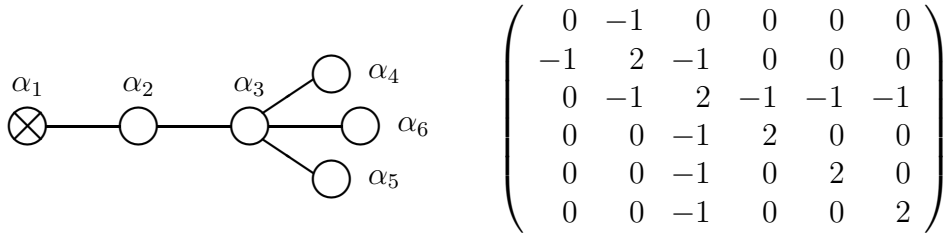
\begin{center}
  \begin{picture}(150,20)
    \thicklines
    \put(0,0){\greydot} \put(42,0){\circle{14}} \put(84,0){\circle{14}} 
    \put(7,0){\line(1,0){28}} \put(49,0){\line(1,0){28}}
    \put(84,0){\vertexdn} \put(126,0){\circle{14}} \put(91,0){\line(1,0){28}}
    \put(0,15){\makebox(0,0){$\alpha_1$}} 
    \put(42,15){\makebox(0,0){$\alpha_2$}} 
    \put(84,15){\makebox(0,0){$\alpha_3$}} 
    \put(135,20){\makebox(0,0){$\alpha_4$}} 
    \put(135,-20){\makebox(0,0){$\alpha_5$}} 
    \put(146,0){\makebox(0,0){$\alpha_6$}} 
  \end{picture}
  \qquad $\left( 
  \begin{array}{rrrrrr} 
    0 & -1 & 0 & 0 & 0 & 0 \\ 
    -1 & 2 & -1 & 0 & 0 & 0 \\
    0 & -1 & 2 & -1 & -1 & -1 \\ 
    0 & 0 & -1 & 2 & 0 & 0 \\
    0 & 0 & -1 & 0 & 2 & 0 \\
    0 & 0 & -1 & 0 & 0 & 2 \\
  \end{array} 
  \right)$
\end{center}
$\bullet$ Simple root system $\Delta^0 = \{ \alpha_1 = K^{-}-\delta+\eps_1,
\alpha_2 = \delta-\eps_2-K^{-}, \alpha_3 = \eps_2-\eps_3, \alpha_4 =
\eps_3-\eps_4, \alpha_5 = \eps_3+\eps_4, \alpha_6 = K^{+}-\eps_1-\eps_2 \}$
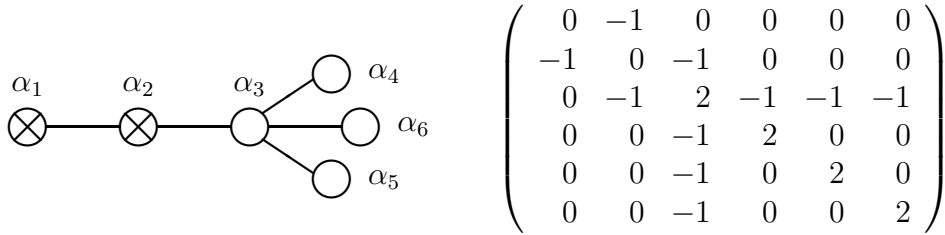
\begin{center}
  \begin{picture}(150,20)
    \thicklines
    \put(0,0){\greydot} \put(42,0){\greydot} \put(84,0){\circle{14}} 
    \put(7,0){\line(1,0){28}} \put(49,0){\line(1,0){28}}
    \put(84,0){\vertexdn} \put(126,0){\circle{14}} \put(91,0){\line(1,0){28}}
    \put(0,15){\makebox(0,0){$\alpha_1$}} 
    \put(42,15){\makebox(0,0){$\alpha_2$}} 
    \put(84,15){\makebox(0,0){$\alpha_3$}} 
    \put(135,20){\makebox(0,0){$\alpha_4$}} 
    \put(135,-20){\makebox(0,0){$\alpha_5$}} 
    \put(146,0){\makebox(0,0){$\alpha_6$}} 
  \end{picture}
  \qquad $\left( 
  \begin{array}{rrrrrr} 
    0 & -1 & 0 & 0 & 0 & 0 \\ 
    -1 & 0 & -1 & 0 & 0 & 0 \\
    0 & -1 & 2 & -1 & -1 & -1 \\ 
    0 & 0 & -1 & 2 & 0 & 0 \\
    0 & 0 & -1 & 0 & 2 & 0 \\
    0 & 0 & -1 & 0 & 0 & 2 \\
  \end{array} 
  \right)$
\end{center}
$\bullet$ Simple root system $\Delta^0 = \{ \alpha_1 = \eps_1-\eps_2,
\alpha_2 = K^{-}-\delta+\eps_2, \alpha_3 = \delta-\eps_3-K^{-}, \alpha_4 =
\eps_3-\eps_4, \alpha_5 = \eps_3+\eps_4, \alpha_6 = K^{+}-\eps_1-\eps_2 \}$
\begin{center}
  \begin{picture}(150,20)
    \thicklines
    \put(0,0){\circle{14}} \put(42,0){\greydot} \put(84,0){\greydot} 
    \put(7,0){\line(1,0){28}} \put(49,0){\line(1,0){28}}
    \put(84,0){\vertexdn} \put(126,0){\circle{14}} \put(91,0){\line(1,0){28}}
    \put(0,15){\makebox(0,0){$\alpha_1$}} 
    \put(42,15){\makebox(0,0){$\alpha_2$}} 
    \put(84,15){\makebox(0,0){$\alpha_3$}} 
    \put(135,20){\makebox(0,0){$\alpha_4$}} 
    \put(135,-20){\makebox(0,0){$\alpha_5$}} 
    \put(146,0){\makebox(0,0){$\alpha_6$}} 
  \end{picture}
  \qquad $\left( 
  \begin{array}{rrrrrr} 
    2 & -1 & 0 & 0 & 0 & 0 \\ 
    -1 & 0 & 1 & 0 & 0 & 0 \\
    0 & -1 & 0 & -1 & -1 & -1 \\ 
    0 & 0 & -1 & 2 & 0 & 0 \\
    0 & 0 & -1 & 0 & 2 & 0 \\
    0 & 0 & -1 & 0 & 0 & 2 \\
  \end{array} 
  \right)$
\end{center}
$\bullet$ Simple root system $\Delta^0 = \{ \alpha_1 = \eps_1-\eps_2,
\alpha_2 = \eps_2-\eps_3, \alpha_3 = K^{-}-\delta+\eps_3, \alpha_4 =
\delta-\eps_4-K^{-}, \alpha_5 = \delta+\eps_4-K^{-}, \alpha_6 =
K^{+}-K^{-}+\delta-\eps_1-\eps_2-\eps_3 \}$
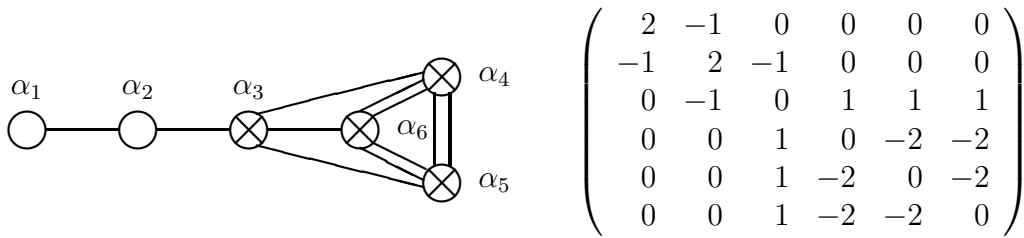
\begin{center}
  \begin{picture}(180,20)
    \thicklines
    \put(0,0){\circle{14}} \put(42,0){\circle{14}} 
    \put(84,0){\greydot} \put(126,0){\greydot}
    \put(7,0){\line(1,0){28}} \put(49,0){\line(1,0){28}}
    \put(91,0){\line(1,0){28}}
    \put(126,0){\vertexgreydn} 
    \put(126,7){\line(2,1){24}}\put(126,-7){\line(2,-1){24}}
    \put(87,6){\line(4,1){63}}\put(87,-6){\line(4,-1){63}}
    \put(0,15){\makebox(0,0){$\alpha_1$}} 
    \put(42,15){\makebox(0,0){$\alpha_2$}} 
    \put(84,15){\makebox(0,0){$\alpha_3$}} 
    \put(177,20){\makebox(0,0){$\alpha_4$}} 
    \put(177,-20){\makebox(0,0){$\alpha_5$}} 
    \put(146,0){\makebox(0,0){$\alpha_6$}} 
  \end{picture}
  \qquad $\left( 
  \begin{array}{rrrrrr} 
    2 & -1 & 0 & 0 & 0 & 0 \\ 
    -1 & 2 & -1 & 0 & 0 & 0 \\
    0 & -1 & 0 & 1 & 1 & 1 \\ 
    0 & 0 & 1 & 0 & -2 & -2 \\
    0 & 0 & 1 & -2 & 0 & -2 \\
    0 & 0 & 1 & -2 & -2 & 0 \\
  \end{array} 
  \right)$
\end{center}
$\bullet$ Simple root system $\Delta^0 = \{ \alpha_1 = \eps_1-\eps_2,
\alpha_2 = \eps_2-\eps_3, \alpha_3 = \eps_3-\eps_4, \alpha_4 =
K^{-}-\delta+\eps_4, \alpha_5 = 2\delta-2K^{-}, \alpha_6 =
K^{+}-2K^{-}+2\delta-\eps_1-\eps_2-\eps_3-\eps_4 \}$
\begin{center}
  \begin{picture}(180,20)
    \thicklines
    \put(0,0){\circle{14}} \put(42,0){\circle{14}} 
    \put(84,0){\circle{14}} \put(126,0){\greydot}
    \put(7,0){\line(1,0){28}} \put(49,0){\line(1,0){28}}
    \put(91,0){\line(1,0){28}}
    \put(126,0){\vertexdn} 
    \put(126,7){\line(3,2){25}}\put(126,-7){\line(3,-2){25}}
    \put(154,-14){\line(0,1){28}} \put(160,-14){\line(0,1){28}}
    \put(157,-13){\line(0,1){26}} 
    \put(0,15){\makebox(0,0){$\alpha_1$}} 
    \put(42,15){\makebox(0,0){$\alpha_2$}} 
    \put(84,15){\makebox(0,0){$\alpha_3$}} 
    \put(177,20){\makebox(0,0){$\alpha_5$}} 
    \put(177,-20){\makebox(0,0){$\alpha_6$}} 
    \put(142,0){\makebox(0,0){$\alpha_4$}} 
  \end{picture}
  \qquad $\left( 
  \begin{array}{rrrrrr} 
    2 & -1 & 0 & 0 & 0 & 0 \\ 
    -1 & 2 & -1 & 0 & 0 & 0 \\
    0 & -1 & 2 & -1 & 0 & 0 \\ 
    0 & 0 & -1 & 0 & 2 & 2 \\
    0 & 0 & 0 & 2 & -4 & -6 \\
    0 & 0 & 0 & 2 & -6 & -4 \\
  \end{array} 
  \right)$
\end{center}

\medskip

The explicit form of the non Serre type supplementary relations for HKM
superalgebras is not known yet, at least for the vertices which are not of
finite nor of affine type. However, another alternative way of describing
the HKM superalgebras is to consider \emph{all} inequivalent Dynkin
diagrams and write the usual Serre relations
(\ref{eq:serre1})--(\ref{eq:serre2}) (of course this leads to redundant
information). Indeed, the non Serre type relations become Serre relations
after a generalized Weyl reflection with respect to an appropriate
isotropic odd root \cite{GL97}.

Note nevertheless that in the case of HKM superalgebras, one may produce by
generalized Weyl transformations some exotic simple root systems
corresponding to Cartan matrices with non integer (rational) entries
associated to weird non standard Dynkin diagrams (this was \emph{not} the
case in the above example). As in the usual case, one gets supplementary
non Serre type relations, but now also associated to non isotropic even
simple roots.

\subsection{Rank two HKM superalgebras}

Clearly any rank 2 HKM superalgebra is described by a Dynkin--Kac diagram of
the form
\begin{center}
\begin{picture}(50,0) \thicklines 
  \put(0,0){\circle*{7}} \put(42,0){\circle*{7}} \put(7,0){\line(1,0){28}} 
 \end{picture}
\end{center}
where the dot can be a white, black or grey dot. Both dots cannot be either
white, as this diagram describes KM algebras, or grey, as this diagram is
isomorphic to the diagram, in the non distinguished basis, of $sl(1/2)$.
The Dynkin--Kac diagrams corresponding to rank 2 finite and affine KM
superalgebras, up to generalized Weyl transformations, are listed below:

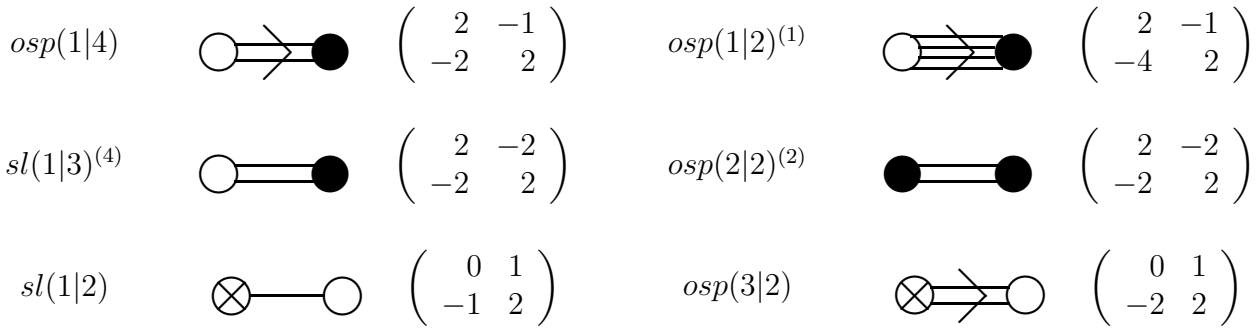
\begin{tabular}{ccccccc}
$osp(1|4)$ 
&\hspace*{12pt}&
\begin{picture}(50,20) \thicklines
  \put(0,0){\circle{14}} \put(42,0){\circle*{14}} \put(0,0){\vertexbn}
\end{picture}
\quad $\left( 
\begin{array}{rr} 
  2 & -1 \\ 
  -2 & 2 \\
\end{array} 
\right)$
&\hspace*{12pt}&
$osp(1|2)^{(1)}$ 
&\hspace*{12pt}&
\begin{picture}(50,20) \thicklines
  \put(0,0){\circle{14}} \put(42,0){\circle*{14}} \put(0,0){\vertexbbn}
\end{picture}
\quad $\left( 
\begin{array}{rr} 
  2 & -1 \\ 
  -4 & 2 \\
\end{array} 
\right)$
\\
\\
$sl(1|3)^{(4)}$ 
&\hspace*{12pt}&
\begin{picture}(50,20) \thicklines
  \put(0,0){\circle{14}} \put(42,0){\circle*{14}} 
  \put(6,-3){\line(1,0){30}}\put(6,3){\line(1,0){30}}
\end{picture}
\quad $\left( 
\begin{array}{rr} 
  2 & -2 \\ 
  -2 & 2 \\
\end{array} 
\right)$
&\hspace*{12pt}&
$osp(2|2)^{(2)}$ 
&\hspace*{12pt}&
\begin{picture}(50,20) \thicklines
  \put(0,0){\circle*{14}} \put(42,0){\circle*{14}} 
  \put(6,-3){\line(1,0){30}}\put(6,3){\line(1,0){30}}
\end{picture}
\quad $\left( 
\begin{array}{rr} 
  2 & -2 \\ 
  -2 & 2 \\
\end{array} 
\right)$
\\
\\
$sl(1|2)$
&\hspace*{12pt}&
\begin{picture}(50,20) \thicklines
  \put(0,0){\greydot} \put(42,0){\circle{14}} \put(7,0){\line(1,0){28}}
\end{picture}
\quad $\left( 
\begin{array}{rr} 
  0 & 1 \\ 
  -1 & 2 \\
\end{array} 
\right)$
&\hspace*{12pt}&
$osp(3|2)$
&\hspace*{12pt}&
\begin{picture}(50,20) \thicklines
  \put(0,0){\greydot} \put(42,0){\circle{14}} \put(0,0){\vertexbn}
\end{picture}
\quad $\left( 
\begin{array}{rr} 
  0 & 1 \\ 
  -2 & 2 \\
\end{array} 
\right)$
\end{tabular}

\vspace*{12pt}

\noindent In \cite{Kac68,VdL85}, it is proven that any superalgebra
associated to a $2 \!\times\! 2$ matrix not appearing in the above list is
of \emph{infinite growth}. It follows:

\begin{theorem}
  The hyperbolic Kac--Moody superalgebras of rank two are infinite in
  number. Their generalized Cartan matrix and Dynkin diagram are, up to
  generalized Weyl transformations, reducible to one of the following list:
  \\
  $\bullet$ with $\ZZ_{2}$-gradation $\tau = \{1\}$
  \\
  \\
  \begin{tabular}{l}
  $\left( 
  \begin{array}{rr} 
    2 & -k \\ 
    -k' & 2 \\
  \end{array} 
  \right)$ \quad with $(k,k') = (1,k')$, $k' \le 4$, $(k,k') = (3,1)$ or
  $k,k' \in \ZZ_{>0}$, $kk' > 4$
  \\
  \\
  $\left( 
  \begin{array}{rr} 
    0 & 1 \\ 
    -k & 2 \\
  \end{array} 
  \right)$ \quad with $k \in \ZZ_{>0}$, $k > 2$
  \end{tabular}
  \\
  \\
  $\bullet$ with $\ZZ_{2}$-gradation $\tau = \{1,2\}$
  \\
  \\
  \begin{tabular}{l}
    $\left( 
    \begin{array}{rr} 
      2 & -k \\ 
      -k' & 2 \\
    \end{array} 
    \right)$ \quad with $(k,k') = (1,k')$, $k' \le 4$ or $k,k' \in
    \ZZ_{>0}$, $kk' > 4$ 
    \\
    \\
    $\left( 
    \begin{array}{rr} 
      0 & 1 \\ 
      -k & 2 \\
    \end{array} 
    \right)$ \quad with $k \in \ZZ_{>0}$, $k \ne 2$
  \end{tabular}
  \\
  \\
  We denote them as $BW(k,k')$, $GW(k)$, $BB(k,k')$ and $GB(k)$, with
  corresponding Dynkin diagrams:
  
  \begin{center}
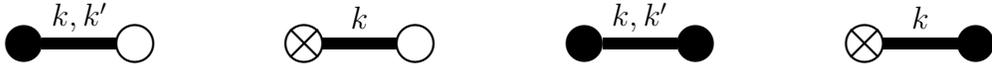
  \begin{tabular}{lllllll}
    \begin{picture}(70,0) \thicklines 
      \put(0,0){\circle*{14}} \put(42,0){\circle{14}}
      \linethickness{4pt} \put(7,0){\line(1,0){28}} 
      \put(21,10){\makebox(0,0){$k,k'$}}
    \end{picture}
    &\hspace*{12pt}&
    \begin{picture}(70,0) \thicklines
      \put(0,0){\greydot} \put(42,0){\circle{14}} 
      \linethickness{4pt} \put(7,0){\line(1,0){28}}
      \put(21,10){\makebox(0,0){$k$}}
    \end{picture}
    &\hspace*{12pt}&
    \begin{picture}(70,0)
      \thicklines \put(0,0){\circle*{14}} \put(42,0){\circle*{14}}
      \linethickness{4pt} \put(7,0){\line(1,0){28}} 
      \put(21,10){\makebox(0,0){$k,k'$}}
    \end{picture}
    &\hspace*{12pt}&
    \begin{picture}(70,0) \thicklines
      \put(0,0){\greydot} \put(42,0){\circle*{14}} 
      \linethickness{4pt} \put(7,0){\line(1,0){28}}
      \put(21,10){\makebox(0,0){$k$}}
    \end{picture}
  \end{tabular}
  \end{center}
\end{theorem}

\noindent
Note that in order to write only one type of diagram, we have not strictly 
followed the rules given in section 2. \\
The simple root systems of the HKM superalgebras $BW(k,k')$ and $BB(k,k')$ 
are given by
\begin{equation}
  \alpha = \sum_{i=1}^{n} k_{i} \, \eps_{i} + k_{\alpha} \, K^{+} \qquad
  \mbox{and} \qquad \alpha' = -\sum_{i=1}^{n'} k'_{i} \, \eps_{i} -
  k'_{\alpha} \, K^{-}
  \label{eq:}
\end{equation}
where $k_{i}, k'_{i}, k_{\alpha}, k'_{\alpha} \in \ZZ_{\ge 0}$ satisfy
$\displaystyle kk' = k_{\alpha} k'_{\alpha} + \sum_{i=1}^{\min(n,n')} k_{i}
k'_{i}$, $\displaystyle k^2 = \sum_{i=1}^{n} k_{i}^{2}$ and $\displaystyle
{k'}^2 = \sum_{i=1}^{n'} {k'}_{i}^{2}$, while for the HKM superalgebras
$GW(k)$ and $GB(k)$ the simple roots are $\alpha = -k \, K^{-}$ and 
$\alpha' = \eps_{1}-\eps_{2}+K^{+}$.

\section{Subalgebras of hyperbolic KM superalgebras}

Let $\cG$ be a Kac--Moody superalgebra, and consider its canonical root
decomposition
\begin{displaymath}
  \cG = \cH \oplus \bigoplus_{\alpha \in \Delta} \cG_{\alpha}
\end{displaymath}
where $\cH$ is the Cartan subalgebra of $\cG$ and $\Delta$ its 
corresponding root system.  \\
A sub(super)algebra $\cG'$ of $\cG$ is called regular if $\cG'$ has the
root decomposition
\begin{displaymath}
  \cG' = \cH' \oplus \bigoplus_{\alpha' \in \Delta'} \cG'_{\alpha'}
\end{displaymath}
where $\cH' \subset \cH$ and $\Delta \subset \Delta'$.

Consider a HKM superalgebra $\cG$. Deleting a dot in the distinguished
Dynkin diagram of $\cG$ leads to regular sub(super)algebras of $\cG$ of
finite or affine type by definition. In Appendix C, we list these regular
sub(super)algebras corresponding to the Dynkin diagrams of Appendix A. 
Note that in several cases, the diagram of the subsuperalgebra is not the 
distinguished one.

\medskip

A sub(super)algebra $\cG'$ of $\cG$ is called singular if it is not
regular. The folding method allows one to obtain some singular
sub(super)algebras of the HKM superalgebras. Let $\cG$ be a HKM
superalgebra with a distinguished Dynkin diagram exhibiting a $\ZZ_{N}$
symmetry. This $\ZZ_{N}$ symmetry is generated by an automorphism $\tau$
of order $N$ ($\tau^N = 1$) acting on the root system. The automorphism
$\tau$ can be lifted at the algebra level by setting $\tau(e_{\alpha}) =
e_{\tau(\alpha)}$ for a generator $e_{\alpha}$ associated to a simple root
$\alpha$. The symmetry of the Dynkin diagram induces a direct construction
of the sub(super)algebra $\cG'$ invariant under the $\cG$ 
automorphism associated to $\tau$. Indeed, if the simple root $\alpha$ is
transformed into $\tau(\alpha)$, then $\alpha' = \alpha + \tau(\alpha) +
\ldots + \tau^{N-1}(\alpha)$ is $\tau$-invariant since $\tau^N = 1$, and
appears as a simple root of $\cG'$ associated to the generator $e_{\alpha'}
= e_{\alpha} + e_{\tau(\alpha)} + \ldots + e_{\tau^{N-1}(\alpha)}$, where
$e_{\tau^k(\alpha)}$ is the generator corresponding to the root
$\tau^k(\alpha)$ ($k=0,\ldots,N-1$). A Dynkin diagram of $\cG'$ will
therefore be obtained by folding the $\ZZ_N$-symmetric Dynkin diagram of
$\cG$, that is by transforming each $N$-uple $(\alpha, \tau(\alpha), \dots,
\tau^{N-1}(\alpha))$ into the root $\alpha' = \alpha + \tau(\alpha) +
\ldots + \tau^{N-1}(\alpha)$ of $\cG'$. It is easy to convince oneself that
for $\cG'$ the defining relations (\ref{eq:serre1})--(\ref{eq:serrerel}) of
a HKM superalgebra hold (be aware that, in particular for the Serre
relations (\ref{eq:serrerel}), the entries of the Cartan matrix are now
those of $\cG'$).

\medskip

We present in Table 1 the list of HKM superalgebras $\cG$ to which the
folding procedure can be applied and the corresponding singular
subsuperalgebras $\cG'$. Note that in general the obtained singular
subsuperalgebras are also HKM superalgebras. However, in the case of the
HKM superalgebra \#6 of rank 6, one obtains for $\cG'$ the simple Lie
superalgebra $F(4)$ (note that for affine Lie (super)algebras the folding
procedure always leads to (super)algebras of affine type). This is due to
the fact that for HKM superalgebras the root system contains \emph{two}
isotropic roots whose scalar product is not trivial.

\begin{remark}
  The folding procedure cannot be applied to the rank four HKM
  superalgebras labelled by the numbers \#59 to \#62, despite the apparent
  $\ZZ_{2}$-symmetry of the distinguished Dynkin diagram, as the
  $\ZZ_{2}$-grading of the invariant generators would not be respected.
\end{remark}

\begin{center}
  Table 1: Folding of the HKM superalgebras
\begin{tabular}{ccccc}
    \Big. rank of $\cG$ & $\cG$ label & order of $\tau$ & rank of $\cG'$ &
    $\cG'$ label \\
    3 & \#24 & 2 & 2 & $BW(4,2)$ \\
    3 & \#25 & 2 & 2 & $BW(2,4)$ \\
    3 & \#27 & 2 & 2 & $BB(2,4)$ \\
    3 & \#28 & 2 & 2 & $BB(4,2)$ \\
    3 & \#31 & 2 & 2 & $BW(8,1)$ \\
    4 & \#33 & 2 & 4 & \#37 \\
    4 & \#34 & 2 & 4 & \#39 \\
    4 & \#40 & 2 & 3 & \#57 \\
    4 & \#41 & 2 & 3 & \#5 \\
    4 & \#42 & 2 & 3 & \#21 \\
    4 & \#44 & 2 & 3 & \#26 \\
    4 & \#45 & 2 & 3 & \#15 \\
    4 & \#46 & 2 & 3 & \#13 \\
    4 & \#47 & 2 & 3 & \#14 \\
    4 & \#48 & 2 & 3 & \#34 \\
    4 & \#55 & 2 & 3 & \#52 \\
    4 & \#56 & 2 & 3 & \#54 \\
    4 & \#57 & 2 & 3 & \#53 \\
    4 & \#58 & 2 & 3 & \#47 \\
  \end{tabular}
\end{center}

\newpage

\begin{center}
  Table 1: Folding of the HKM superalgebras (cont'd)
  \begin{tabular}{ccccc}
    \Big. rank of $\cG$ & $\cG$ label & order of $\tau$ & rank of $\cG'$ &
    $\cG'$ label \\
    4 & \#63 & 2 & 3 & \#37 \\
    4 & \#64 & 2 & 3 & \#5 \\
    4 & \#65 & 2 & 3 & \#1 \\
    4 & \#66 & 2 & 3 & \#9 \\
    4 & \#67 & 2 & 3 & \#6 \\
    4 & \#68 & 2 & 3 & \#24 \\
    4 & \#68 & 2 & 4 & \#28 \\
    4 & \#69 & 2 & 4 & \#7 \\
    4 & \#70 & 2 & 4 & \#42 \\
    4 & \#71 & 2 & 4 & \#51 \\
    4 & \#72 & 2 & 4 & \#50 \\
    4 & \#73 & 2 & 4 & \#58 \\
    5 & \#28 & 2 & 4 & \#14 \\
    5 & \#30 & 2 & 4 & \#8 \\
    5 & \#31 & 2 & 4 & \#9 \\
    5 & \#32 & 2 & 4 & \#11 \\
    5 & \#33 & 2 & 4 & \#5 \\
    5 & \#34 & 2 & 4 & \#7 \\
    5 & \#35 & 3 & 3 & \#2 \\
    5 & \#36 & 2 & 4 & \#1 \\
    5 & \#37 & 2 & 4 & \#13 \\
    5 & \#38 & 3 & 3 & \#36 \\
    6 & \#6 & 2 & 4 & F(4) \\
    6 & \#12 & 2 & 4 & \#1 \\
    6 & \#14 & 3 & 4 & \#21 \\
    6 & \#15 & 3 & 4 & \#26 \\
  \end{tabular}
\end{center}

\appendix

\section*{A. Dynkin diagrams of the hyperbolic superalgebras}

\subsubsection*{Rank 3 hyperbolic superalgebras (87 diagrams)}

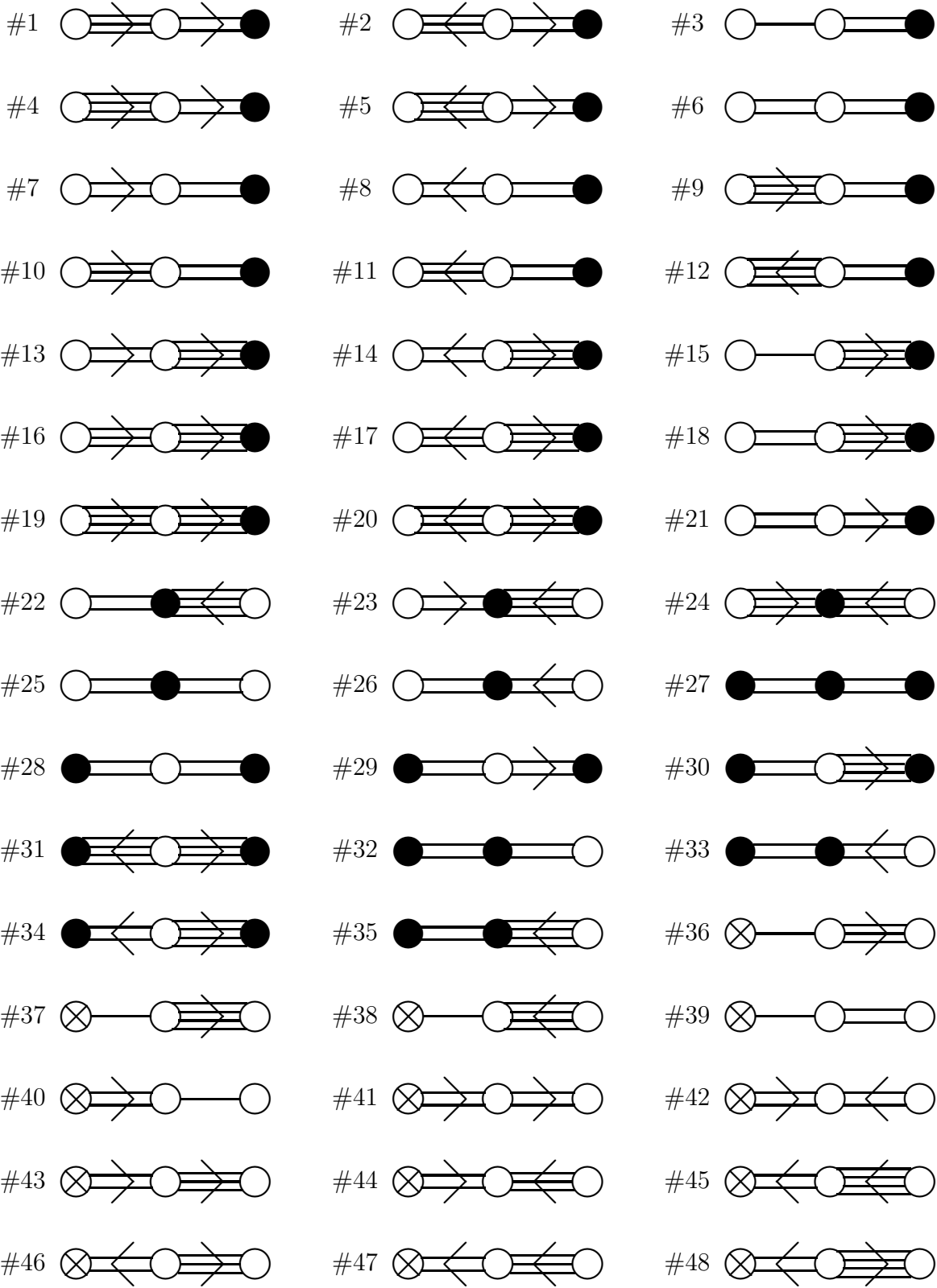
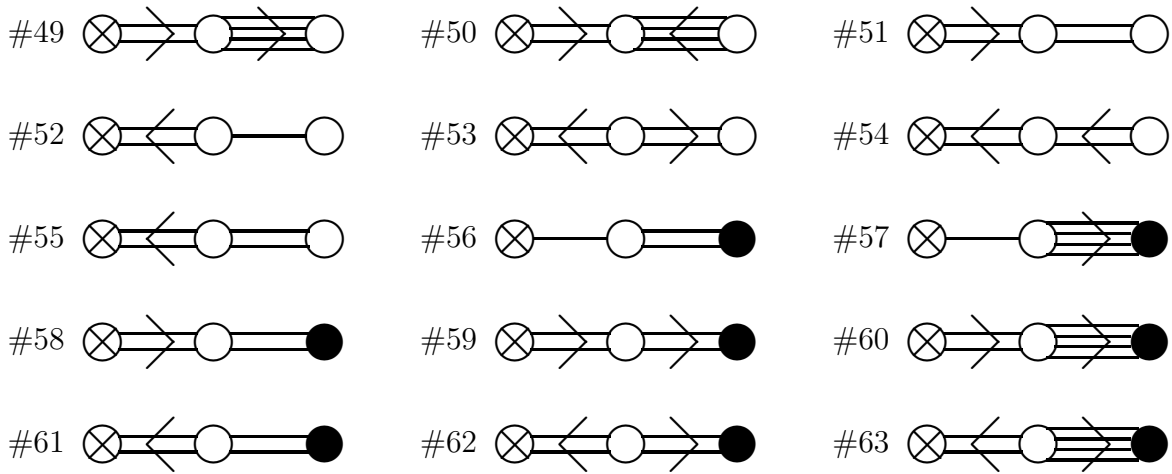
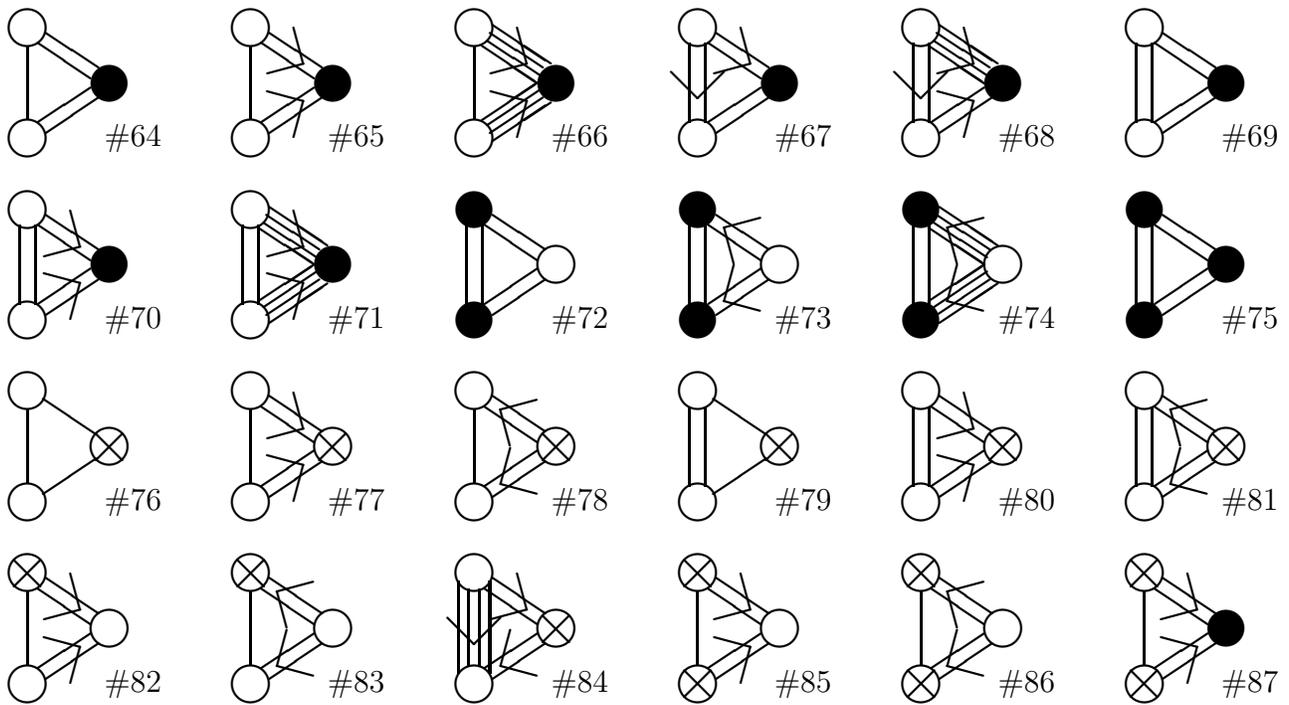
\begin{center}
\begin{tabular}{ccccc}
\begin{picture}(90,20) \thicklines 
  \put(0,0){\circle{14}} \put(42,0){\circle{14}} \put(84,0){\circle*{14}}
  \put(0,0){\vertexgbn} \put(42,0){\vertexbn}
  \put(-25,0){\makebox(0,0){\#1}}
\end{picture}
&\hspace*{42pt}& 
\begin{picture}(90,20) \thicklines 
  \put(0,0){\circle{14}} \put(42,0){\circle{14}} \put(84,0){\circle*{14}}
  \put(0,0){\vertexgcn} \put(42,0){\vertexbn}
  \put(-25,0){\makebox(0,0){\#2}}
\end{picture}
&\hspace*{42pt}& 
\begin{picture}(90,20) \thicklines 
  \put(0,0){\circle{14}} \put(42,0){\circle{14}} \put(84,0){\circle*{14}}
  \put(7,0){\line(1,0){28}}
  \put(48,-3){\line(1,0){30}}\put(48,3){\line(1,0){30}}
  \put(-25,0){\makebox(0,0){\#3}}
\end{picture}
\\
\\
\begin{picture}(90,20) \thicklines 
  \put(0,0){\circle{14}} \put(42,0){\circle{14}} \put(84,0){\circle*{14}}
  \put(0,0){\vertexbbn} \put(42,0){\vertexbn}
  \put(-25,0){\makebox(0,0){\#4}}
\end{picture}
&\hspace*{42pt}& 
\begin{picture}(90,20) \thicklines 
  \put(0,0){\circle{14}} \put(42,0){\circle{14}} \put(84,0){\circle*{14}}
  \put(0,0){\vertexccn} \put(42,0){\vertexbn}
  \put(-25,0){\makebox(0,0){\#5}}
\end{picture}
&\hspace*{42pt}& 
\begin{picture}(90,20) \thicklines 
  \put(0,0){\circle{14}} \put(42,0){\circle{14}} \put(84,0){\circle*{14}} 
  \put(6,-3){\line(1,0){30}}\put(6,3){\line(1,0){30}}
  \put(48,-3){\line(1,0){30}}\put(48,3){\line(1,0){30}}
  \put(-25,0){\makebox(0,0){\#6}}
\end{picture}
\\
\\
\end{tabular}

\begin{tabular}{ccccc}
\begin{picture}(90,20) \thicklines 
  \put(0,0){\greydot} \put(42,0){\circle{14}} \put(84,0){\circle{14}} 
  \put(0,0){\vertexbn} \put(42,0){\vertexbbn}
  \put(-25,0){\makebox(0,0){\#49}}
\end{picture}
&\hspace*{42pt}& 
\begin{picture}(90,20) \thicklines 
  \put(0,0){\greydot} \put(42,0){\circle{14}} \put(84,0){\circle{14}} 
  \put(0,0){\vertexbn} \put(42,0){\vertexccn}
  \put(-25,0){\makebox(0,0){\#50}}
\end{picture}
&\hspace*{42pt}& 
\begin{picture}(90,20) \thicklines 
  \put(0,0){\greydot} \put(42,0){\circle{14}} \put(84,0){\circle{14}} 
  \put(0,0){\vertexbn} 
  \put(48,-3){\line(1,0){30}}\put(48,3){\line(1,0){30}}
  \put(-25,0){\makebox(0,0){\#51}}
\end{picture}
\\
\\
\begin{picture}(90,20) \thicklines 
  \put(0,0){\greydot} \put(42,0){\circle{14}} \put(84,0){\circle{14}} 
  \put(0,0){\vertexcn} \put(49,0){\line(1,0){28}}
  \put(-25,0){\makebox(0,0){\#52}}
\end{picture}
&\hspace*{42pt}& 
\begin{picture}(90,20) \thicklines 
  \put(0,0){\greydot} \put(42,0){\circle{14}} \put(84,0){\circle{14}} 
  \put(0,0){\vertexcn} \put(42,0){\vertexbn}
  \put(-25,0){\makebox(0,0){\#53}}
\end{picture}
&\hspace*{42pt}& 
\begin{picture}(90,20) \thicklines 
  \put(0,0){\greydot} \put(42,0){\circle{14}} \put(84,0){\circle{14}} 
  \put(0,0){\vertexcn} \put(42,0){\vertexcn}
  \put(-25,0){\makebox(0,0){\#54}}
\end{picture}
\\
\\
\end{tabular}
\end{center}

\begin{tabular}{ccccccccccc}
\begin{picture}(50,50) \thicklines 
  \put(0,21){\circle{14}} \put(0,-21){\circle{14}} \put(31,0){\circle*{14}}
  \put(4,15){\line(3,-2){20}} \put(7,19){\line(3,-2){20}}
  \put(4,-15){\line(3,2){20}} \put(7,-19){\line(3,2){20}}
  \put(0,-14){\line(0,1){28}}
  \put(40,-21){\makebox(0,0){\#64}}
\end{picture}
&\hspace*{10.5pt}& 
\begin{picture}(50,50) \thicklines 
  \put(0,21){\circle{14}} \put(0,-21){\circle{14}} \put(31,0){\circle*{14}}
  \put(4,15){\line(3,-2){20}} \put(7,19){\line(3,-2){20}}
  \put(4,-15){\line(3,2){20}} \put(7,-19){\line(3,2){20}}
  \put(0,-14){\line(0,1){28}} \put(20,7){\arrowse} \put(20,-7){\arrowne}
  \put(40,-21){\makebox(0,0){\#65}}
\end{picture}
&\hspace*{10.5pt}& 
\begin{picture}(50,50) \thicklines 
  \put(0,21){\circle{14}} \put(0,-21){\circle{14}} \put(31,0){\circle*{14}}
  \put(0,-14){\line(0,1){28}} \put(3,14){\line(3,-2){21}}
  \put(5,16){\line(3,-2){20}} \put(6,19){\line(3,-2){20}}
  \put(7,22){\line(3,-2){22}} \put(3,-14){\line(3,2){21}}
  \put(5,-16){\line(3,2){20}} \put(6,-19){\line(3,2){20}}
  \put(7,-22){\line(3,2){22}} \put(20,7){\arrowse} \put(20,-7){\arrowne}
  \put(40,-21){\makebox(0,0){\#66}}
\end{picture}
&\hspace*{10.5pt}& 
\begin{picture}(50,50) \thicklines 
  \put(0,21){\circle{14}} \put(0,-21){\circle{14}} \put(31,0){\circle*{14}}
  \put(0,21){\vertexdowndbl} \put(4,15){\line(3,-2){20}}
  \put(7,19){\line(3,-2){20}} \put(4,-15){\line(3,2){20}}
  \put(7,-19){\line(3,2){20}} \put(20,7){\arrowse}
  \put(40,-21){\makebox(0,0){\#67}}
\end{picture}
&\hspace*{10.5pt}& 
\begin{picture}(50,50) \thicklines 
  \put(0,21){\circle{14}} \put(0,-21){\circle{14}} \put(31,0){\circle*{14}}
  \put(0,21){\vertexdowndbl} \put(3,14){\line(3,-2){21}}
  \put(5,16){\line(3,-2){20}} \put(6,19){\line(3,-2){20}}
  \put(7,22){\line(3,-2){22}} \put(4,-15){\line(3,2){20}}
  \put(7,-19){\line(3,2){20}} \put(20,7){\arrowse} \put(20,-7){\arrowne}
  \put(40,-21){\makebox(0,0){\#68}}
\end{picture}
&\hspace*{10.5pt}& 
\begin{picture}(50,50) \thicklines 
  \put(0,21){\circle{14}} \put(0,-21){\circle{14}} \put(31,0){\circle*{14}}
  \put(4,15){\line(3,-2){20}} \put(7,19){\line(3,-2){20}}
  \put(4,-15){\line(3,2){20}} \put(7,-19){\line(3,2){20}}
  \put(-3,-15){\line(0,1){30}}\put(3,-15){\line(0,1){30}}
  \put(40,-21){\makebox(0,0){\#69}}
\end{picture}
\\
\end{tabular}

\clearpage

\subsubsection*{Rank 4 hyperbolic superalgebras (73 diagrams)}

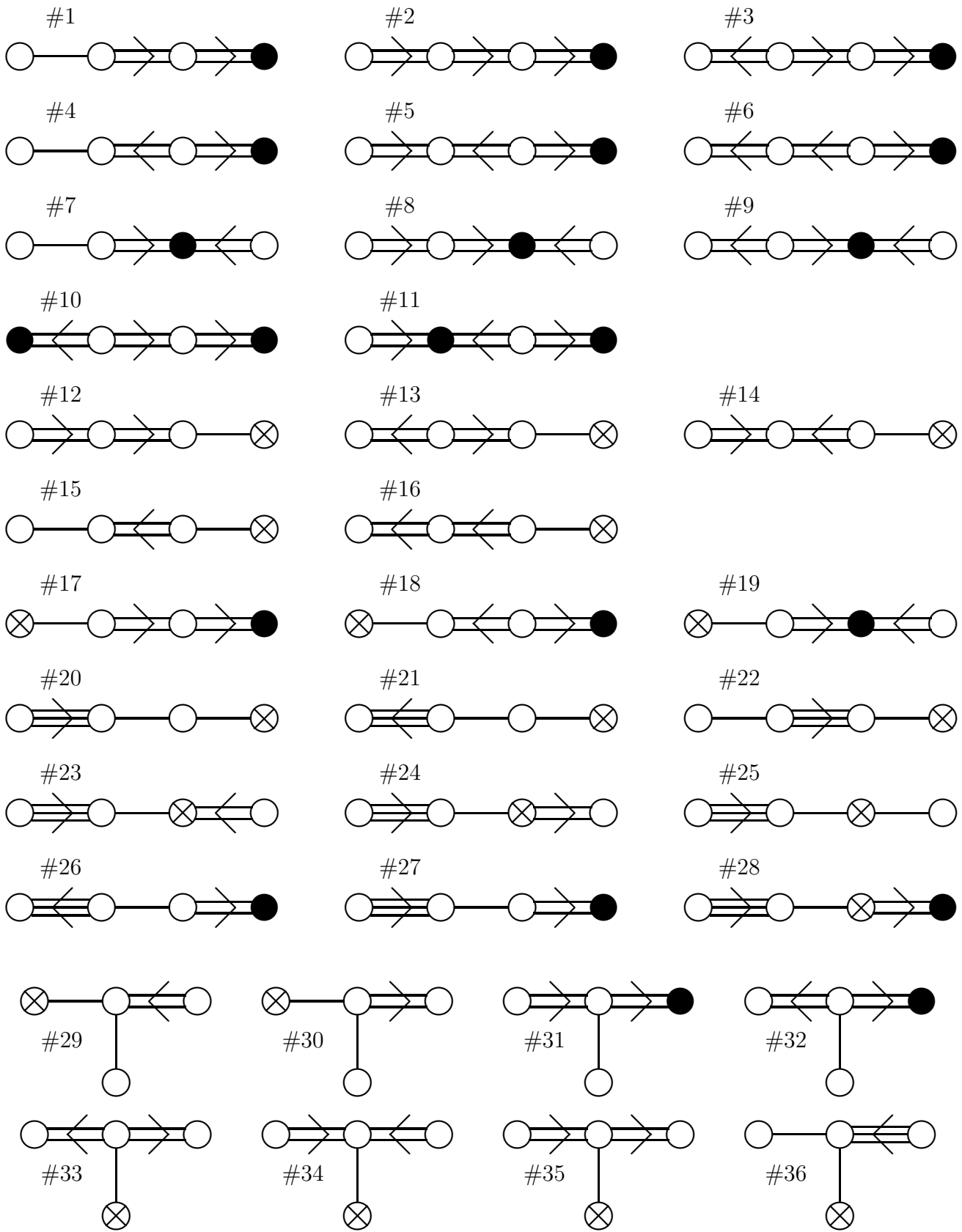
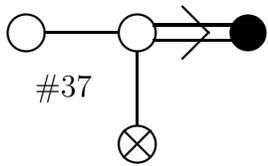
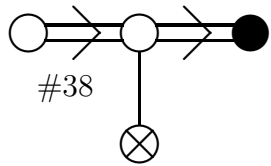
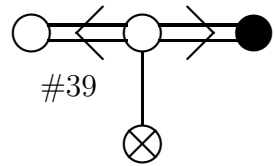
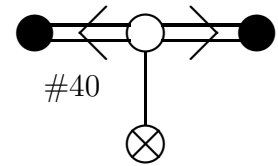
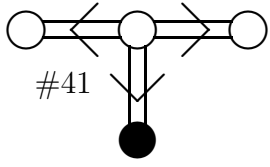
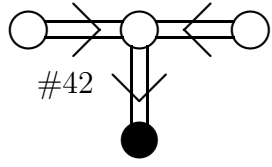
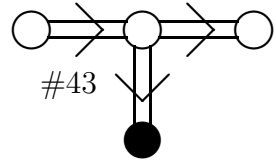
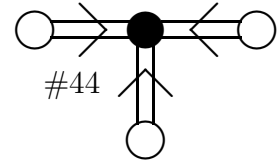
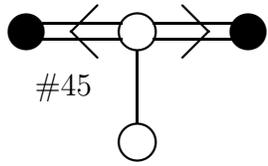
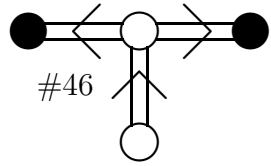
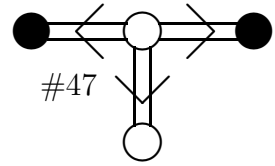
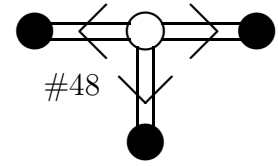
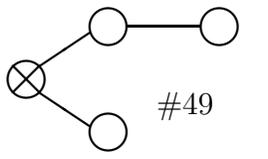
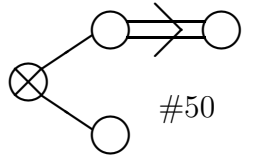
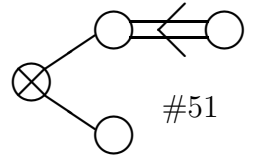
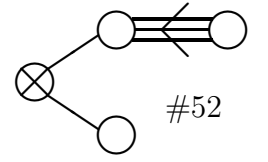
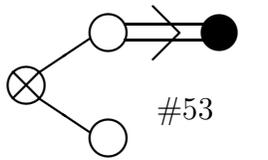
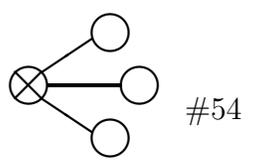
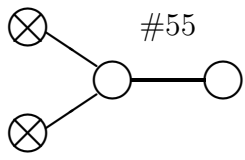
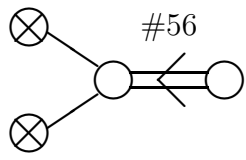
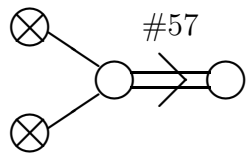
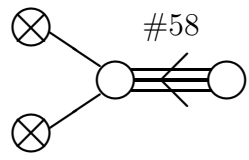
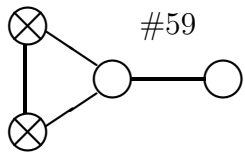
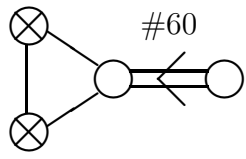
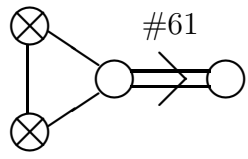
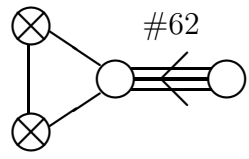
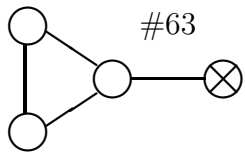
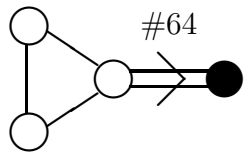
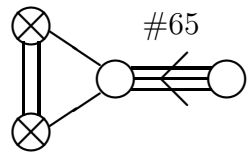
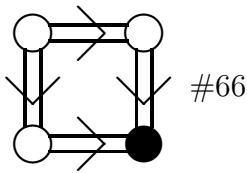
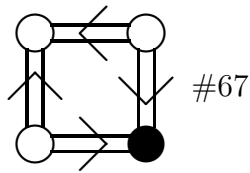
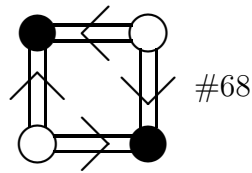
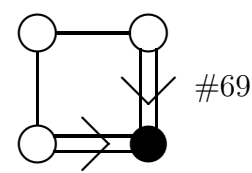
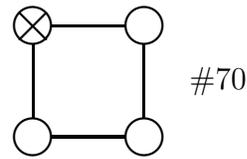
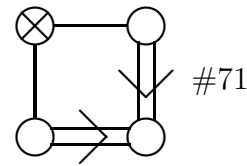
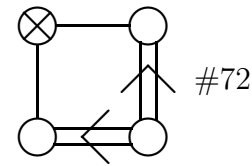
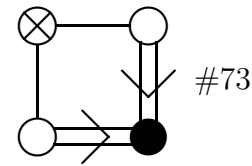
\begin{center}
\begin{tabular}{ccccc}
\begin{picture}(130,30) \thicklines 
  \put(0,0){\circle{14}} \put(42,0){\circle{14}} \put(84,0){\circle{14}}
  \put(126,0){\circle*{14}} \put(7,0){\line(1,0){28}} \put(42,0){\vertexbn}
  \put(84,0){\vertexbn}
  \put(21,20){\makebox(0,0){\#1}}
\end{picture}
&\hspace*{21pt}& 
\begin{picture}(130,30) \thicklines 
  \put(0,0){\circle{14}} \put(42,0){\circle{14}} \put(84,0){\circle{14}}
  \put(126,0){\circle*{14}} \put(0,0){\vertexbn} \put(42,0){\vertexbn}
  \put(84,0){\vertexbn}
  \put(21,20){\makebox(0,0){\#2}}
\end{picture}
&\hspace*{21pt}& 
\begin{picture}(130,30) \thicklines 
  \put(0,0){\circle{14}} \put(42,0){\circle{14}} \put(84,0){\circle{14}}
  \put(126,0){\circle*{14}} \put(0,0){\vertexcn} \put(42,0){\vertexbn}
  \put(84,0){\vertexbn}
  \put(21,20){\makebox(0,0){\#3}}
\end{picture}
\\
\\
\begin{picture}(130,30) \thicklines 
  \put(0,0){\circle{14}} \put(42,0){\circle{14}} \put(84,0){\circle{14}}
  \put(126,0){\circle*{14}} \put(7,0){\line(1,0){28}} \put(42,0){\vertexcn}
  \put(84,0){\vertexbn}
  \put(21,20){\makebox(0,0){\#4}}
\end{picture}
&\hspace*{21pt}& 
\begin{picture}(130,30) \thicklines 
  \put(0,0){\circle{14}} \put(42,0){\circle{14}} \put(84,0){\circle{14}}
  \put(126,0){\circle*{14}} \put(0,0){\vertexbn} \put(42,0){\vertexcn}
  \put(84,0){\vertexbn}
  \put(21,20){\makebox(0,0){\#5}}
\end{picture}
&\hspace*{21pt}& 
\begin{picture}(130,30) \thicklines 
  \put(0,0){\circle{14}} \put(42,0){\circle{14}} \put(84,0){\circle{14}}
  \put(126,0){\circle*{14}} \put(0,0){\vertexcn} \put(42,0){\vertexcn}
  \put(84,0){\vertexbn}
  \put(21,20){\makebox(0,0){\#6}}
\end{picture}
\\
\\
\end{tabular}
\end{center}

\begin{tabular}{ccccccc}
\begin{picture}(90,30) \thicklines 
  \put(0,0){\greydot} \put(42,0){\circle{14}} \put(84,0){\circle{14}}
  \put(7,0){\line(1,0){28}} \put(42,0){\vertexcn}
  \put(42,-42){\circle{14}}\put(42,-7){\line(0,-1){28}}
  \put(14,-21){\makebox(0,0){\#29}}  
\end{picture}
&\hspace*{10.5pt}& 
\begin{picture}(90,30) \thicklines 
  \put(0,0){\greydot} \put(42,0){\circle{14}} \put(84,0){\circle{14}}
  \put(7,0){\line(1,0){28}} \put(42,0){\vertexbn}
  \put(42,-42){\circle{14}}\put(42,-7){\line(0,-1){28}}
  \put(14,-21){\makebox(0,0){\#30}}  
\end{picture}
&\hspace*{10.5pt}& 
\begin{picture}(90,30) \thicklines 
  \put(0,0){\circle{14}} \put(42,0){\circle{14}} \put(84,0){\circle*{14}}
  \put(0,0){\vertexbn} \put(42,0){\vertexbn}
  \put(42,-42){\circle{14}}\put(42,-7){\line(0,-1){28}}
  \put(14,-21){\makebox(0,0){\#31}}  
\end{picture}
&\hspace*{10.5pt}& 
\begin{picture}(90,30) \thicklines 
  \put(0,0){\circle{14}} \put(42,0){\circle{14}} \put(84,0){\circle*{14}}
  \put(0,0){\vertexcn} \put(42,0){\vertexbn}
  \put(42,-42){\circle{14}}\put(42,-7){\line(0,-1){28}}
  \put(14,-21){\makebox(0,0){\#32}}  
\end{picture}
\\
\\
\end{tabular}

\begin{tabular}{ccccccc}
\begin{picture}(90,0) \thicklines 
  \put(0,0){\circle{14}} \put(42,0){\circle{14}} \put(84,0){\circle*{14}}
  \put(7,0){\line(1,0){28}} \put(42,0){\vertexbn}
  \put(42,-42){\greydot}\put(42,-7){\line(0,-1){28}}
  \put(14,-21){\makebox(0,0){\#37}}  
\end{picture}
&\hspace*{10.5pt}& 
\begin{picture}(90,0) \thicklines 
  \put(0,0){\circle{14}} \put(42,0){\circle{14}} \put(84,0){\circle*{14}}
  \put(0,0){\vertexbn} \put(42,0){\vertexbn}
  \put(42,-42){\greydot}\put(42,-7){\line(0,-1){28}}
  \put(14,-21){\makebox(0,0){\#38}}  
\end{picture}
&\hspace*{10.5pt}& 
\begin{picture}(90,0) \thicklines 
  \put(0,0){\circle{14}} \put(42,0){\circle{14}} \put(84,0){\circle*{14}}
  \put(0,0){\vertexcn} \put(42,0){\vertexbn}
  \put(42,-42){\greydot}\put(42,-7){\line(0,-1){28}}
  \put(14,-21){\makebox(0,0){\#39}}  
\end{picture}
&\hspace*{10.5pt}& 
\begin{picture}(90,0) \thicklines 
  \put(0,0){\circle*{14}} \put(42,0){\circle{14}} \put(84,0){\circle*{14}}
  \put(0,0){\vertexcn} \put(42,0){\vertexbn}
  \put(42,-42){\greydot}\put(42,-7){\line(0,-1){28}}
  \put(14,-21){\makebox(0,0){\#40}}  
\end{picture}
\\
\\
\end{tabular}

\begin{tabular}{ccccccc}
\begin{picture}(90,90) \thicklines 
  \put(0,0){\greydot} \put(73,20){\circle{14}}
  \put(38,20){\line(1,0){28}} \put(0,0){\vertexdn}
  \put(60,-10){\makebox(0,0){\#49}}
\end{picture}
&\hspace*{10.5pt}& 
\begin{picture}(90,90) \thicklines 
  \put(0,0){\greydot} \put(73,20){\circle{14}}
  \put(31,20){\vertexbn} \put(0,0){\vertexdn}
  \put(60,-10){\makebox(0,0){\#50}}
\end{picture}
&\hspace*{10.5pt}& 
\begin{picture}(90,90) \thicklines 
  \put(0,0){\greydot} \put(73,20){\circle{14}}
  \put(31,20){\vertexcn} \put(0,0){\vertexdn}
  \put(60,-10){\makebox(0,0){\#51}}
\end{picture}
&\hspace*{10.5pt}& 
\begin{picture}(90,90) \thicklines 
  \put(0,0){\greydot} \put(73,20){\circle{14}}
  \put(31,20){\vertexgcn} \put(0,0){\vertexdn}
  \put(60,-10){\makebox(0,0){\#52}}
\end{picture}
\\
\\
\begin{picture}(90,50) \thicklines 
  \put(0,0){\greydot} \put(73,20){\circle*{14}}
  \put(31,20){\vertexbn} \put(0,0){\vertexdn}
  \put(60,-10){\makebox(0,0){\#53}}
\end{picture}
&\hspace*{10.5pt}& 
\begin{picture}(90,50) \thicklines 
  \put(0,0){\greydot} \put(42,0){\circle{14}} 
  \put(0,0){\vertexdn} \put(7,0){\line(1,0){28}}
  \put(70,-10){\makebox(0,0){\#54}}
\end{picture}
&\hspace*{10.5pt}& 
&\hspace*{10.5pt}& 
\\
\\
\end{tabular}

\begin{center}
\begin{tabular}{ccccccc}
\begin{picture}(50,50) \thicklines 
  \put(0,21){\circle{14}} \put(42,21){\circle{14}} \put(0,-21){\circle{14}}
  \put(42,-21){\circle*{14}} \put(0,21){\vertexbn} \put(0,-21){\vertexbn}
  \put(0,21){\vertexdowndbl} \put(42,21){\vertexdowndbl}
  \put(70,0){\makebox(0,0){\#66}}  
\end{picture}
&\hspace*{49pt}& 
\begin{picture}(50,50) \thicklines 
  \put(0,21){\circle{14}} \put(42,21){\circle{14}} \put(0,-21){\circle{14}}
  \put(42,-21){\circle*{14}} \put(0,21){\vertexcn} \put(0,-21){\vertexbn}
  \put(0,-21){\vertexupdbl} \put(42,21){\vertexdowndbl}
  \put(70,0){\makebox(0,0){\#67}}  
\end{picture}
&\hspace*{49pt}& 
\begin{picture}(50,50) \thicklines 
  \put(0,21){\circle*{14}} \put(42,21){\circle{14}}
  \put(0,-21){\circle{14}} \put(42,-21){\circle*{14}} \put(0,21){\vertexcn}
  \put(0,-21){\vertexbn} \put(0,-21){\vertexupdbl}
  \put(42,21){\vertexdowndbl}
  \put(70,0){\makebox(0,0){\#68}}  
\end{picture}
&\hspace*{49pt}& 
\begin{picture}(50,50) \thicklines 
  \put(0,21){\circle{14}} \put(42,21){\circle{14}} \put(0,-21){\circle{14}}
  \put(42,-21){\circle*{14}} \put(7,21){\line(1,0){28}}
  \put(0,-21){\vertexbn} \put(0,-14){\line(0,1){28}}
  \put(42,21){\vertexdowndbl}
  \put(70,0){\makebox(0,0){\#69}}  
\end{picture}
\\
\\
\end{tabular}
\end{center}

\clearpage

\subsubsection*{Rank 5 hyperbolic superalgebras (38 diagrams)}

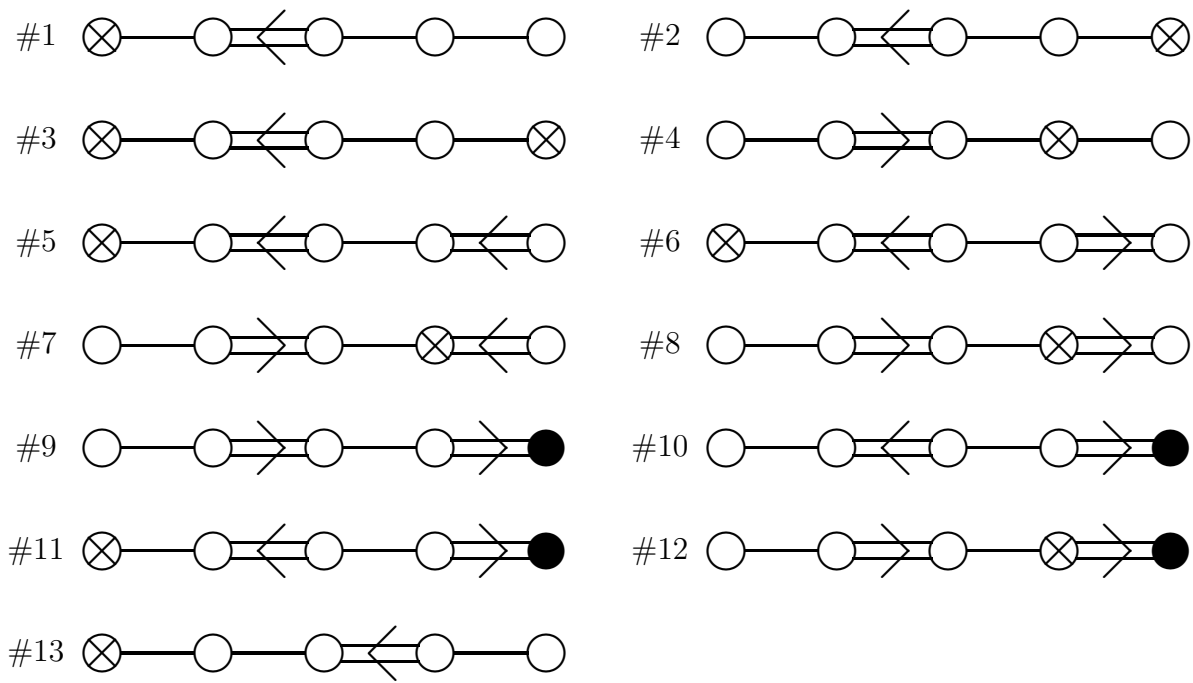
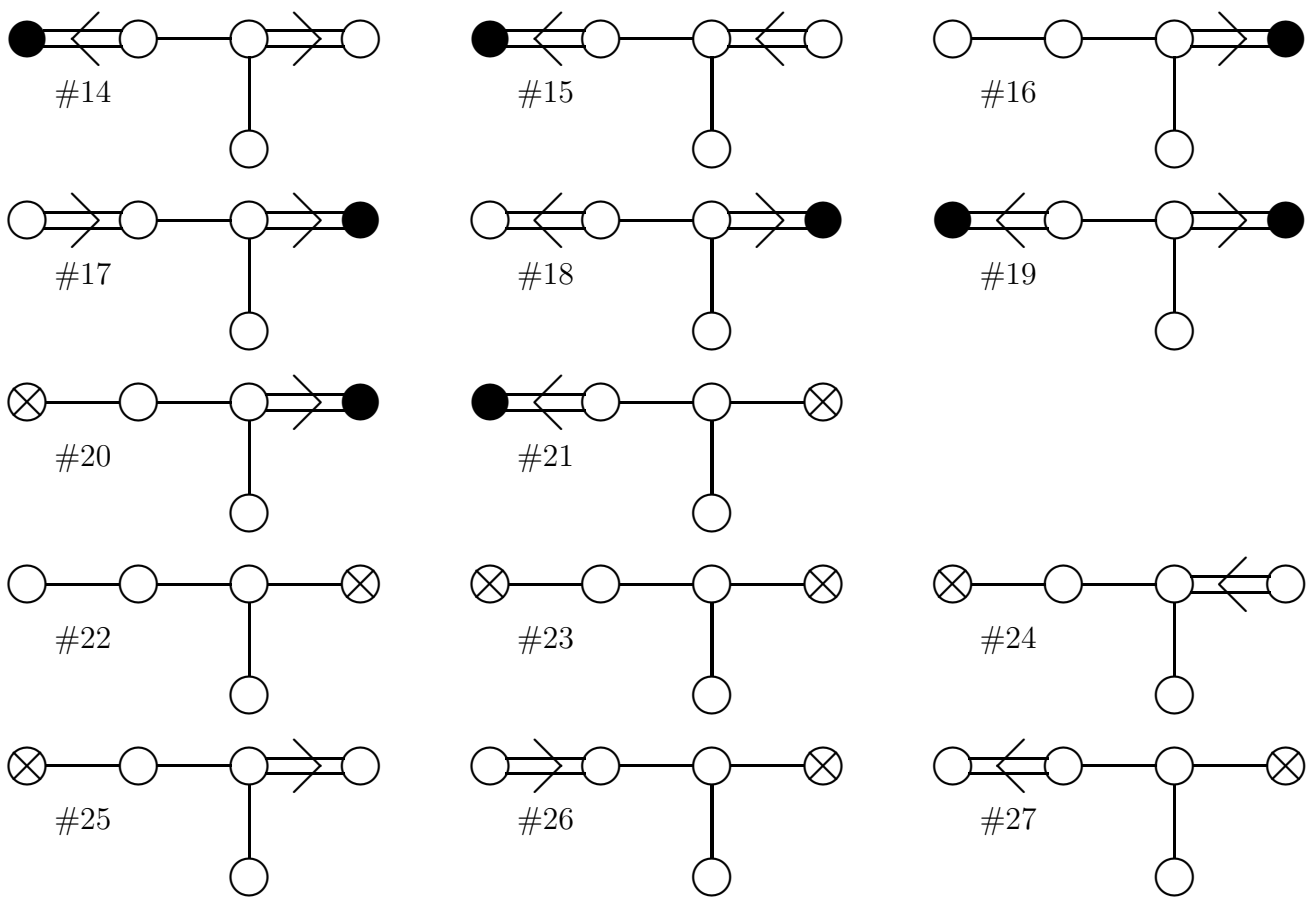
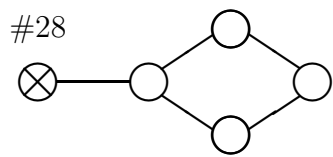
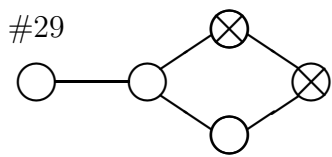
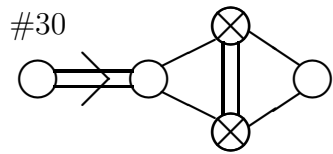
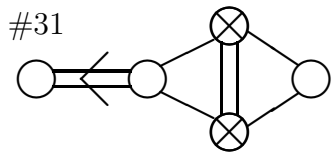
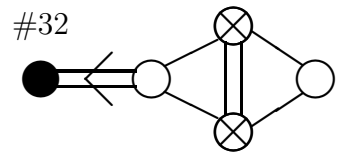
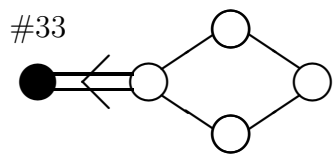
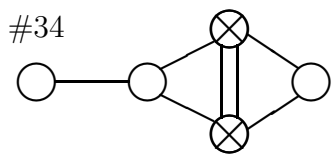
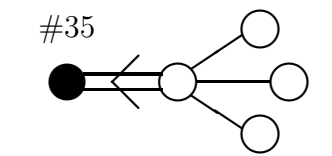
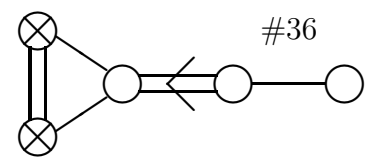
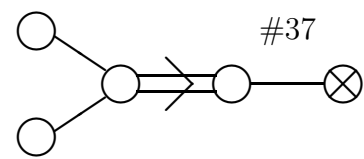
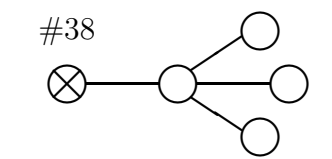
\begin{center}
\begin{tabular}{ccc}
\begin{picture}(170,20) \thicklines 
  \put(0,0){\greydot} \put(42,0){\circle{14}} \put(84,0){\circle{14}}
  \put(126,0){\circle{14}} \put(168,0){\circle{14}} \put(7,0){\line(1,0){28}}
  \put(42,0){\vertexcn} \put(91,0){\line(1,0){28}}
  \put(133,0){\line(1,0){28}}
  \put(-25,0){\makebox(0,0){\#1}}
\end{picture}
&\hspace*{42pt}& 
\begin{picture}(170,20) \thicklines 
  \put(0,0){\circle{14}} \put(42,0){\circle{14}} \put(84,0){\circle{14}}
  \put(126,0){\circle{14}} \put(168,0){\greydot} \put(7,0){\line(1,0){28}}
  \put(42,0){\vertexcn} \put(91,0){\line(1,0){28}}
  \put(133,0){\line(1,0){28}}
  \put(-25,0){\makebox(0,0){\#2}}
\end{picture}
\\
\\
\begin{picture}(170,20) \thicklines 
  \put(0,0){\greydot} \put(42,0){\circle{14}} \put(84,0){\circle{14}}
  \put(126,0){\circle{14}} \put(168,0){\greydot} \put(7,0){\line(1,0){28}}
  \put(42,0){\vertexcn} \put(91,0){\line(1,0){28}}
  \put(133,0){\line(1,0){28}}
  \put(-25,0){\makebox(0,0){\#3}}
\end{picture}
&\hspace*{42pt}& 
\begin{picture}(170,20) \thicklines 
  \put(0,0){\circle{14}} \put(42,0){\circle{14}} \put(84,0){\circle{14}}
  \put(126,0){\greydot} \put(168,0){\circle{14}} \put(7,0){\line(1,0){28}}
  \put(42,0){\vertexbn} \put(91,0){\line(1,0){28}}
  \put(133,0){\line(1,0){28}}
  \put(-25,0){\makebox(0,0){\#4}}
\end{picture}
\\
\\
\begin{picture}(170,20) \thicklines 
  \put(0,0){\greydot} \put(42,0){\circle{14}} \put(84,0){\circle{14}}
  \put(126,0){\circle{14}} \put(168,0){\circle{14}}
  \put(7,0){\line(1,0){28}} \put(42,0){\vertexcn}
  \put(91,0){\line(1,0){28}} \put(126,0){\vertexcn}
  \put(-25,0){\makebox(0,0){\#5}}
\end{picture}
&\hspace*{42pt}& 
\begin{picture}(170,20) \thicklines 
  \put(0,0){\greydot} \put(42,0){\circle{14}} \put(84,0){\circle{14}}
  \put(126,0){\circle{14}} \put(168,0){\circle{14}}
  \put(7,0){\line(1,0){28}} \put(42,0){\vertexcn}
  \put(91,0){\line(1,0){28}} \put(126,0){\vertexbn}
  \put(-25,0){\makebox(0,0){\#6}}
\end{picture}
\\
\\
\end{tabular}
\end{center}

\begin{tabular}{ccccc}
\begin{picture}(130,20) \thicklines 
  \put(0,0){\circle*{14}} \put(42,0){\circle{14}} \put(84,0){\circle{14}}
  \put(126,0){\circle{14}} \put(0,0){\vertexcn} \put(49,0){\line(1,0){28}}
  \put(84,0){\vertexbn}
  \put(84,-42){\circle{14}}\put(84,-7){\line(0,-1){28}}
  \put(21,-21){\makebox(0,0){\#14}}  
\end{picture}
&\hspace*{21pt}& 
\begin{picture}(130,20) \thicklines 
  \put(0,0){\circle*{14}} \put(42,0){\circle{14}} \put(84,0){\circle{14}}
  \put(126,0){\circle{14}} \put(0,0){\vertexcn} \put(49,0){\line(1,0){28}}
  \put(84,0){\vertexcn}
  \put(84,-42){\circle{14}}\put(84,-7){\line(0,-1){28}}
  \put(21,-21){\makebox(0,0){\#15}}  
\end{picture}
&\hspace*{21pt}& 
\begin{picture}(130,20) \thicklines 
  \put(0,0){\circle{14}} \put(42,0){\circle{14}} \put(84,0){\circle{14}}
  \put(126,0){\circle*{14}} \put(7,0){\line(1,0){28}}
  \put(49,0){\line(1,0){28}} \put(84,0){\vertexbn}
  \put(84,-42){\circle{14}}\put(84,-7){\line(0,-1){28}}
  \put(21,-21){\makebox(0,0){\#16}}  
\end{picture}
\\
\\
\end{tabular}

\begin{tabular}{ccccc}
\begin{picture}(110,50) \thicklines 
  \put(0,0){\greydot} \put(42,0){\circle{14}} \put(104,0){\circle{14}}
  \put(7,0){\line(1,0){28}} \put(42,0){\vertexdn} \put(74,0){\vertexvn}
  \put(0,20){\makebox(0,0){\#28}}  
\end{picture}
&\hspace*{21pt}& 
\begin{picture}(110,50) \thicklines 
  \put(0,0){\circle{14}} \put(42,0){\circle{14}} \put(104,0){\greydot}
  \put(7,0){\line(1,0){28}} \put(42,0){\vertexdn} \put(74,0){\vertexvn}
  \put(68,15){\line(1,1){10}}\put(68,25){\line(1,-1){10}}
  \put(0,20){\makebox(0,0){\#29}}  
\end{picture}
&\hspace*{21pt}& 
\\
\\
\begin{picture}(110,50) \thicklines 
  \put(0,0){\circle{14}} \put(42,0){\circle{14}} \put(104,0){\circle{14}}
  \put(0,0){\vertexbn} \put(42,0){\vertexgreydn} \put(74,0){\vertexvn}
  \put(0,20){\makebox(0,0){\#30}}  
\end{picture}
&\hspace*{21pt}& 
\begin{picture}(110,50) \thicklines 
  \put(0,0){\circle{14}} \put(42,0){\circle{14}} \put(104,0){\circle{14}}
  \put(0,0){\vertexcn} \put(42,0){\vertexgreydn} \put(74,0){\vertexvn}
  \put(0,20){\makebox(0,0){\#31}}  
\end{picture}
&\hspace*{21pt}& 
\begin{picture}(110,50) \thicklines 
  \put(0,0){\circle*{14}} \put(42,0){\circle{14}} \put(104,0){\circle{14}}
  \put(0,0){\vertexcn} \put(42,0){\vertexgreydn} \put(74,0){\vertexvn}
  \put(0,20){\makebox(0,0){\#32}}  
\end{picture}
\\
\\
\end{tabular}

\clearpage

\subsubsection*{Rank 6 hyperbolic superalgebras (15 diagrams)}

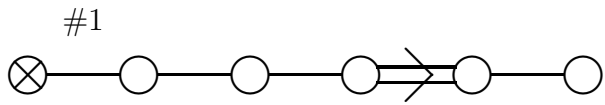
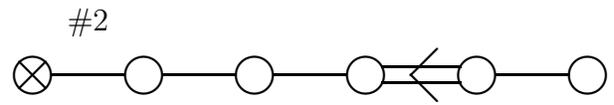
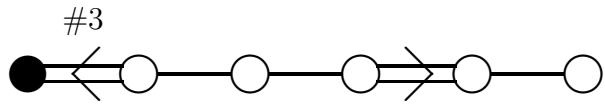
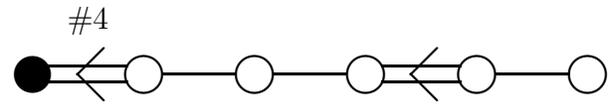
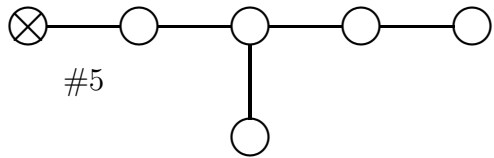
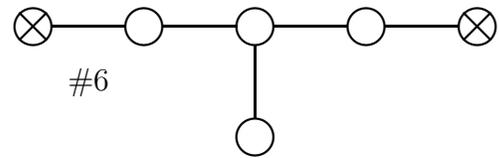
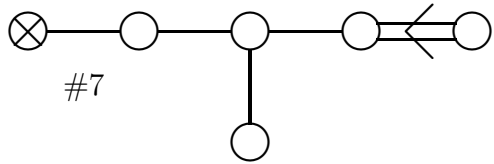
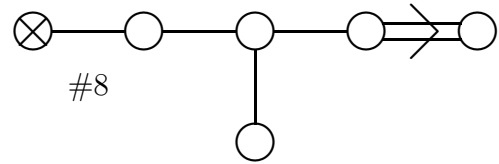
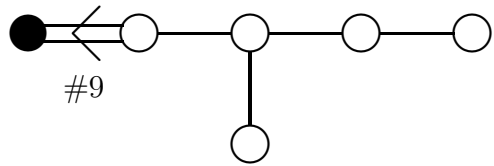
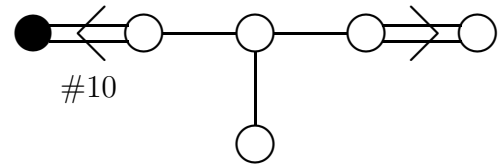
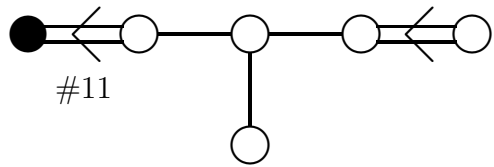
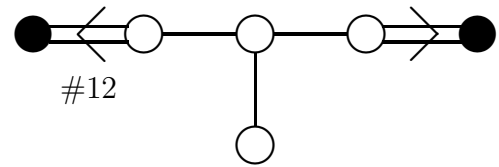
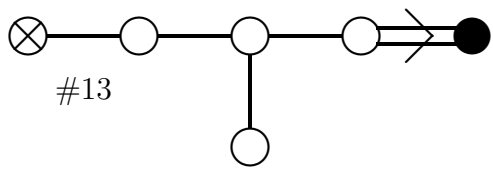
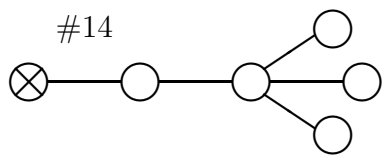
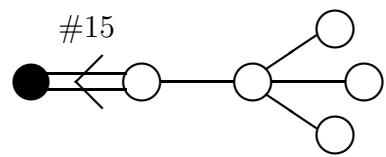
\begin{center}
\begin{tabular}{ccc}
\begin{picture}(210,30) \thicklines 
  \put(0,0){\greydot} \put(42,0){\circle{14}} \put(84,0){\circle{14}}
  \put(126,0){\circle{14}} \put(168,0){\circle{14}}
  \put(210,0){\circle{14}} \put(7,0){\line(1,0){28}}
  \put(49,0){\line(1,0){28}} \put(91,0){\line(1,0){28}}
  \put(126,0){\vertexbn} \put(175,0){\line(1,0){28}}
  \put(21,20){\makebox(0,0){\#1}}
\end{picture}
&\qquad\qquad&
\begin{picture}(210,30) \thicklines 
  \put(0,0){\greydot} \put(42,0){\circle{14}} \put(84,0){\circle{14}}
  \put(126,0){\circle{14}} \put(168,0){\circle{14}}
  \put(210,0){\circle{14}} \put(7,0){\line(1,0){28}}
  \put(49,0){\line(1,0){28}} \put(91,0){\line(1,0){28}}
  \put(126,0){\vertexcn} \put(175,0){\line(1,0){28}}
  \put(21,20){\makebox(0,0){\#2}}
\end{picture}
\\
\\
\begin{picture}(210,30) \thicklines 
  \put(0,0){\circle*{14}} \put(42,0){\circle{14}} \put(84,0){\circle{14}}
  \put(126,0){\circle{14}} \put(168,0){\circle{14}}
  \put(210,0){\circle{14}} \put(0,0){\vertexcn} \put(49,0){\line(1,0){28}}
  \put(91,0){\line(1,0){28}} \put(126,0){\vertexbn}
  \put(175,0){\line(1,0){28}}
  \put(21,20){\makebox(0,0){\#3}}
\end{picture}
&\qquad\qquad&
\begin{picture}(210,30) \thicklines 
  \put(0,0){\circle*{14}} \put(42,0){\circle{14}} \put(84,0){\circle{14}}
  \put(126,0){\circle{14}} \put(168,0){\circle{14}}
  \put(210,0){\circle{14}} \put(0,0){\vertexcn} \put(49,0){\line(1,0){28}}
  \put(91,0){\line(1,0){28}} \put(126,0){\vertexcn}
  \put(175,0){\line(1,0){28}}
  \put(21,20){\makebox(0,0){\#4}}
\end{picture}
\\
\\
\end{tabular}
\end{center}

\begin{center}
\begin{tabular}{ccc}
\begin{picture}(170,20) \thicklines 
  \put(0,0){\greydot} \put(42,0){\circle{14}} \put(84,0){\circle{14}}
  \put(126,0){\circle{14}} \put(168,0){\circle{14}}
  \put(84,-42){\circle{14}}\put(84,-7){\line(0,-1){28}}
  \put(7,0){\line(1,0){28}} \put(49,0){\line(1,0){28}}
  \put(91,0){\line(1,0){28}} \put(133,0){\line(1,0){28}}
  \put(21,-21){\makebox(0,0){\#5}}  
\end{picture}
&\hspace*{42pt}&
\begin{picture}(170,20) \thicklines 
  \put(0,0){\greydot} \put(42,0){\circle{14}} \put(84,0){\circle{14}}
  \put(126,0){\circle{14}} \put(168,0){\greydot}
  \put(84,-42){\circle{14}}\put(84,-7){\line(0,-1){28}}
  \put(7,0){\line(1,0){28}} \put(49,0){\line(1,0){28}}
  \put(91,0){\line(1,0){28}} \put(133,0){\line(1,0){28}}
  \put(21,-21){\makebox(0,0){\#6}}  
\end{picture}
\\
\\
\begin{picture}(170,50) \thicklines 
  \put(0,0){\greydot} \put(42,0){\circle{14}} \put(84,0){\circle{14}}
  \put(126,0){\circle{14}} \put(168,0){\circle{14}}
  \put(84,-42){\circle{14}}\put(84,-7){\line(0,-1){28}}
  \put(7,0){\line(1,0){28}} \put(49,0){\line(1,0){28}}
  \put(91,0){\line(1,0){28}} \put(126,0){\vertexcn}
  \put(21,-21){\makebox(0,0){\#7}}  
\end{picture}
&\hspace*{42pt}&
\begin{picture}(170,50) \thicklines 
  \put(0,0){\greydot} \put(42,0){\circle{14}} \put(84,0){\circle{14}}
  \put(126,0){\circle{14}} \put(168,0){\circle{14}}
  \put(84,-42){\circle{14}}\put(84,-7){\line(0,-1){28}}
  \put(7,0){\line(1,0){28}} \put(49,0){\line(1,0){28}}
  \put(91,0){\line(1,0){28}} \put(126,0){\vertexbn}
  \put(21,-21){\makebox(0,0){\#8}}  
\end{picture}
\\
\\
\end{tabular}
\end{center}

\clearpage

\section*{B. Simple root systems}

We describe in this section the simple root systems corresponding to the
Dynkin diagrams of Appendix A (as usual, the parametrization is not
unique). We give below the conventions used to describe the simple root
systems depending on the topology of the considered Dynkin diagrams. In any
case, the simple roots are written in terms of orthogonal vectors
$\eps_{i}$, $\delta_{i}$, $K^{+}$ and $K^{-}$ such that
$(\eps_{i},\eps_{j}) = 1$, $(\delta_{i},\delta_{j}) = -1$, $(K^{+},K^{-}) =
1$ and all other scalar products are zero. It is also convenient to
introduce $\widetilde\delta = \delta_{1}+\delta_{2}+\delta_{3}$ which
satisfies $(\widetilde\delta,\widetilde\delta) = -3$ and
$(\widetilde\delta,\eps_{i}) = (\widetilde\delta,K^{+}) =
(\widetilde\delta,K^{-}) = 0$.

\subsubsection*{Rank 3 hyperbolic superalgebras}

Conventions for the simple root systems $\Delta^{0} = \{ \alpha_{1},
\alpha_{2}, \alpha_{3} \}$:

\begin{center}
\begin{tabular}{ccc}
\begin{picture}(90,40) \thicklines 
  \put(0,20){\circle*{7}} \put(42,20){\circle*{7}} \put(84,20){\circle*{7}}
  \put(7,20){\line(1,0){28}} \put(49,20){\line(1,0){28}}
  \put(0,30){\makebox(0,0){$\alpha_{1}$}}
  \put(42,30){\makebox(0,0){$\alpha_{2}$}}
  \put(84,30){\makebox(0,0){$\alpha_{3}$}}
\end{picture}
&\hspace*{84pt}&
\begin{picture}(50,40) \thicklines 
  \put(0,41){\circle*{7}} \put(0,-1){\circle*{7}} \put(31,20){\circle*{7}}
  \put(6,38){\line(3,-2){20}} \put(6,2){\line(3,2){20}}
  \put(0,6){\line(0,1){28}}
  \put(-15,41){\makebox(0,0){$\alpha_{1}$}}
  \put(-15,-1){\makebox(0,0){$\alpha_{2}$}}
  \put(46,20){\makebox(0,0){$\alpha_{3}$}}
\end{picture}
\end{tabular}
\end{center}

\vspace*{6pt}

\noindent
\#1 : $\Delta^{0} = \{ \;2\eps_{1}-\eps_{2}-\eps_{3}-K^{-}, 
\;\eps_{2}-\eps_{1}, \;2K^{+}+\eps_{1} \; \}$ \\
\#2 : $\Delta^{0} = \{ \;\sfrac{1}{3}(2\eps_{1}-\eps_{2}-\eps_{3})-K^{-}, 
\;\eps_{2}-\eps_{1}, \;\sfrac{2}{3} K^{+}+\eps_{1} \; \}$ \\
\#3 : $\Delta^{0} = \{ \;-\half(\eps_{1}+\eps_{2}+\eps_{3}+\eps_{4})-K^{-},
\;\eps_{1}, \;-\eps_{1}+\half K^{+} \; \}$ \\
\#4 : $\Delta^{0} = \{ \;-2\eps_{1}+2\eps_{2}-K^{-}, \;\eps_{1}-\eps_{2},
\;2K^{+}+\eps_{2} \; \}$ \\
\#5 : $\Delta^{0} = \{ \;-\half(\eps_{1}-\eps_{2})-K^{-},
\;\eps_{1}-\eps_{2}, \;\half K^{+}+\eps_{2} \; \}$ \\
\#6 : $\Delta^{0} = \{ \;-\eps_{1}-K^{-}, \;\eps_{1}, \;-\eps_{1}+K^{+} \;
\}$ \\
\#7 : $\Delta^{0} = \{ \;-\eps_{1}+\eps_{2}-K^{-}, \;\eps_{1},
\;-\eps_{1}+K^{+} \; \}$ \\
\#8 : $\Delta^{0} = \{ \;\half(-\eps_{1}+\eps_{2})-K^{-}, \;\eps_{1},
\;-\eps_{1}+\half K^{+} \; \}$ \\
\#9 : $\Delta^{0} = \{ \;-2\eps_{1}-K^{-}, \;\eps_{1}, \;-\eps_{1}+2K^{+}
\; \}$ \\
\#10 : $\Delta^{0} = \{
\;-\half(3\eps_{1}+\eps_{2}+\eps_{3}+\eps_{4})-\sfrac{3}{2} K^{-},
\;\eps_{1}, \;-\eps_{1}+K^{+} \; \}$ \\
\#11 : $\Delta^{0} = \{ \;-\sfrac{1}{6}
(3\eps_{1}+\eps_{2}+\eps_{3}+\eps_{4})-\half K^{-}, \;\eps_{1},
\;-\eps_{1}+K^{+} \; \}$ \\
\#12 : $\Delta^{0} = \{ \;-\half\eps_{1}-K^{-}, \;\eps_{1},
\;-\eps_{1}+\half K^{+} \; \}$ \\
\#13 : $\Delta^{0} = \{ \;2\eps_{1}-2\eps_{2}-K^{-}, \;2\eps_{2}, 
\;2K^{+}-\eps_{2} \; \}$ \\
\#14 : $\Delta^{0} = \{ \;\eps_{1}-\eps_{2}-K^{-}, \;2\eps_{2}, 
\;K^{+}-\eps_{2} \; \}$ \\
\#15 : $\Delta^{0} = \{ \;\eps_{1}+\eps_{2}+\eps_{3}+\eps_{4}-K^{-}, 
\;-2\eps_{1}, \;K^{+}+\eps_{1} \; \}$ \\
\#16 : $\Delta^{0} = \{ \;2\eps_{1}-\eps_{2}-\eps_{3}-K^{-},
\;\eps_{2}-\eps_{1}, \;\sfrac{3}{2} K^{+}+\half(\eps_{1}-\eps_{2}) \; \}$
\\
\#17 : $\Delta^{0} = \{ \;\sfrac{1}{3}(2\eps_{1}-\eps_{2}-\eps_{3})-K^{-},
\;\eps_{2}-\eps_{1}, \;\half K^{+}+\half(\eps_{1}-\eps_{2}) \; \}$ \\
\#18 : $\Delta^{0} = \{ \;2\eps_{1}-K^{-}, \;-2\eps_{1}, \;2K^{+}+\eps_{1}
\; \}$ \\
\#19 : $\Delta^{0} = \{ \;-4\eps_{1}-K^{-}, \;2\eps_{1}, \;-\eps_{1}+4K^{+}
\; \}$ \\
\#20 : $\Delta^{0} = \{ \;-\eps_{1}-K^{-}, \;2\eps_{1}, \;K^{+}-\eps_{1} \;
\}$ \\
\#21 : $\Delta^{0} = \{ \;-\eps_{1}+\eps_{2}-K^{-}, \;\eps_{1}-\eps_{2},
\;K^{+}+\eps_{2} \; \}$ \\
\#22 : $\Delta^{0} = \{ \;-\eps_{1}-2K^{-}, \;\eps_{1}, \;-2\eps_{1}+K^{+}
\; \}$ \\
\#23 : $\Delta^{0} = \{ \;\eps_{1}-\eps_{2}-2K^{-}, \;\eps_{2},
\;K^{+}-2\eps_{2} \; \}$ \\
\#24 : $\Delta^{0} = \{ \;-2\eps_{1}-2K^{-}, \;\eps_{1}, \;2K^{+}-2\eps_{1}
\; \}$ \\
\#25 : $\Delta^{0} = \{ \;-\eps_{1}-K^{-}, \;\eps_{1}, \;-\eps_{1}+K^{+} \;
\}$ \\
\#26 : $\Delta^{0} = \{ \;-\eps_{1}-K^{-}, \;\eps_{1},
\;-\eps_{1}+\eps_{2}+K^{+} \; \}$ \\
\#27 : $\Delta^{0} = \{ \;-\eps_{1}-K^{-}, \;\eps_{1}, \;-\eps_{1}+K^{+} \;
\}$ \\
\#28 : $\Delta^{0} = \{ \;-\eps_{1}-K^{-}, \;\eps_{1}, \;-\eps_{1}+K^{+} \;
\}$ \\
\#29 : $\Delta^{0} = \{ \;-\eps_{1}-\half K^{-}, \;\eps_{1},
\;-\half(\eps_{1}-\eps_{2})+K^{+} \; \}$ \\
\#30 : $\Delta^{0} = \{ \;-\eps_{1}-\half K^{-}, \;\eps_{1},
\;-\half\eps_{1}+K^{+} \; \}$ \\
\#31 : $\Delta^{0} = \{ \;-\eps_{1}-K^{-}, \;2\eps_{1}, \;K^{+}-\eps_{1} \;
\}$ \\
\#32 : $\Delta^{0} = \{ \;-\eps_{1}-K^{-}, \;\eps_{1}, \;-\eps_{1}+K^{+} \;
\}$ \\
\#33 : $\Delta^{0} = \{ \;-\eps_{1}-K^{-}, \;\eps_{1},
\;-\eps_{1}+\eps_{2}+K^{+} \; \}$ \\
\#34 : $\Delta^{0} = \{ \;-\eps_{1}-\half K^{-}, \;\eps_{1}-\eps_{2}, 
\;-\half(\eps_{1}-\eps_{2})+K^{+} \; \}$ \\
\#35 : $\Delta^{0} = \{ \;-\eps_{1}-2K^{-}, \;\eps_{1}, \;-2\eps_{1}+K^{+}
\; \}$ \\
\#36 : $\Delta^{0} = \{ \;\delta_{1}-\eps_{2}-K^{-}, \;\eps_{2}-\eps_{1},
\;\sfrac{1}{3} K^{+}+\sfrac{1}{3}(2\eps_{1}-\eps_{2}-\eps_{3}) \; \}$ \\
\#37 : $\Delta^{0} = \{ \;\delta_{1}-\eps_{1}-K^{-}, \;\eps_{1}-\eps_{2},
\;-\half(\eps_{1}-\eps_{2})+\half K^{+} \; \}$ \\
\#38 : $\Delta^{0} = \{ \;\delta_{1}-\eps_{1}-K^{-}, \;\eps_{1}-\eps_{2},
\;-2\eps_{1}+2\eps_{2}+2K^{+} \; \}$ \\
\#39 : $\Delta^{0} = \{ \;\delta_{1}-\eps_{1}-K^{-}, \;\eps_{1}-\eps_{2},
\;-\eps_{1}+\eps_{2}+K^{+} \}$ \\
\#40 : $\Delta^{0} = \{ \;\eps_{1}-\delta_{1}-K^{-}, \;-\eps_{1},
\;\half(\eps_{1}+\eps_{2}+\eps_{3}+\eps_{4})+\half K^{+} \; \}$ \\
\#41 : $\Delta^{0} = \{ \;\eps_{1}-\delta_{1}-K^{-}, \;-\eps_{1},
\;\half(\eps_{1}-\eps_{2})+\half K^{+} \; \}$ \\
\#42 : $\Delta^{0} = \{ \;\eps_{1}-\delta_{1}-K^{-}, \;-\eps_{1},
\;\eps_{1}-\eps_{2}+K^{+} \; \}$ \\
\#43 : $\Delta^{0} = \{ \;\delta_{1}-\eps_{1}-K^{-}, \;\eps_{1},
\;-\half\eps_{1}+\sfrac{1}{6} (\eps_{2}+\eps_{3}+\eps_{4})+\half K^{+} \;
\}$ \\
\#44 : $\Delta^{0} = \{ \;\delta_{1}-\eps_{1}-K^{-}, \;\eps_{1},
\;-\sfrac{3}{2} \eps_{1}+\half(\eps_{2}+\eps_{3}+\eps_{4})+\sfrac{3}{2}
K^{+} \; \}$ \\
\#45 : $\Delta^{0} = \{ \;\eps_{1}-\delta_{1}-4K^{-}, \;-2\eps_{1},
\;4\eps_{1}+K^{+} \; \}$ \\
\#46 : $\Delta^{0} = \{ \;\delta_{1}-\eps_{1}-K^{-}, \;2\eps_{1},
\;-\eps_{1}+\sfrac{1}{3} (\eps_{2}+\eps_{3}+\eps_{4})+K^{+} \; \}$ \\
\#47 : $\Delta^{0} = \{ \;\delta_{1}-\eps_{1}-K^{-}, \;\eps_{1},
\;-3\eps_{1}+\eps_{2}+\eps_{3}+\eps_{4}+3K^{+} \; \}$ \\
\#48 : $\Delta^{0} = \{ \;\eps_{1}-\delta_{1}-K^{-}, \;-2\eps_{1},
\;\eps_{1}+K^{+} \; \}$ \\
\#49 : $\Delta^{0} = \{ \;\eps_{1}-\delta_{1}-\half K^{-}, \;-\eps_{1},
\;\half\eps_{1}+K^{+} \; \}$ \\
\#50 : $\Delta^{0} = \{ \;\eps_{1}-\delta_{1}-2K^{-}, \;-\eps_{1},
\;2\eps_{1}+K^{+} \; \}$ \\
\#51 : $\Delta^{0} = \{ \;\eps_{1}-\eps_{2}-\delta_{1}+\delta_{2}-2K^{-},
\;-\eps_{1}+\eps_{2}, \;\eps_{1}-\eps_{2}+K^{+} \; \}$ \\
\#52 : $\Delta^{0} = \{ \;\eps_{1}-\delta_{1}-K^{-}, \;-2\eps_{1},
\;\eps_{1}+\eps_{2}+\eps_{3}+\eps_{4}+K^{+} \; \}$ \\
\#53 : $\Delta^{0} = \{ \;\eps_{1}-\delta_{1}-K^{-}, \;-2\eps_{1},
\;\eps_{1}-\eps_{2}+K^{+} \; \}$ \\
\#54 : $\Delta^{0} = \{ \;\eps_{1}-\delta_{1}-2K^{-}, \;-2\eps_{1},
\;2\eps_{1}-2\eps_{2}+K^{+} \; \}$ \\
\#55 : $\Delta^{0} = \{ \;\eps_{1}-\delta_{1}-2K^{-}, \;-2\eps_{1},
\;2\eps_{1}+K^{+} \; \}$ \\
\#56 : $\Delta^{0} = \{ \;\delta_{1}-\eps_{1}-K^{-}, \;\eps_{1}-\eps_{2},
\;-\eps_{1}+\eps_{2}+K^{+} \; \}$ \\
\#57 : $\Delta^{0} = \{ \;\delta_{1}-\eps_{1}-\half K^{-},
\;\eps_{1}-\eps_{2}, \;-\half(\eps_{1}-\eps_{2})+K^{+} \; \}$ \\
\#58 : $\Delta^{0} = \{ \;\eps_{1}-\delta_{1}-K^{-}, \;-\eps_{1},
\;\eps_{1}+K^{+} \; \}$ \\
\#59 : $\Delta^{0} = \{ \;\eps_{1}-\delta_{1}-K^{-}, \;-\eps_{1},
\;\half(\eps_{1}-\eps_{2})+\half K^{+} \; \}$ \\
\#60 : $\Delta^{0} = \{ \;\eps_{1}-\delta_{1}-\half K^{-}, \;-\eps_{1},
\;\half\eps_{1}+K^{+} \; \}$ \\
\#61 : $\Delta^{0} = \{ \;\eps_{1}-\delta_{1}-2K^{-}, \;-2\eps_{1},
\;2\eps_{1}+K^{+} \; \}$ \\
\#62 : $\Delta^{0} = \{ \;\eps_{1}-\delta_{1}-K^{-}, \;-2\eps_{1},
\;\eps_{1}-\eps_{2}+K^{+} \; \}$ \\
\#63 : $\Delta^{0} = \{ \;\eps_{1}-\delta_{1}-K^{-}, \;-2\eps_{1},
\;\eps_{1}-\eps_{2}+K^{+} \; \}$ \\
\#64 : $\Delta^{0} = \{ \;-\eps_{1}+\eps_{2}-3K^{-},
\;-\eps_{1}+\eps_{2}+K^{+}, \;\eps_{1}-\eps_{2} \; \}$ \\
\#65 : $\Delta^{0} = \{ \;\eps_{1}+\eps_{2}-3K^{-},
\;\eps_{1}+\eps_{2}+K^{+}, \;-\eps_{1} \; \}$ \\
\#66 : $\Delta^{0} = \{ \;-2\eps_{1}-3K^{-}, \;-2\eps_{1}+2K^{+},
\;\eps_{1} \; \}$ \\
\#67 : $\Delta^{0} = \{ \;\eps_{1}-\eps_{2}-2K^{-}, \;-\eps_{2}+K^{+},
\;\eps_{2} \; \}$ \\
\#68 : $\Delta^{0} = \{ \;-2\eps_{2}-2K^{-}, \;\eps_{1}-\eps_{2}+2K^{+}, 
\;\eps_{2} \; \}$ \\
\#69 : $\Delta^{0} = \{ \;\eps_{1}-\eps_{2}-2K^{-},
\;\eps_{1}-\eps_{2}+2K^{+}, \;-\eps_{1}+\eps_{2} \; \}$ \\
\#70 : $\Delta^{0} = \{ \;\eps_{1}-\eps_{2}-2K^{-},
\;\eps_{1}-\eps_{2}+2K^{+}, \;-\eps_{1} \; \}$ \\
\#71 : $\Delta^{0} = \{ \;\eps_{1}-\eps_{2}-2K^{-},
\;\eps_{1}-\eps_{2}+2K^{+}, \;-\half(\eps_{1}-\eps_{2}) \; \}$ \\
\#72 : $\Delta^{0} = \{ \;\eps_{1}-2K^{-}, \;\eps_{1}+K^{+}, \;-\eps_{1} \;
\}$ \\
\#73 : $\Delta^{0} = \{ \;\eps_{1}, \;-\eps_{1}-2K^{-},
\;\eps_{2}-\eps_{1}+K^{+} \; \}$ \\
\#74 : $\Delta^{0} = \{ \;\eps_{1}, \;-\eps_{1}-2K^{-},
\;-2\eps_{1}+2K^{+} \; \}$ \\
\#75 : $\Delta^{0} = \{ \;\eps_{1}-2K^{-}, \;\eps_{1}+K^{+}, \;-\eps_{1} \;
\}$ \\
\#76 : $\Delta^{0} = \{ \;\eps_{1}-\eps_{2}-K^{-},
\;\eps_{1}+\eps_{2}+K^{+}, \;\delta_{1}-\eps_{1} \; \}$ \\
\#77 : $\Delta^{0} = \{ \;2\eps_{1}-3K^{-}, \;2\eps_{1}+2K^{+},
\;\delta_{1}-\eps_{1} \; \}$ \\
\#78 : $\Delta^{0} = \{ \;\eps_{1}-\sfrac{3}{2} K^{-}, \;\eps_{1}+K^{+},
\;\delta_{1}-\eps_{1} \; \}$ \\
\#79 : $\Delta^{0} = \{ \;\eps_{1}-\eps_{2}-2K^{-},
\;\eps_{1}-\eps_{2}+2K^{+}, \;\delta_{1}-\eps_{1} \; \}$ \\
\#80 : $\Delta^{0} = \{ \;2\eps_{1}-4K^{-}, \;2\eps_{1}+2K^{+},
\;\delta_{1}-\eps_{1} \; \}$ \\
\#81 : $\Delta^{0} = \{ \;\eps_{1}-2K^{-}, \;\eps_{1}+K^{+},
\;\delta_{1}-\eps_{1} \; \}$ \\
\#82 : $\Delta^{0} = \{ \;\eps_{1}-\delta_{1}-2K^{-},
\;\eps_{1}-\eps_{2}+K^{+}, \;-\eps_{1} \; \}$ \\
\#83 : $\Delta^{0} = \{ \;\eps_{1}-\delta_{1}-2K^{-},
\;\eps_{1}-\eps_{2}+K^{+}, \;-2\eps_{1} \; \}$ \\
\#84 : $\Delta^{0} = \{ \;2\eps_{1}-2K^{-}, \;\eps_{1}+2K^{+},
\;\delta_{1}-\eps_{1} \; \}$ \\
\#85 : $\Delta^{0} = \{ \;\eps_{1}-\delta_{1}+K^{-},
\;\eps_{1}-\delta_{1}+K^{+}, \;-\eps_{1} \; \}$ \\
\#86 : $\Delta^{0} = \{ \;\eps_{1}-\delta_{1}+K^{-},
\;\eps_{1}-\delta_{1}+K^{+}, \;-2\eps_{1} \; \}$ \\
\#87 : $\Delta^{0} = \{ \;\eps_{1}+\delta_{1}-K^{-},
\;\eps_{1}+\delta_{1}+K^{+}, \;-\eps_{1} \; \}$ \\

\vspace*{-15pt}

\subsubsection*{Rank 4 hyperbolic superalgebras}

Conventions for the simple root systems $\Delta^{0} = \{ \alpha_{1},
\alpha_{2}, \alpha_{3}, \alpha_{4} \}$:

\begin{tabular}{ccccc}
\begin{picture}(130,60) \thicklines 
  \put(0,40){\circle*{7}} \put(42,40){\circle*{7}} \put(84,40){\circle*{7}}
  \put(126,40){\circle*{7}} \put(7,40){\line(1,0){28}}
  \put(49,40){\line(1,0){28}} \put(91,40){\line(1,0){28}}
  \put(0,50){\makebox(0,0){$\alpha_{1}$}}
  \put(42,50){\makebox(0,0){$\alpha_{2}$}}
  \put(84,50){\makebox(0,0){$\alpha_{3}$}}
  \put(126,50){\makebox(0,0){$\alpha_{4}$}}
\end{picture}
&\hspace*{18pt}&
\begin{picture}(90,60) \thicklines 
  \put(0,40){\circle*{7}} \put(42,40){\circle*{7}} \put(84,40){\circle*{7}}
  \put(7,40){\line(1,0){28}} \put(49,40){\line(1,0){28}}
  \put(42,-2){\circle*{7}}\put(42,33){\line(0,-1){28}}
  \put(0,50){\makebox(0,0){$\alpha_{1}$}}
  \put(42,50){\makebox(0,0){$\alpha_{2}$}}
  \put(84,50){\makebox(0,0){$\alpha_{3}$}}
  \put(57,-2){\makebox(0,0){$\alpha_{4}$}}
\end{picture}
&\hspace*{18pt}&
\begin{picture}(90,60) \thicklines 
  \put(10,40){\circle*{7}} \put(10,0){\circle*{7}} \put(42,20){\circle*{7}}
  \put(84,20){\circle*{7}} \put(17,38){\line(3,-2){19}}
  \put(17,2){\line(3,2){19}} \put(49,20){\line(1,0){28}}
  \put(-5,40){\makebox(0,0){$\alpha_{1}$}}
  \put(-5,0){\makebox(0,0){$\alpha_{2}$}}
  \put(42,30){\makebox(0,0){$\alpha_{3}$}}
  \put(84,30){\makebox(0,0){$\alpha_{4}$}}
\end{picture}
\\
\begin{picture}(75,60) \thicklines 
  \put(0,20){\circle*{7}} \put(73,40){\circle*{7}} 
  \put(31,40){\circle*{7}} \put(31,0){\circle*{7}} 
  \put(5,25){\line(3,2){19}} \put(5,15){\line(3,-2){19}} 
  \put(38,40){\line(1,0){28}}
  \put(0,30){\makebox(0,0){$\alpha_{1}$}}
  \put(31,-10){\makebox(0,0){$\alpha_{2}$}}
  \put(31,50){\makebox(0,0){$\alpha_{3}$}}
  \put(73,50){\makebox(0,0){$\alpha_{4}$}}
\end{picture}
&\hspace*{18pt}&
\begin{picture}(60,60) \thicklines 
  \put(0,40){\circle*{7}} \put(42,40){\circle*{7}} \put(0,-2){\circle*{7}}
  \put(42,-2){\circle*{7}} \put(7,40){\line(1,0){28}}
  \put(7,-2){\line(1,0){28}} \put(0,5){\line(0,1){28}}
  \put(42,5){\line(0,1){28}}
  \put(-15,40){\makebox(0,0){$\alpha_{1}$}}
  \put(57,40){\makebox(0,0){$\alpha_{2}$}}
  \put(-15,-2){\makebox(0,0){$\alpha_{3}$}}
  \put(57,-2){\makebox(0,0){$\alpha_{4}$}}
\end{picture}
\end{tabular}

\vspace*{18pt}

\noindent
\#1 : $\Delta^{0} = \{ \;\eps_{1}+\eps_{2}+\eps_{3}+\eps_{4}-K^{-},
\;-2\eps_{1}, \;\eps_{1}-\eps_{2}, \;K^{+}+\eps_{2} \; \}$ \\
\#2 : $\Delta^{0} = \{ \;2\eps_{1}+2\eps_{2}-K^{-}, \;-2\eps_{1}, 
\;\eps_{1}-\eps_{2}, \;2K^{+}+\eps_{2} \; \}$ \\
\#3 : $\Delta^{0} = \{ \;\eps_{1}+\eps_{2}-K^{-}, \;-2\eps_{1}, 
\;\eps_{1}-\eps_{2}, \;K^{+}+\eps_{2} \; \}$ \\
\#4 : $\Delta^{0} = \{ \;\half(\eps_{1}+\eps_{2}+\eps_{3}+\eps_{4})-K^{-}, 
\;-\eps_{1}, \;\eps_{1}-\eps_{2}, \;\half K^{+}+\eps_{2} \; \}$ \\
\#5 : $\Delta^{0} = \{ \;\eps_{1}+\eps_{2}-K^{-}, \;-\eps_{1},
\;\eps_{1}-\eps_{2}, \;K^{+}+\eps_{2} \; \}$ \\
\#6 : $\Delta^{0} = \{ \;\half(\eps_{1}+\eps_{2})-K^{-}, \;-\eps_{1},
\;\eps_{1}-\eps_{2}, \;\half K^{+}+\eps_{2} \; \}$ \\
\#7 : $\Delta^{0} = \{ \;\eps_{1}-\eps_{2}-K^{-}, \;\eps_{2}-\eps_{3}, 
\;\eps_{3}, \;K^{+}-\eps_{2}-\eps_{3} \; \}$ \\
\#8 : $\Delta^{0} = \{ \;-2\eps_{1}-K^{-}, \;\eps_{1}-\eps_{2}, \;\eps_{2},
\;2K^{+}-\eps_{1}-\eps_{2} \; \}$ \\
\#9 : $\Delta^{0} = \{ \;-\eps_{1}-K^{-}, \;\eps_{1}-\eps_{2}, \;\eps_{2},
\;K^{+}-\eps_{1}-\eps_{2} \; \}$ \\
\#10 : $\Delta^{0} = \{ \;\eps_{1}+\eps_{2}-K^{-}, \;-2\eps_{1}, 
\;\eps_{1}-\eps_{2}, \;K^{+}+\eps_{2} \; \}$ \\
\#11 : $\Delta^{0} = \{ \;\eps_{1}+\eps_{2}-K^{-}, \;-\eps_{1},
\;\eps_{1}-\eps_{2}, \;K^{+}+\eps_{2} \; \}$ \\
\#12 : $\Delta^{0} = \{ \;-2\eps_{1}-2\eps_{2}-K^{-}, \;2\eps_{2},
\;\eps_{1}-\eps_{2}, \;\delta_{1}-\eps_{1}+2K^{+} \; \}$ \\
\#13 : $\Delta^{0} = \{ \;-\eps_{1}-\eps_{2}-K^{-}, \;2\eps_{2},
\;\eps_{1}-\eps_{2}, \;\delta_{1}-\eps_{1}+K^{+} \; \}$ \\
\#14 : $\Delta^{0} = \{ \;-\eps_{1}-\eps_{2}-K^{-}, \;\eps_{2},
\;\eps_{1}-\eps_{2}, \;K^{+}+\delta_{1}-\eps_{1} \; \}$ \\
\#15 : $\Delta^{0} = \{
\;-\half(\eps_{1}+\eps_{2}+\eps_{3}+\eps_{4})+K^{-}, \;\eps_{2},
\;\eps_{1}-\eps_{2}, \;\half K^{+}+\eps_{2}-\delta_{1} \; \}$ \\
\#16 : $\Delta^{0} = \{ \;\half(\eps_{1}+\eps_{2})-K^{-}, \;-\eps_{1},
\;\eps_{1}-\eps_{2}, \;\half K^{+}+\eps_{2}-\delta_{1} \; \}$ \\
\#17 : $\Delta^{0} = \{ \;\delta_{1}-\eps_{1}-K^{-}, \;\eps_{1}-\eps_{2},
\;\eps_{2}, \;-\half(\eps_{1}+\eps_{2}) + \half K^{+} \; \}$ \\
\#18 : $\Delta^{0} = \{ \;\delta_{1}-\eps_{1}-K^{-}, \;\eps_{1}-\eps_{2},
\;2\eps_{2}, \;-\eps_{1}-\eps_{2}+K^{+} \; \}$ \\
\#19 : $\Delta^{0} = \{ \;\delta_{1}-\eps_{1}-K^{-}, \;\eps_{1}-\eps_{2},
\;\eps_{2}, \;K^{+}-\eps_{1}-\eps_{2} \; \}$ \\
\#20 : $\Delta^{0} = \{ \;2\eps_{3}-\eps_{1}-\eps_{2}-K^{-},
\;\eps_{2}-\eps_{3}, \;\eps_{1}-\eps_{2}, \;K^{+}-\eps_{1}+\delta_{1} \;
\}$ \\
\#21 : $\Delta^{0} = \{ \;\sfrac{1}{3}(2\eps_{3}-\eps_{1}-\eps_{2})-K^{-}, 
\;\eps_{2}-\eps_{3}, \;\eps_{1}-\eps_{2}, \;\sfrac{1}{3} 
K^{+}-\eps_{1}+\delta_{1} \; \}$ \\
\#22 : $\Delta^{0} = \{ \;2\eps_{3}-\eps_{1}-\eps_{2}-K^{-},
\;2\eps_{2}-\eps_{1}-\eps_{3}-K^{-}, \;\eps_{1}-\eps_{2},
\;\delta_{1}-\eps_{1}+K^{+} \; \}$ \\
\#23 : $\Delta^{0} = \{ \;2\eps_{2}-\eps_{1}-\eps_{3}-K^{-},
\;\eps_{1}-\eps_{2}, \;\delta_{1}-\eps_{1}+K^{+}, \;-2\delta_{1} \; \}$ \\
\#24 : $\Delta^{0} = \{ \;2\eps_{2}-\eps_{1}-\eps_{3}-K^{-},
\;\eps_{1}-\eps_{2}, \;\delta_{1}-\eps_{1}+K^{+}, \;-\delta_{1} \; \}$ \\
\#25 : $\Delta^{0} = \{ \;2\eps_{2}-\eps_{1}-\eps_{3}-K^{-},
\;\eps_{1}-\eps_{2}, \;\delta_{1}-\eps_{1}+K^{+}, \;\delta_{2}-\delta_{1}
\; \}$ \\
\#26 : $\Delta^{0} = \{ \;\sfrac{1}{3}(2\eps_{3}-\eps_{1}-\eps_{2})-K^{-},
\;\eps_{2}-\eps_{3}, \;\eps_{1}-\eps_{2}, \;\sfrac{1}{3} K^{+}-\eps_{1} \;
\}$ \\
\#27 : $\Delta^{0} = \{ \;2\eps_{3}-\eps_{1}-\eps_{2}-K^{-},
\;\eps_{2}-\eps_{3}, \;\eps_{1}-\eps_{2}, \;K^{+}-\eps_{1} \; \}$ \\
\#28 : $\Delta^{0} = \{ \;2\eps_{2}-\eps_{1}-\eps_{3}-K^{-},
\;\eps_{1}-\eps_{2}, \;\delta_{1}-\eps_{1}+K^{+}, \;-\delta_{1} \; \}$ \\
\#29 : $\Delta^{0} = \{ \;\delta_{1}-\eps_{2}-K^{-}, \;\eps_{2}-\eps_{3},
\;2\eps_{3}, \;K^{+}+\eps_{1}-\eps_{2} \; \}$ \\
\#30 : $\Delta^{0} = \{ \;\delta_{1}-\eps_{2}-K^{-}, \;\eps_{2}-\eps_{3},
\;\eps_{3}, \;K^{+}+\eps_{1}-\eps_{2} \;\}$ \\
\#31 : $\Delta^{0} = \{ \;-2\eps_{2}-K^{-}, \;\eps_{2}-\eps_{3},
\;\eps_{3}, \;2K^{+}+\eps_{1}-\eps_{2} \; \}$ \\
\#32 : $\Delta^{0} = \{ \;-\eps_{2}-K^{-}, \;\eps_{2}-\eps_{3}, \;\eps_{3},
\;K^{+}+\eps_{1}-\eps_{2} \; \}$ \\
\#33 : $\Delta^{0} = \{ \;-\eps_{1}-K^{-}, \;\eps_{1}-\eps_{2}, \;\eps_{2},
\;K^{+}+\delta_{1}-\eps_{1} \; \}$ \\
\#34 : $\Delta^{0} = \{ \;-2\eps_{1}-K^{-}, \;\eps_{1}-\eps_{2},
\;2\eps_{2}, \;2K^{+}+\delta_{1}-\eps_{1} \; \}$ \\
\#35 : $\Delta^{0} = \{ \;-2\eps_{1}-K^{-}, \;\eps_{1}-\eps_{2},
\;\eps_{2}, \;2K^{+}+\delta_{1}-\eps_{1} \; \}$ \\
\#36 : $\Delta^{0} = \{ \;\eps_{1}-\eps_{2}, \;\eps_{2}-\eps_{3},
\;2\eps_{3}-\eps_{1}-\eps_{2}-K^{-}, \;2K^{+}+\eps_{3}-\delta_{1} \; \}$ \\
\#37 : $\Delta^{0} = \{ \;\eps_{1}-\eps_{2}-K^{-}, \;\eps_{2}-\eps_{3},
\;\eps_{3}, \;K^{+}+\delta_{1}-\eps_{2} \; \}$ \\
\#38 : $\Delta^{0} = \{ \;-2\eps_{1}-K^{-}, \;\eps_{1}-\eps_{2},
\;\eps_{2}, \;2K^{+}+\delta_{1}-\eps_{1} \; \}$ \\
\#39 : $\Delta^{0} = \{ \;-\eps_{1}-K^{-}, \;\eps_{1}-\eps_{2}, \;\eps_{2},
\;K^{+}+\delta_{1}-\eps_{1} \; \}$ \\
\#40 : $\Delta^{0} = \{ \;-\eps_{1}-K^{-}, \;\eps_{1}-\eps_{2}, \;\eps_{2},
\;K^{+}+\delta_{1}-\eps_{1} \; \}$ \\
\#41 : $\Delta^{0} = \{ \;-\eps_{1}-K^{-}, \;\eps_{1}-\eps_{2}, \;\eps_{2}, 
\;K^{+}-\eps_{1} \; \}$ \\
\#42 : $\Delta^{0} = \{ \;-2\eps_{1}-K^{-}, \;\eps_{1}-\eps_{2},
\;2\eps_{2}, \;2K^{+}-\eps_{1} \; \}$ \\
\#43 : $\Delta^{0} = \{ \;-2\eps_{1}-K^{-}, \;\eps_{1}-\eps_{2},
\;\eps_{2}, \;2K^{+}-\eps_{1} \; \}$ \\
\#44 : $\Delta^{0} = \{ \;\eps_{1}-\eps_{2}-K^{-}, \;\eps_{2}, 
\;-\eps_{1}-\eps_{2}, \;2K^{+}+\eps_{1}-\eps_{2} \; \}$ \\
\#45 : $\Delta^{0} = \{ \;-\eps_{2}-K^{-}, \;\eps_{2}-\eps_{3}, \;\eps_{3},
\;K^{+}+\eps_{1}-\eps_{2} \; \}$ \\
\#46 : $\Delta^{0} = \{ \;-\eps_{1}-2K^{-}, \;\eps_{1}-\eps_{2},
\;\eps_{2}, \;K^{+}-2\eps_{1} \; \}$ \\
\#47 : $\Delta^{0} = \{ \;-\eps_{1}-K^{-}, \;\eps_{1}-\eps_{2}, \;\eps_{2},
\;K^{+}-\eps_{1} \; \}$ \\
\#48 : $\Delta^{0} = \{ \;-\eps_{1}-K^{-}, \;\eps_{1}-\eps_{2}, \;\eps_{2},
\;K^{+}-\eps_{1} \; \}$ \\
\#49 : $\Delta^{0} = \{ \;\delta_{1}-\eps_{1}, \;\eps_{1}+\eps_{2}-K^{-}, 
\;\eps_{1}-\eps_{2}, \;\eps_{2}-\eps_{3}+K^{+} \; \}$ \\
\#50 : $\Delta^{0} = \{ \;\delta_{1}-\eps_{1}, \;\eps_{1}+\eps_{2}-K^{-}, 
\;\eps_{1}-\eps_{2}, \;\eps_{2}+K^{+} \; \}$ \\
\#51 : $\Delta^{0} = \{ \;\delta_{1}-\eps_{1}, \;\eps_{1}+\eps_{2}-2K^{-}, 
\;\eps_{1}-\eps_{2}, \;2\eps_{2}+K^{+} \; \}$ \\
\#52 : $\Delta^{0} = \{ \;\delta_{1}-\eps_{1}, \;\eps_{1}+\eps_{2}-K^{-}, 
\;\eps_{1}-\eps_{2}, \;2\eps_{2}-\eps_{1}-\eps_{3}+K^{+} \; \}$ \\
\#53 : $\Delta^{0} = \{ \;\delta_{1}-\eps_{1}, \;\eps_{1}+\eps_{2}-K^{-}, 
\;\eps_{1}-\eps_{2}, \;\eps_{2}+K^{+} \; \}$ \\
\#54 : $\Delta^{0} = \{ \;\delta_{1}-\eps_{1}-K^{-}, \;\eps_{1}-\eps_{2}, 
\;\eps_{1}+\eps_{2}, \;\eps_{3}-\eps_{4}+K^{+} \; \}$ \\
\#55 : $\Delta^{0} = \{ \;\delta_{1}-\eps_{1}-K^{-},
\;\delta_{2}-\eps_{1}+K^{+}, \;\eps_{1}-\eps_{2}, \;\eps_{2}-\eps_{3} \; \}$ \\
\#56 : $\Delta^{0} = \{ \;\delta_{1}-\eps_{1}-K^{-},
\;\delta_{2}-\eps_{1}+K^{+}, \;\eps_{1}-\eps_{2}, \;2\eps_{2} \; \}$ \\
\#57 : $\Delta^{0} = \{ \;\delta_{1}-\eps_{1}-K^{-},
\;\delta_{2}-\eps_{1}+K^{+}, \;\eps_{1}-\eps_{2}, \;\eps_{2} \; \}$ \\
\#58 : $\Delta^{0} = \{ \;\delta_{1}-\eps_{1}-K^{-},
\;-\delta_{2}+\eps_{2}-2K^{-}, \;\eps_{1}-\eps_{2},
\;K^{+}+2\eps_{2}-\eps_{1}-\eps_{3} \; \}$ \\
\#59 : $\Delta^{0} = \{ \;\delta_{1}-\eps_{1}, \;\eps_{2}-\delta_{1}-K^{-},
\;\eps_{1}-\eps_{2}, \;K^{+}+\eps_{2}-\eps_{3} \; \}$ \\
\#60 : $\Delta^{0} = \{ \;\delta_{1}-\eps_{1}, \;\eps_{2}-\delta_{1}-2K^{-},
\;\eps_{1}-\eps_{2}, \;K^{+}+2\eps_{2} \; \}$ \\
\#61 : $\Delta^{0} = \{ \;\delta_{1}-\eps_{1}, \;\eps_{2}-\delta_{1}-K^{-},
\;\eps_{1}-\eps_{2}, \;K^{+}+\eps_{2} \; \}$ \\
\#62 : $\Delta^{0} = \{ \;\delta_{1}-\eps_{2}-K^{-}, \;\eps_{3}-\delta_{1}-2K^{-}, 
\;\eps_{2}-\eps_{3}, \;K^{+}+2\eps_{3}-\eps_{1}-\eps_{2} \; \}$ \\
\#63 : $\Delta^{0} = \{ \;\eps_{1}-\eps_{2}, \;\eps_{3}-\eps_{1}-K^{-},
\;\eps_{2}-\eps_{3}, \;K^{+}+\eps_{3}-\delta_{1} \; \}$ \\
\#64 : $\Delta^{0} = \{ \;\eps_{1}-\eps_{2}, \;\eps_{3}-\eps_{1}-K^{-},
\;\eps_{2}-\eps_{3}, \;K^{+}+\eps_{3} \; \}$ \\
\#65 : $\Delta^{0} = \{ \;\delta_{1}-\eps_{2}-K^{-}, \;-\delta_{1}-\eps_{2}-K^{-}, 
\;\eps_{2}-\eps_{3}, \;K^{+}+2\eps_{3}-\eps_{1}-\eps_{2} \; \}$ \\
\#66 : $\Delta^{0} = \{ \;-2\eps_{1}-K^{-}, \;\eps_{1}-\eps_{2},
\;\eps_{1}+\eps_{2}, \;2K^{+}-\eps_{1} \; \}$ \\
\#67 : $\Delta^{0} = \{ \;-\eps_{1}-K^{-}, \;\eps_{1}-\eps_{2},
\;\eps_{1}+\eps_{2}, \;K^{+}-\eps_{1} \; \}$ \\
\#68 : $\Delta^{0} = \{ \;-\eps_{1}-K^{-}, \;\eps_{1}-\eps_{2},
\;\eps_{1}+\eps_{2}, \;K^{+}-\eps_{1} \; \}$ \\
\#69 : $\Delta^{0} = \{ \;\eps_{1}-\eps_{2}-K^{-}, \;\eps_{2}-\eps_{3}, 
\;\eps_{2}+\eps_{3}, \;K^{+}-\eps_{2} \; \}$ \\
\#70 : $\Delta^{0} = \{ \;\delta_{1}-\eps_{1}-K^{-}, \;\eps_{1}-\eps_{2},
\;\eps_{1}+\eps_{2}, \;K^{+}-\eps_{1}+\eps_{3} \; \}$ \\
\#71 : $\Delta^{0} = \{ \;\delta_{1}-\eps_{1}-K^{-}, \;\eps_{1}-\eps_{2},
\;\eps_{1}+\eps_{2}, \;K^{+}-\eps_{1} \; \}$ \\
\#72 : $\Delta^{0} = \{ \;\delta_{1}-\eps_{1}-2K^{-}, \;\eps_{1}-\eps_{2},
\;\eps_{1}+\eps_{2}, \;K^{+}-2\eps_{1} \; \}$ \\
\#73 : $\Delta^{0} = \{ \;\delta_{1}-\eps_{1}-K^{-}, \;\eps_{1}-\eps_{2},
\;\eps_{1}+\eps_{2}, \;K^{+}-\eps_{1} \; \}$ \\

\vspace*{-15pt}

\subsubsection*{Rank 5 hyperbolic superalgebras}

Conventions for the simple root systems $\Delta^{0} = \{ \alpha_{1},
\alpha_{2}, \alpha_{3}, \alpha_{4}, \alpha_{5} \}$:

\begin{center}
\begin{tabular}{ccccc}
\begin{picture}(170,60) \thicklines 
  \put(0,40){\circle*{7}} \put(42,40){\circle*{7}} \put(84,40){\circle*{7}}
  \put(126,40){\circle*{7}} \put(168,40){\circle*{7}}
  \put(7,40){\line(1,0){28}} \put(49,40){\line(1,0){28}}
  \put(91,40){\line(1,0){28}} \put(133,40){\line(1,0){28}}
  \put(0,50){\makebox(0,0){$\alpha_{1}$}}
  \put(42,50){\makebox(0,0){$\alpha_{2}$}}
  \put(84,50){\makebox(0,0){$\alpha_{3}$}}
  \put(126,50){\makebox(0,0){$\alpha_{4}$}}
  \put(168,50){\makebox(0,0){$\alpha_{5}$}}
\end{picture}
&\hspace*{21pt}&
\begin{picture}(130,60) \thicklines 
  \put(0,40){\circle*{7}} \put(42,40){\circle*{7}} \put(84,40){\circle*{7}}
  \put(126,40){\circle*{7}} \put(7,40){\line(1,0){28}}
  \put(49,40){\line(1,0){28}} \put(91,40){\line(1,0){28}}
  \put(84,-2){\circle*{7}}\put(84,33){\line(0,-1){28}}
  \put(0,50){\makebox(0,0){$\alpha_{1}$}}
  \put(42,50){\makebox(0,0){$\alpha_{2}$}}
  \put(84,50){\makebox(0,0){$\alpha_{3}$}}
  \put(126,50){\makebox(0,0){$\alpha_{4}$}}
  \put(99,-2){\makebox(0,0){$\alpha_{5}$}}
\end{picture}
\\
\begin{picture}(120,50) \thicklines 
  \put(0,20){\circle*{7}} \put(42,20){\circle*{7}} \put(104,20){\circle*{7}}
  \put(73,40){\circle*{7}} \put(73,0){\circle*{7}}
  \put(7,20){\line(1,0){28}} \put(47,25){\line(3,2){19}}
  \put(47,15){\line(3,-2){19}} \put(80,38){\line(3,-2){19}}
  \put(80,2){\line(3,2){19}}
  \put(0,30){\makebox(0,0){$\alpha_{1}$}}
  \put(42,30){\makebox(0,0){$\alpha_{2}$}}
  \put(73,50){\makebox(0,0){$\alpha_{3}$}}
  \put(73,-10){\makebox(0,0){$\alpha_{4}$}}
  \put(119,20){\makebox(0,0){$\alpha_{5}$}}
\end{picture}
&\hspace*{21pt}&
\begin{picture}(130,50) \thicklines 
  \put(10,40){\circle*{7}} \put(10,0){\circle*{7}} \put(42,20){\circle*{7}}
  \put(84,20){\circle*{7}} \put(126,20){\circle*{7}}
  \put(17,38){\line(3,-2){19}} \put(17,2){\line(3,2){19}}
  \put(49,20){\line(1,0){28}} \put(91,20){\line(1,0){28}}
  \put(-5,40){\makebox(0,0){$\alpha_{1}$}}
  \put(-5,0){\makebox(0,0){$\alpha_{2}$}}
  \put(42,30){\makebox(0,0){$\alpha_{3}$}}
  \put(84,30){\makebox(0,0){$\alpha_{4}$}}
  \put(126,30){\makebox(0,0){$\alpha_{5}$}}
\end{picture}
&\hspace*{12pt}&
\begin{picture}(100,50) \thicklines 
  \put(0,20){\circle*{7}} \put(42,20){\circle*{7}} \put(84,20){\circle*{7}}
  \put(73,40){\circle*{7}} \put(73,0){\circle*{7}}
  \put(7,20){\line(1,0){28}} \put(47,25){\line(3,2){19}}
  \put(47,15){\line(3,-2){19}} \put(49,20){\line(1,0){28}} 
  \put(0,30){\makebox(0,0){$\alpha_{1}$}}
  \put(42,30){\makebox(0,0){$\alpha_{2}$}}
  \put(73,50){\makebox(0,0){$\alpha_{3}$}}
  \put(73,-10){\makebox(0,0){$\alpha_{4}$}}
  \put(99,20){\makebox(0,0){$\alpha_{5}$}}
\end{picture}
\end{tabular}
\end{center}

\vspace*{6pt}

\noindent 
\#1 : $\Delta^{0} = \{
\;\half(\widetilde\delta-\eps_{1}-\eps_{2}-\eps_{3})-K^{-}, \;\eps_{3},
\;\eps_{2}-\eps_{3}, \;\eps_{1}-\eps_{2}, \;\half K^{+}+\eps_{4}-\eps_{1}
\; \}$ \\
\#2 : $\Delta^{0} = \{ \;\half(\eps_{4}-\eps_{1}-\eps_{2}-\eps_{3})-K^{-},
\;\eps_{3}, \;\eps_{2}-\eps_{3}, \;\eps_{1}-\eps_{2},
\;\half K^{+}-\eps_{1}+\delta_{1} \; \}$ \\
\#3 : $\Delta^{0} = \{
\;\half(\widetilde\delta-\eps_{1}-\eps_{2}-\eps_{3})-K^{-}, \;\eps_{3},
\;\eps_{2}-\eps_{3}, \;\eps_{1}-\eps_{2}, \;\half K^{+}+\delta_{4}-\eps_{1}
\; \}$ \\
\#4 : $\Delta^{0} = \{ \;\eps_{1}-\eps_{2}-K^{-}, \;\eps_{2}-\eps_{3},
\;\eps_{3}, \;\half(\widetilde\delta-\eps_{1}-\eps_{2}-\eps_{3}),
\;K^{+}+\eps_{1} \; \}$ \\
\#5 : $\Delta^{0} = \{
\;\half(\widetilde\delta-\eps_{1}-\eps_{2}-\eps_{3})-K^{-}, \;\eps_{3},
\;\eps_{2}-\eps_{3}, \;\eps_{1}-\eps_{2}, \;K^{+}-2\eps_{1} \; \}$ \\
\#6 : $\Delta^{0} = \{
\;\half(\widetilde\delta-\eps_{1}-\eps_{2}-\eps_{3})-K^{-}, \;\eps_{3},
\;\eps_{2}-\eps_{3}, \;\eps_{1}-\eps_{2}, \;\half K^{+}-\eps_{1} \; \}$ \\
\#7 : $\Delta^{0} = \{ \;\eps_{1}+\eps_{2}+\eps_{3}+\eps_{4}-K^{-},
\;-2\eps_{1}, \;\eps_{1}-\eps_{2}, \;K^{+}+\eps_{2}-\delta_{1},
\;2\delta_{1} \; \}$ \\
\#8 : $\Delta^{0} = \{ \;\eps_{1}+\eps_{2}+\eps_{3}+\eps_{4}-K^{-},
\;-2\eps_{1}, \;\eps_{1}-\eps_{2}, \;K^{+}+\eps_{2}-\delta_{1},
\;\delta_{1} \; \}$ \\
\#9 : $\Delta^{0} = \{ \;\eps_{1}+\eps_{2}+\eps_{3}+\eps_{4}-K^{-},
\;-2\eps_{1}, \;\eps_{1}-\eps_{2}, \;\eps_{2}-\eps_{3}, \;K^{+}+\eps_{3} \;
\}$ \\
\#10 : $\Delta^{0} = \{ \;\half(\eps_{4}-\eps_{1}-\eps_{2}-\eps_{3})-K^{-},
\;\eps_{3}, \;\eps_{2}-\eps_{3}, \;\eps_{1}-\eps_{2}, \;\half
K^{+}-\eps_{1} \; \}$ \\
\#11 : $\Delta^{0} = \{
\;\half(\widetilde\delta-\eps_{1}-\eps_{2}-\eps_{3})-K^{-}, \;\eps_{3},
\;\eps_{2}-\eps_{3}, \;\eps_{1}-\eps_{2}, \;\half K^{+}-\eps_{1} \; \}$ \\
\#12 : $\Delta^{0} = \{ \;\eps_{1}+\eps_{2}+\eps_{3}+\eps_{4}-K^{-},
\;-2\eps_{1}, \;\eps_{1}-\eps_{2}, \;K^{+}+\eps_{2}-\delta_{1},
\;\delta_{1} \; \}$ \\
\#13 : $\Delta^{0} = \{ \;\delta_{1}-\eps_{1}-K^{-}, \;\eps_{1}-\eps_{2},
\;\eps_{2}-\eps_{3}, \;2\eps_{3}, \;
\;K^{+}-\eps_{1}-\eps_{2}-\eps_{3}-\eps_{4} \}$ \\
\#14 : $\Delta^{0} = \{ \;-\eps_{1}-K^{-}, \;\eps_{1}-\eps_{2}, 
\;\eps_{2}-\eps_{3}, \;\eps_{3}, \;K^{+}-\eps_{1}-\eps_{2} \; \}$ \\
\#15 : $\Delta^{0} = \{ \;-\eps_{1}-K^{-}, \;\eps_{1}-\eps_{2}, 
\;\eps_{2}-\eps_{3}, \;2\eps_{3}, \;K^{+}-\eps_{1}-\eps_{2} \; \}$ \\
\#16 : $\Delta^{0} = \{ \;\eps_{1}-\eps_{2}-K^{-}, \;\eps_{2}-\eps_{3}, 
\;\eps_{3}-\eps_{4}, \;\eps_{4}, \;K^{+}-\eps_{2}-\eps_{3} \; \}$ \\
\#17 : $\Delta^{0} = \{ \;-2\eps_{1}-K^{-}, \;\eps_{1}-\eps_{2}, 
\;\eps_{2}-\eps_{3}, \;\eps_{3}, \;2K^{+}-\eps_{1}-\eps_{2} \; \}$ \\
\#18 : $\Delta^{0} = \{ \;-\eps_{1}-K^{-}, \;\eps_{1}-\eps_{2}, 
\;\eps_{2}-\eps_{3}, \;\eps_{3}, \;K^{+}-\eps_{1}-\eps_{2} \; \}$ \\
\#19 : $\Delta^{0} = \{ \;-\eps_{1}-K^{-}, \;\eps_{1}-\eps_{2}, 
\;\eps_{2}-\eps_{3}, \;\eps_{3}, \;K^{+}-\eps_{1}-\eps_{2} \; \}$ \\
\#20 : $\Delta^{0} = \{ \;\delta_{1}-\eps_{1}-K^{-}, \;\eps_{1}-\eps_{2},
\;\eps_{2}-\eps_{3}, \;\eps_{3}, \;K^{+}-\eps_{1}-\eps_{2} \; \}$ \\
\#21 : $\Delta^{0} = \{ \;-\eps_{1}-K^{-}, \;\eps_{1}-\eps_{2},
\;\eps_{2}-\eps_{3}, \;\eps_{3}-\delta_{1}, \;K^{+}-\eps_{1}-\eps_{2} \;
\}$ \\
\#22 : $\Delta^{0} = \{ \;\eps_{1}-\eps_{2}-K^{-}, \;\eps_{2}-\eps_{3},
\;\eps_{3}-\eps_{4}, \;\eps_{4}-\delta_{1}, \;K^{+}-\eps_{2}-\eps_{3}
\; \}$ \\
\#23 : $\Delta^{0} = \{ \;\delta_{1}-\eps_{1}-K^{-}, \;\eps_{1}-\eps_{2},
\;\eps_{2}-\eps_{3}, \;\eps_{3}-\delta_{2}, \;K^{+}-\eps_{1}-\eps_{2} \;
\}$ \\
\#24 : $\Delta^{0} = \{ \;\delta_{1}-\eps_{1}-K^{-}, \;\eps_{1}-\eps_{2}, 
\;\eps_{2}-\eps_{3}, \;2\eps_{3}, \;K^{+}-\eps_{1}-\eps_{2} \; \}$ \\
\#25 : $\Delta^{0} = \{ \;\delta_{1}-\eps_{1}-K^{-}, \;\eps_{1}-\eps_{2},
\;\eps_{2}-\eps_{3}, \;\eps_{3}, \;K^{+}-\eps_{1}-\eps_{2} \; \}$ \\
\#26 : $\Delta^{0} = \{ \;-2\eps_{1}-K^{-}, \;\eps_{1}-\eps_{2},
\;\eps_{2}-\eps_{3}, \;\eps_{3}-\delta_{1}, \;2K^{+}-\eps_{1}-\eps_{2}
\; \}$ \\
\#27 : $\Delta^{0} = \{ \;-\eps_{1}-K^{-}, \;\eps_{1}-\eps_{2},
\;\eps_{2}-\eps_{3}, \;\eps_{3}-\delta_{1}, \;K^{+}-\eps_{1}-\eps_{2} \;
\}$ \\
\#28 : $\Delta^{0} = \{ \;\eps_{4}-\delta_{1}-K^{-}, \;\eps_{3}-\eps_{4},
\;\eps_{2}-\eps_{3}, \;K^{+}+\eps_{4}-\eps_{1}, \;\eps_{1}-\eps_{2} \; \}$ \\
\#29 : $\Delta^{0} = \{ \;-\eps_{1}-\eps_{2}-K^{-}, \;\eps_{2}-\eps_{3},
\;\eps_{3}-\delta_{1}, \;\eps_{1}-\eps_{2}, \;K^{+}+\delta_{1}-\eps_{1} \;
\}$ \\
\#30 : $\Delta^{0} = \{ \;-2\eps_{2}-K^{-}, \;\eps_{1}+\eps_{2},
\;\delta_{1}-\eps_{1}, \;-\delta_{1}-\eps_{1}, \;2K^{+}+\eps_{1}-\eps_{2}
\; \}$ \\
\#31 : $\Delta^{0} = \{ \;-\eps_{2}-K^{-}, \;\eps_{1}+\eps_{2},
\;\delta_{1}-\eps_{1}, \;-\delta_{1}-\eps_{1}, \;K^{+}+\eps_{1}-\eps_{2} \;
\}$ \\
\#32 : $\Delta^{0} = \{ \;-\eps_{2}-K^{-}, \;\eps_{1}+\eps_{2},
\;\delta_{1}-\eps_{1}, \;-\delta_{1}-\eps_{1}, \;K^{+}+\eps_{1}-\eps_{2} \;
\}$ \\
\#33 : $\Delta^{0} = \{ \;\eps_{4}-K^{-}, \;\eps_{3}-\eps_{4}, 
\;\eps_{2}-\eps_{3}, \;K^{+}+\eps_{4}-\eps_{1}, \;\eps_{1}-\eps_{2} \; \}$ \\
\#34 : $\Delta^{0} = \{ \;\eps_{3}-\eps_{2}-K^{-}, \;\eps_{1}+\eps_{2},
\;\delta_{1}-\eps_{1}, \;-\delta_{1}-\eps_{1}, \;K^{+}+\eps_{1}-\eps_{2} \;
\}$ \\
\#35 : $\Delta^{0} = \{ \;-\eps_{2}-K^{-}, \;\eps_{2}-\eps_{3}, 
\;\eps_{3}-\eps_{4}, \;\eps_{3}+\eps_{4}, \;K^{+}+\eps_{1}-\eps_{2} \; \}$ \\
\#36 : $\Delta^{0} = \{ \;\delta_{1}-\eps_{1}-K^{-},
\;-\delta_{1}-\eps_{1}-K^{-}, \;\eps_{1}-\eps_{2}, \;2\eps_{2},
\;-\eps_{1}-\eps_{2}-\eps_{3}-\eps_{4}+K^{+} \; \}$ \\
\#37 : $\Delta^{0} = \{ \;\eps_{1}-\eps_{2}, \;K^{+}-\eps_{1}-\eps_{2},
\;\eps_{2}-\eps_{3}, \;\eps_{3},
\;\half(\widetilde\delta-\eps_{1}-\eps_{2}-\eps_{3})-K^{-} \; \}$ \\
\#38 : $\Delta^{0} = \{ \;\delta_{1}-\eps_{2}-K^{-}, \;\eps_{2}-\eps_{3},
\;\eps_{3}-\eps_{4}, \;\eps_{3}+\eps_{4}, \;K^{+}+\eps_{1}-\eps_{2} \; \}$ \\

\vspace*{-15pt}

\subsubsection*{Rank 6 hyperbolic superalgebras}

Conventions for the simple root systems $\Delta^{0} = \{ \alpha_{1},
\alpha_{2}, \alpha_{3}, \alpha_{4}, \alpha_{5}, \alpha_{6} \}$:

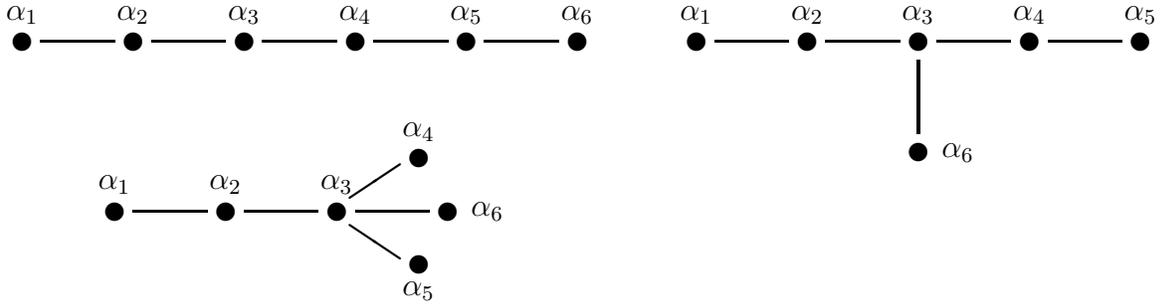
\begin{center}
\begin{tabular}{ccccc}
\begin{picture}(210,60) \thicklines 
  \put(0,40){\circle*{7}} \put(42,40){\circle*{7}} \put(84,40){\circle*{7}}
  \put(126,40){\circle*{7}} \put(168,40){\circle*{7}}
  \put(210,40){\circle*{7}} \put(7,40){\line(1,0){28}}
  \put(49,40){\line(1,0){28}} \put(91,40){\line(1,0){28}}
  \put(133,40){\line(1,0){28}} \put(175,40){\line(1,0){28}}
  \put(0,50){\makebox(0,0){$\alpha_{1}$}}
  \put(42,50){\makebox(0,0){$\alpha_{2}$}}
  \put(84,50){\makebox(0,0){$\alpha_{3}$}}
  \put(126,50){\makebox(0,0){$\alpha_{4}$}}
  \put(168,50){\makebox(0,0){$\alpha_{5}$}}
  \put(210,50){\makebox(0,0){$\alpha_{6}$}}
\end{picture}
&\hspace*{21pt}&
\begin{picture}(170,60) \thicklines 
  \put(0,40){\circle*{7}} \put(42,40){\circle*{7}} \put(84,40){\circle*{7}}
  \put(126,40){\circle*{7}} \put(168,40){\circle*{7}}
  \put(7,40){\line(1,0){28}} \put(49,40){\line(1,0){28}}
  \put(91,40){\line(1,0){28}} \put(133,40){\line(1,0){28}}
  \put(84,-2){\circle*{7}}\put(84,33){\line(0,-1){28}}
  \put(0,50){\makebox(0,0){$\alpha_{1}$}}
  \put(42,50){\makebox(0,0){$\alpha_{2}$}}
  \put(84,50){\makebox(0,0){$\alpha_{3}$}}
  \put(126,50){\makebox(0,0){$\alpha_{4}$}}
  \put(168,50){\makebox(0,0){$\alpha_{5}$}}
  \put(99,-2){\makebox(0,0){$\alpha_{6}$}}
\end{picture}
\\
\begin{picture}(140,40) \thicklines 
  \put(0,20){\circle*{7}} \put(42,20){\circle*{7}} \put(84,20){\circle*{7}}
  \put(126,20){\circle*{7}} \put(115,40){\circle*{7}}
  \put(115,0){\circle*{7}} \put(7,20){\line(1,0){28}}
  \put(89,25){\line(3,2){19}} \put(89,15){\line(3,-2){19}}
  \put(49,20){\line(1,0){28}} \put(91,20){\line(1,0){28}}
  \put(0,30){\makebox(0,0){$\alpha_{1}$}}
  \put(42,30){\makebox(0,0){$\alpha_{2}$}}
  \put(84,30){\makebox(0,0){$\alpha_{3}$}}
  \put(115,50){\makebox(0,0){$\alpha_{4}$}}
  \put(115,-10){\makebox(0,0){$\alpha_{5}$}}
  \put(141,20){\makebox(0,0){$\alpha_{6}$}}
\end{picture}
\end{tabular}
\end{center}

\vspace*{6pt}

\noindent
\#1 : $\Delta^{0} = \{ \;-\eps_{1}+\delta_{1}-\half K^{-},
\;\eps_{1}-\eps_{2}, \;\eps_{2}-\eps_{3}, \;\eps_{3}-\eps_{4}, \;\eps_{4},
\;K^{+}-\half(\eps_{1}+\eps_{2}+\eps_{3}+\eps_{4}) \; \}$ \\
\#2 : $\Delta^{0} = \{ \;\delta_{1}-\eps_{1}-K^{-}, \;\eps_{1}-\eps_{2},
\;\eps_{2}-\eps_{3}, \;\eps_{3}-\eps_{4}, \;2\eps_{4},
\;K^{+}-(\eps_{1}+\eps_{2}+\eps_{3}+\eps_{4}) \; \}$ \\
\#3 : $\Delta^{0} = \{ \;-\eps_{1}-\half K^{-}, \;\eps_{1}-\eps_{2},
\;\eps_{2}-\eps_{3}, \;\eps_{3}-\eps_{4}, \;\eps_{4},
\;K^{+}-\half(\eps_{1}+\eps_{2}+\eps_{3}+\eps_{4}) \; \}$ \\
\#4 : $\Delta^{0} = \{ \;-\eps_{1}-K^{-}, \;\eps_{1}-\eps_{2},
\;\eps_{2}-\eps_{3}, \;\eps_{3}-\eps_{4}, \;2\eps_{4},
\;K^{+}-(\eps_{1}+\eps_{2}+\eps_{3}+\eps_{4}) \; \}$ \\
\#5 : $\Delta^{0} = \{ \;\delta_{1}-\eps_{1}-K^{-}, \;\eps_{1}-\eps_{2},
\;\eps_{2}-\eps_{3}, \;\eps_{3}-\eps_{4}, \;\eps_{4}-\eps_{5},
\;K^{+}-\eps_{1}-\eps_{2} \; \}$ \\
\#6 : $\Delta^{0} = \{
\;\half(\widetilde\delta-\eps_{1}-\eps_{2}-\eps_{3})-3K^{-},
\;\eps_{3}+\eps_{4}, \;\eps_{2}-\eps_{3}, \;\eps_{3}-\eps_{4}, 
\;\half(-\widetilde\delta-\eps_{1}-\eps_{2}-\eps_{3})+\half K^{+},
\;\eps_{1}-\eps_{2} \; \}$ \\
\#7 : $\Delta^{0} = \{ \;\delta_{1}-\eps_{1}-K^{-}, \;\eps_{1}-\eps_{2},
\;\eps_{2}-\eps_{3}, \;\eps_{3}-\eps_{4}, \;2\eps_{4},
\;K^{+}-\eps_{1}-\eps_{2} \; \}$ \\
\#8 : $\Delta^{0} = \{ \;\delta_{1}-\eps_{1}-K^{-}, \;\eps_{1}-\eps_{2},
\;\eps_{2}-\eps_{3}, \;\eps_{3}-\eps_{4}, \;\eps_{4},
\;K^{+}-\eps_{1}-\eps_{2} \; \}$ \\
\#9 : $\Delta^{0} = \{ \;-\eps_{1}-K^{-}, \;\eps_{1}-\eps_{2}, 
\;\eps_{2}-\eps_{3}, \;\eps_{3}-\eps_{4}, \;\eps_{4}-\eps_{5}, 
\;K^{+}-\eps_{1}-\eps_{2} \; \}$ \\
\#10 : $\Delta^{0} = \{ \;-\eps_{1}-K^{-}, \;\eps_{1}-\eps_{2}, 
\;\eps_{2}-\eps_{3}, \;\eps_{3}-\eps_{4}, \;\eps_{4}, 
\;K^{+}-\eps_{1}-\eps_{2} \; \}$ \\
\#11 : $\Delta^{0} = \{ \;-\eps_{1}-K^{-}, \;\eps_{1}-\eps_{2}, 
\;\eps_{2}-\eps_{3}, \;\eps_{3}-\eps_{4}, \;2\eps_{4}, 
\;K^{+}-\eps_{1}-\eps_{2} \; \}$ \\
\#12 : $\Delta^{0} = \{ \;-\eps_{1}-K^{-}, \;\eps_{1}-\eps_{2},
\;\eps_{2}-\eps_{3}, \;\eps_{3}-\eps_{4}, \;\eps_{4},
\;K^{+}-\eps_{1}-\eps_{2} \; \}$ \\
\#13 : $\Delta^{0} = \{ \;\delta_{1}-\eps_{1}-K^{-}, \;\eps_{1}-\eps_{2},
\;\eps_{2}-\eps_{3}, \;\eps_{3}-\eps_{4}, \;\eps_{4},
\;K^{+}-\eps_{1}-\eps_{2} \; \}$ \\
\#14 : $\Delta^{0} = \{ \;\delta_{1}-\eps_{1}-K^{-}, \;\eps_{1}-\eps_{2},
\;\eps_{2}-\eps_{3}, \;\eps_{3}-\eps_{4}, \;\eps_{3}+\eps_{4},
\;K^{+}-\eps_{1}-\eps_{2} \; \}$ \\
\#15 : $\Delta^{0} = \{ \;-\eps_{1}-K^{-}, \;\eps_{1}-\eps_{2},
\;\eps_{2}-\eps_{3}, \;\eps_{3}-\eps_{4}, \;\eps_{3}+\eps_{4},
\;K^{+}-\eps_{1}-\eps_{2} \; \}$ \\

\section*{C. Subalgebras of the hyperbolic KM superalgebras}

\subsubsection*{Rank 3 hyperbolic superalgebras}

\hspace*{-6pt}%
\begin{tabular}{ll} 
\#1 : $G_{2}$, $osp(1|2) \oplus sl(2)$, $osp(1|4)$ 
& \#2 : $G_{2}$, $osp(1|2) \oplus sl(2)$, $osp(1|4)$ \\
\#3 : $sl(1|3)^{(4)}$, $osp(1|2) \oplus sl(2)$, $sl(3)$ 
& \#4 : $osp(1|4)$, $osp(1|2) \oplus sl(2)$, $sl(3)^{(2)}$ \\
\#5 : $osp(1|4)$, $osp(1|2) \oplus sl(2)$, $sl(3)^{(2)}$ 
& \#6 : $sl(1|3)^{(4)}$, $osp(1|2) \oplus sl(2)$, $sl(2)^{(1)}$ \\
\#7 : $sl(1|3)^{(4)}$, $osp(1|2) \oplus sl(2)$, $sp(4)$ 
& \#8 : $sl(1|3)^{(4)}$, $osp(1|2) \oplus sl(2)$, $sp(4)$ \\
\#9 : $sl(1|3)^{(4)}$, $osp(1|2) \oplus sl(2)$, $sl(3)^{(2)}$ 
& \#10 : $sl(1|3)^{(4)}$, $osp(1|2) \oplus sl(2)$, $G_{2}$ \\
\#11 : $sl(1|3)^{(4)}$, $osp(1|2) \oplus sl(2)$, $G_{2}$ 
& \#12 : $sl(1|3)^{(4)}$, $osp(1|2) \oplus sl(2)$, $sl(3)^{(2)}$ \\
\#13 : $osp(1|2)^{(1)}$, $osp(1|2) \oplus sl(2)$, $sp(4)$ 
& \#14 : $osp(1|2)^{(1)}$, $osp(1|2) \oplus sl(2)$, $sp(4)$ \\
\#15 : $osp(1|2)^{(1)}$, $osp(1|2) \oplus sl(2)$, $sl(3)$ 
& \#16 : $osp(1|2)^{(1)}$, $osp(1|2) \oplus sl(2)$, $G_{2}$ \\
\#17 : $osp(1|2)^{(1)}$, $osp(1|2) \oplus sl(2)$, $G_{2}$ 
& \#18 : $osp(1|2)^{(1)}$, $osp(1|2) \oplus sl(2)$, $sl(2)^{(1)}$ \\
\#19 : $osp(1|2)^{(1)}$, $osp(1|2) \oplus sl(2)$, $sl(3)^{(2)}$ 
& \#20 : $osp(1|2)^{(1)}$, $osp(1|2) \oplus sl(2)$, $sl(3)^{(2)}$ \\
\#21 : $osp(1|4)$, $osp(1|2) \oplus sl(2)$, $sl(2)^{(1)}$ 
& \#22 : $osp(1|2)^{(1)}$, $2\, sl(2)$, $sl(1|3)^{(4)}$ \\
\#23 : $osp(1|2)^{(1)}$, $2\, sl(2)$, $osp(1|4)$  
& \#24 : $osp(1|2)^{(1)}$, $2\, sl(2)$ \\
\#25 : $sl(1|3)^{(4)}$, $2\, sl(2)$ 
& \#26 : $sl(1|3)^{(4)}$, $2\, sl(2)$, $osp(1|4)$ \\
\#27 : $osp(2|2)^{(2)}$, $2\, osp(1|2)$ 
& \#28 : $sl(1|3)^{(4)}$, $2\, osp(1|2)$ \\
\#29 : $osp(1|4)$, $2\, osp(1|2)$, $sl(1|3)^{(4)}$ 
& \#30 :$osp(1|2)^{(1)}$, $2\, osp(1|2)$, $sl(1|3)^{(4)}$ \\
\#31 : $osp(1|2)^{(1)}$, $2\, osp(1|2)$ 
& \#32 : $sl(1|3)^{(4)}$, $osp(1|2) \oplus sl(2)$, $osp(2|2)^{(2)}$ \\
\#33 : $osp(1|4)$, $osp(1|2) \oplus sl(2)$, $osp(2|2)^{(2)}$ 
& \#34 : $osp(1|2)^{(1)}$, $2\, osp(1|2)$, $osp(1|4)$ \\
\#35 : $osp(1|2)^{(1)}$, $osp(1|2) \oplus sl(2)$, $osp(2|2)^{(2)}$ 
& \#36 : $G_{2}$, $sl(1|1) \oplus sl(2)$, $sl(1|2)$ \\
\#37 : $sl(3)^{(2)}$, $sl(1|1) \oplus sl(2)$, $sl(1|2)$ 
& \#38 : $sl(3)^{(2)}$, $sl(1|1) \oplus sl(2)$, $sl(1|2)$ \\
\#39 : $sl(2)^{(1)}$, $sl(1|1) \oplus sl(2)$, $sl(1|2)$ 
& \#40 : $sl(3)$, $sl(1|1) \oplus sl(2)$, $osp(3|2)$ \\
\#41 : $sp(4)$, $sl(1|1) \oplus sl(2)$, $osp(3|2)$ 
& \#42 : $sp(4)$, $sl(1|1) \oplus sl(2)$, $osp(3|2)$ \\
\#43 : $G_{2}$, $sl(1|1) \oplus sl(2)$, $osp(3|2)$ 
& \#44 : $G_{2}$, $sl(1|1) \oplus sl(2)$, $osp(3|2)$ \\
\#45 : $sl(3)^{(2)}$, $sl(1|1) \oplus sl(2)$, $osp(2|2)$ 
& \#46 : $G_{2}$, $sl(1|1) \oplus sl(2)$, $osp(2|2)$ \\
\#47 : $G_{2}$, $sl(1|1) \oplus sl(2)$, $osp(2|2)$ 
& \#48 : $sl(3)^{(2)}$, $sl(1|1) \oplus sl(2)$, $osp(2|2)$ \\
\#49 : $sl(3)^{(2)}$, $sl(1|1) \oplus sl(2)$, $osp(3|2)$ 
& \#50 : $sl(3)^{(2)}$, $sl(1|1) \oplus sl(2)$, $osp(3|2)$ \\
\#51 : $sl(2)^{(1)}$, $sl(1|1) \oplus sl(2)$, $osp(3|2)$ 
& \#52 : $sl(3)$, $sl(1|1) \oplus sl(2)$, $osp(2|2)$ \\
\#53 : $so(5)$, $sl(1|1) \oplus sl(2)$, $osp(2|2)$ 
& \#54 : $sp(4)$, $sl(1|1) \oplus sl(2)$, $osp(2|2)$ \\
\#55 : $sl(2)^{(1)}$, $sl(1|1) \oplus sl(2)$, $osp(2|2)$ 
& \#56 : $sl(1|3)^{(4)}$, $sl(1|1) \oplus osp(1|2)$, $sl(1|2)$ \\
\#57 : $osp(1|2)^{(1)}$, $sl(1|1) \oplus osp(1|2)$, $sl(1|2)$ 
& \#58 : $sl(1|3)^{(4)}$, $sl(1|1) \oplus osp(1|2)$, $osp(3|2)$ \\
\#59 : $osp(1|4)$, $osp(1|2) \oplus sl(1|1)$, $osp(3|2)$
& \#60 : $osp(1|2)^{(1)}$, $osp(1|2) \oplus sl(1|1)$, $osp(3|2)$ \\
\#61 : $sl(1|3)^{(4)}$, $sl(1|1) \oplus osp(1|2)$, $osp(2|2)$ 
& \#62 : $osp(1|4)$, $osp(1|2) \oplus sl(1|1)$, $osp(2|2)$ \\
\#63 : $osp(1|2)^{(1)}$, $osp(1|2) \oplus sl(1|1)$, $osp(2|2)$ 
& \#64 : $sl(1|3)^{(4)}$, $sl(3)$ \\
\#65 : $osp(1|4)$, $sl(3)$ 
& \#66 : $osp(1|2)^{1)}$, $sl(3)$ \\
\#67 : $osp(1|4)$, $sp(4)$, $sl(1|3)^{(4)}$ 
& \#68 : $osp(1|4)$, $osp(1|2)^{(1)}$, $sp(4)$ \\
\#69 : $sl(1|3)^{(4)}$, $sl(2)^{(1)}$ 
& \#70 : $osp(1|4)$, $sl(2)^{(1)}$ \\
\#71 : $osp(1|2)^{(1)}$, $sl(2)^{(1)}$ 
& \#72 : $osp(2|2)^{(2)}$, $sl(1|3)^{(4)}$ \\
\#73 : $osp(2|2)^{(2)}$, $osp(1|4)$ 
& \#74 : $osp(2|2)^{(2)}$, $osp(1|2)^{(1)}$ \\
\#75 : $osp(2|2)^{(2)}$ 
& \#76 : $sl(1|2)$, $sl(3)$ \\
\#77 : $osp(2|2)$, $sl(3)$ 
& \#78 : $osp(3|2)$, $sl(3)$ \\
\#79 : $sl(1|2)$, $sl(2)^{(1)}$ 
& \#80 : $osp(2|2)$, $sl(2)^{(1)}$ \\
\#81 : $osp(3|2)$, $sl(2)^{(1)}$ 
& \#82 : $so(5)$, $sl(1|2)$, $osp(3|2)$ \\
\#83 : $sp(4)$, $sl(1|2)$, $osp(2|2)$ 
& \#84 : $sl(3)^{(2)}$, $osp(3|2)$, $osp(2|2)$ \\
\#85 : $osp(3|2)$, $sl(1|2)$ 
& \#86 : $osp(2|2)$, $sl(1|2)$ \\
\#87 : $osp(3|2)$, $sl(1|2)$ \\
\end{tabular}

\subsubsection*{Rank 4 hyperbolic superalgebras}

\#1 : $osp(1|4)^{(1)}$, $sl(2) \oplus osp(1|4)$, $sl(3) \oplus osp(1|2)$,
$so(7)$ \\
\#2 : $osp(1|4)^{(1)}$, $sl(2) \oplus osp(1|4)$, $so(5) \oplus osp(1|2)$,
$sl(5)^{(2)}$ \\
\#3 : $osp(1|4)^{(1)}$, $sl(2) \oplus osp(1|4)$, $sp(4) \oplus osp(1|2)$,
$sl(4)^{(2)}$ \\
\#4 : $sl(1|5)^{(4)}$, $sl(2) \oplus osp(1|4)$, $sl(3) \oplus osp(1|2)$,
$sp(6)$ \\
\#5 : $sl(1|5)^{(4)}$, $sl(2) \oplus osp(1|4)$, $sp(4) \oplus osp(1|2)$,
$sp(4)^{(1)}$ \\
\#6 : $sl(1|5)^{(4)}$, $sl(2) \oplus osp(1|4)$, $sp(4) \oplus osp(1|2)$,
$sl(5)^{(2)}$ \\
\#7 : $sl(1|4)^{(2)}$, $sl(2) \oplus osp(1|4)$, $sl(3) \oplus sl(2)$,
$osp(1|6)$ \\
\#8 : $sl(1|4)^{(2)}$, $sl(2) \oplus osp(1|4)$, $sp(4) \oplus sl(2)$,
$osp(1|4)^{(1)}$ \\
\#9 : $sl(1|4)^{(2)}$, $sl(2) \oplus osp(1|4)$, $sp(4) \oplus sl(2)$,
$sl(1|5)^{(4)}$ \\
\#10 : $osp(1|4)^{(1)}$, $osp(1|2) \oplus osp(1|4)$, $sl(1|5)^{(4)}$ \\
\#11 : $osp(2|4)^{(2)}$, $sl(2) \oplus osp(1|4)$, $osp(1|4) \oplus
osp(1|2)$, $sl(1|4)^{(2)}$ \\
\#12 : $osp(2|4)$, $sl(2) \oplus sl(1|2)$, $sp(4) \oplus sl(1|1)$,
$sl(5)^{(2)}$ \\
\#13 : $osp(2|4)$, $sl(2) \oplus sl(1|2)$, $so(5) \oplus sl(1|1)$,
$sl(4)^{(2)}$ \\
\#14 : $osp(5|2)$, $sl(2) \oplus sl(1|2)$, $sp(4) \oplus sl(1|1)$,
$sp(4)^{(1)}$ \\
\#15 : $osp(5|2)$, $sl(2) \oplus sl(1|2)$, $sl(3) \oplus sl(1|1)$, $sp(6)$
\\
\#16 : $osp(5|2)$, $sl(2) \oplus sl(1|2)$, $sp(4) \oplus sl(1|1)$,
$sl(5)^{(2)}$ \\
\#17 : $osp(1|4)^{(1)}$, $sl(1|1) \oplus osp(1|4)$, $sl(1|2) \oplus
osp(1|2)$, $osp(5|2)$ \\
\#18 : $sl(1|5)^{(4)}$, $sl(1|1) \oplus osp(1|4)$, $sl(1|2) \oplus
osp(1|2)$, $osp(2|4)$ \\
\#19 : $sl(1|4)^{(2)}$, $sl(1|1) \oplus osp(1|4)$, $sl(1|2) \oplus sl(2)$,
$osp(3|4)$ \\
\#20 : $sl(1|3)$, $sl(2) \oplus sl(1|2)$, $G_{2} \oplus sl(1|1)$,
$D_{4}^{(3)}$ \\
\#21 : $sl(1|3)$, $sl(2) \oplus sl(1|2)$, $G_{2} \oplus sl(1|1)$,
$G_{2}^{(1)}$ \\
\#22 : $G(3)$, $sl(2) \oplus sl(1|2)$, $sl(3) \oplus sl(1|1)$,
$G_{2}^{(1)}$ \\
\#23 : $osp(4|2)$, $sl(2) \oplus osp(2|2)$, $G_{2} \oplus sl(2)$, $G(3)$ \\
\#24 : $osp(3|4)$, $sl(2) \oplus osp(3|2)$, $G_{2} \oplus sl(2)$, $G(3)$ \\
\#25 : $sl(2|2)$, $sl(2) \oplus sl(1|2)$, $G_{2} \oplus sl(2)$, $G(3)$ \\
\#26 : $osp(1|6)$, $sl(2) \oplus osp(1|4)$, $G_{2} \oplus osp(1|2)$,
$G_{2}^{(1)}$ \\
\#27 : $osp(1|6)$, $sl(2) \oplus osp(1|4)$, $G_{2} \oplus osp(1|2)$,
$D_{4}^{(3)}$ \\
\#28 : $osp(5|2)$, $sl(2) \oplus osp(3|2)$, $G_{2} \oplus osp(1|2)$, $G(3)$
\\
\#29 : $sp(6)$, $sl(1|1) \oplus 2\, sl(2)$, $sl(1|3)$, $osp(2|4)$ \\
\#30 : $so(7)$, $sl(1|1) \oplus 2\, sl(2)$, $sl(1|3)$, $osp(5|2)$ \\
\#31 : $osp(1|6)$, $osp(1|2) \oplus 2\, sl(2)$, $sp(6)$, $osp(1|4)^{(1)}$
\\
\#32 : $osp(1|6)$, $osp(1|2) \oplus 2\, sl(2)$, $so(7)$, $sl(1|5)^{(4)}$ \\
\#33 : $osp(5|2)$, $sl(1|1) \oplus 2\, sl(2)$, $sl(4)^{(2)}$ \\
\#34 : $osp(2|4)$, $sl(1|1) \oplus 2\, sl(2)$, $sp(4)^{(1)}$ \\
\#35 : $osp(2|4)$, $sl(1|1) \oplus 2\, sl(2)$, $osp(5|2)$, $sl(5)^{(2)}$ \\
\#36 : $G(3)$, $sl(1|1) \oplus 2\, sl(2)$, $sl(1|3)$, $D_{4}^{(3)}$ \\
\#37 : $osp(3|4)$, $osp(1|2) \oplus sl(1|1) \oplus sl(2)$, $sl(1|3)$,
$osp(1|6)$ \\
\#38 : $osp(3|4)$, $osp(1|2) \oplus sl(1|1) \oplus sl(2)$, $osp(2|4)$,
$osp(1|4)^{(1)}$ \\
\#39 : $osp(3|4)$, $osp(1|2) \oplus sl(1|1) \oplus sl(2)$, $osp(5|2)$,
$sl(1|5)^{(4)}$ \\
\#40 : $osp(3|4)$, $2\, osp(1|2) \oplus sl(1|1)$, $osp(2|4)^{(2)}$ \\
\#41 : $sl(1|5)^{(2)}$, $osp(1|2) \oplus 2\, sl(2)$, $sl(4)^{(2)}$ \\
\#42 : $osp(1|4)^{(1)}$, $osp(1|2) \oplus 2\, sl(2)$, $sp(4)^{(1)}$ \\
\#43 : $osp(1|4)^{(1)}$, $osp(1|2) \oplus 2\, sl(2)$, $sl(1|5)^{(4)}$,
$sl(5)^{(2)}$ \\
\#44 : $sl(1|4)^{(2)}$, $3\, sl(2)$ \\
\#45 : $osp(1|6)$, $2\, osp(1|2) \oplus sl(2)$, $sl(1|5)^{(4)}$ \\
\#46 : $osp(1|4)^{(1)}$, $2\, osp(1|2) \oplus sl(2)$, $sl(1|5)^{(4)}$ \\
\#47 : $osp(2|4)^{(2)}$, $2\, osp(1|2) \oplus sl(2)$, $sl(1|5)^{(4)}$ \\
\#48 : $sl(1|5)^{(4)}$, $3\, osp(1|2)$ \\
\#49 : $sl(1|3)$, $sl(2) \oplus sl(3)$, $sl(1|2) \oplus sl(2)$, $osp(4|2)$
\\
\#50 : $osp(5|2)$, $sl(2) \oplus so(5)$, $sl(1|2) \oplus sl(2)$, $osp(4|2)$
\\
\#51 : $osp(2|4)$, $sl(2) \oplus sp(4)$, $sl(1|2) \oplus sl(2)$, $osp(4|2)$
\\
\#52 : $G(3)$, $sl(2) \oplus G_{2}$, $sl(1|2) \oplus sl(2)$, $osp(4|2)$ \\
\#53 : $osp(3|4)$, $sl(2) \oplus osp(1|4)$, $sl(1|2) \oplus osp(1|2)$,
$osp(4|2)$ \\
\#54 : $osp(4|2)$, $3\, sl(2)$ \\
\hspace*{-6pt}%
\begin{tabular}{lcl} 
\#55 : $sl(1|3)$, $sl(2) \oplus 2\, sl(1|1)$, $sl(2|2)$ 
&\hspace*{18pt}
& \#56 : $osp(2|4)$, $sl(2) \oplus 2\, sl(1|1)$, $sl(2|2)$  \\
\#57 : $osp(5|2)$, $sl(2) \oplus 2\, sl(1|1)$, $sl(2|2)$  
&& \#58 : $G(3)$, $sl(2) \oplus 2\, sl(1|1)$, $sl(2|2)$  \\
\#59 : $sl(1|2)^{(1)}$, $sl(1|2) \oplus sl(2)$, $sl(1|3)$ 
&& \#60 : $sl(1|2)^{(1)}$, $sl(1|2) \oplus sl(2)$, $osp(2|4)$ \\
\#61 : $sl(1|2)^{(1)}$, $sl(1|2) \oplus sl(2)$, $osp(5|2)$ 
&& \#62 : $sl(1|2)^{(1)}$, $sl(1|2) \oplus sl(2)$, $G(3)$ \\
\#63 : $sl(3)^{(1)}$, $sl(3) \oplus sl(1|1)$, $sl(1|3)$ 
&& \#64 : $sl(3)^{(1)}$, $sl(3) \oplus osp(1|2)$, $osp(1|6)$ \\
\#65 : $osp(2|4)$, $osp(2|2) \oplus sl(2)$, $G(3)$ 
&& \#66 : $sl(1|4)^{(2)}$, $osp(1|4)^{(1)}$, $sl(4)^{(2)}$ \\
\#67 : $sl(1|4)^{(2)}$, $sl(1|5)^{(4)}$, $sp(4)^{(1)}$ 
&& \#68 : $sl(1|4)^{(2)}$, $osp(2|4)^{(2)}$ \\
\#69 : $sl(1|4)^{(2)}$, $osp(1|6)$, $sl(4)$ 
&& \#70 : $sl(4)$, $sl(1|3)$, $sl(2|2)$ \\
\#71 : $sp(4)^{(1)}$, $osp(5|2)$, $osp(4|2)$ 
&& \#72 : $sl(4)^{(2)}$, $osp(2|4)$, $osp(4|2)$ \\
\#73 : $sl(1|4)^{(2)}$, $sl(2|2)$, $osp(3|4)$ \\
\end{tabular}

\subsubsection*{Rank 5 hyperbolic superalgebras}

\#1 : $so(9)$, $sl(4) \oplus sl(1|1)$, $sl(3) \oplus sl(1|2)$, $osp(2|4)
\oplus sl(2)$, $F(4)$ \\
\#2 : $osp(7|2)$, $sl(2) \oplus sl(1|3)$, $sl(3) \oplus sl(1|2)$, $sp(6)
\oplus sl(1|1)$, $F_{4}$ \\
\#3 : $osp(7|2)$, $sl(1|3) \oplus sl(1|1)$, $2\, sl(1|2)$, $osp(2|4) \oplus
sl(1|1)$, $F(4)$ \\
\#4 : $osp(4|4)$, $sl(2|2) \oplus sl(2)$, $sl(3) \oplus sl(1|2)$, $so(7)
\oplus sl(2)$, $F(4)$ \\
\#5 : $sl(7)^{(2)}$, $sp(6) \oplus sl(1|1)$, $sp(4) \oplus sl(1|2)$,
$osp(2|4) \oplus sl(2)$, $F(4)$ \\
\#6 : $so(8)^{(2)}$, $so(7) \oplus sl(1|1)$, $so(5) \oplus sl(1|2)$,
$osp(2|4) \oplus sl(2)$, $F(4)$ \\
\#7 : $sl(2|4)^{(2)}$, $osp(4|2) \oplus sl(2)$, $osp(2|2) \oplus sl(3)$,
$so(7) \oplus sl(2)$, $F(4)$ \\
\#8 : $osp(3|4)^{(1)}$, $osp(3|4) \oplus sl(2)$, $osp(3|2) \oplus sl(3)$,
$so(7) \oplus sl(2)$, $F(4)$ \\
\#9 : $osp(1|6)^{(1)}$, $osp(1|6) \oplus sl(2)$, $osp(1|4) \oplus sl(3)$,
$osp(1|2) \oplus so(7)$, $F_{4}$ \\
\#10 : $sl(1|7)^{(4)}$, $osp(1|6) \oplus sl(2)$, $osp(1|4) \oplus sl(3)$,
$osp(1|2) \oplus sp(6)$, $F_{4}$ \\
\#11 : $sl(1|7)^{(4)}$, $osp(1|6) \oplus sl(1|1)$, $osp(1|4) \oplus
sl(1|2)$, $osp(2|4) \oplus osp(1|2)$, $F(4)$ \\
\#12 : $sl(5|2)^{(2)}$, $osp(5|2) \oplus sl(2)$, $osp(3|2) \oplus sl(3)$,
$so(7) \oplus osp(1|2)$, $F(4)$ \\
\#13 : $F_{4}$, $so(7) \oplus sl(1|1)$, $sl(3) \oplus sl(1|2)$, $sl(2)
\oplus sl(1|3)$, $osp(2|6)$ \\
\#14 : $so(7)^{(1)}$, $so(7) \oplus osp(1|2)$, $osp(1|4) \oplus 2\, sl(2)$,
$osp(1|8)$, $sl(1|7)^{(4)}$ \\
\#15 : $sl(6)^{(2)}$, $sp(6) \oplus osp(1|2)$, $osp(1|4) \oplus 2\, sl(2)$,
$osp(1|8)$, $osp(1|6)^{(1)}$ \\
\#16 : $sl(1|6)^{(2)}$, $osp(1|6) \oplus sl(2)$, $sl(3) \oplus sl(2) \oplus
osp(1|2)$, $sl(5)$, $osp(1|8)$ \\
\#17 : $sl(1|6)^{(2)}$, $osp(1|6) \oplus sl(2)$, $sp(4) \oplus osp(1|2)
\oplus sl(2)$, $sp(8)$, $osp(1|6)^{(1)}$ \\
\#18 : $sl(1|6)^{(2)}$, $osp(1|6) \oplus sl(2)$, $so(5) \oplus osp(1|2)
\oplus sl(2)$, $so(9)$, $sl(1|7)^{(4)}$ \\
\#19 : $sl(1|6)^{(2)}$, $osp(1|6) \oplus osp(1|2)$, $osp(1|4) \oplus
osp(1|2) \oplus sl(2)$, $osp(1|8)$, $osp(2|6)^{(2)}$ \\
\#20 : $sl(1|6)^{(2)}$, $osp(1|6) \oplus sl(1|1)$, $sl(1|2) \oplus osp(1|2)
\oplus sl(2)$, $sl(1|4)$, $osp(3|6)$ \\
\#21 : $osp(6|2)$, $osp(1|2) \oplus sl(1|3)$, $osp(1|4) \oplus sl(1|1)
\oplus sl(2)$, $osp(1|8)$, $osp(3|6)$ \\
\#22 : $osp(6|2)$, $sl(2) \oplus sl(1|3)$, $sl(3) \oplus sl(2) \oplus
sl(1|1)$, $sl(5)$, $sl(1|4)$ \\
\#23 : $osp(6|2)$, $sl(1|3) \oplus sl(1|1)$, $sl(1|2) \oplus sl(1|1) \oplus
sl(2)$, $sl(1|4)$, $sl(2|3)$ \\
\#24 : $sl(6)^{(2)}$, $sp(6) \oplus sl(1|1)$, $sl(1|2) \oplus 2\, sl(2)$,
$sl(1|4)$, $osp(2|6)$ \\
\#25 : $so(7)^{(1)}$, $so(7) \oplus sl(1|1)$, $sl(1|2) \oplus 2\, sl(2)$,
$sl(1|4)$, $osp(7|2)$ \\
\#26 : $osp(6|2)$, $sl(2) \oplus sl(1|3)$, $sp(4) \oplus sl(2) \oplus
sl(1|1)$, $sp(8)$, $osp(2|6)$ \\
\#27 : $osp(6|2)$, $sl(2) \oplus sl(1|3)$, $so(5) \oplus sl(2) \oplus
sl(1|1)$, $so(9)$, $osp(7|2)$ \\
\#28 : $sl(4)^{(1)}$, $sl(4) \oplus sl(1|1)$, $sl(1|4)$, $osp(6|2)$ \\
\#29 : $sl(1|3)^{(1)}$, $sl(1|3) \oplus sl(2)$, $sl(1|4)$, $osp(6|2)$ \\
\#30 : $sl(2|4)^{(2)}$, $osp(2|4) \oplus sl(2)$, $osp(4|4)$,
$osp(2|4)^{(1)}$ \\
\#31 : $sl(2|4)^{(2)}$, $osp(2|4) \oplus sl(2)$, $osp(5|4)$,
$sl(5|2)^{(2)}$ \\
\#32 : $sl(2|4)^{(2)}$, $osp(2|4) \oplus osp(1|2)$, $osp(5|4)$,
$osp(3|4)^{(1)}$ \\
\#33 : $sl(4)^{(1)}$, $sl(4) \oplus osp(1|2)$, $sl(1|6)^{(2)}$, $osp(1|8)$
\\
\#34 : $sl(2|4)^{(2)}$, $osp(2|4) \oplus sl(2)$, $sl(2|3)$, $osp(2|6)$ \\
\#35 : $so(8)$, $osp(1|2) \oplus 3\, sl(2)$, $sl(1|6)^{(2)}$ \\
\#36 : $F(4)$, $osp(2|2) \oplus sl(3)$, $osp(2|4) \oplus sl(2)$,
$osp(2|4)^{(1)}$ \\
\#37 : $F(4)$, $sl(1|2) \oplus 2\, sl(2)$, $sl(4) \oplus sl(1|1)$,
$so(7)^{(1)}$ \\
\#38 : $so(8)$, $osp(6|2)$, $sl(1|1) \oplus 3\, sl(2)$ \\

\subsubsection*{Rank 6 hyperbolic superalgebras}

\#1 : $F_{4}^{(1)}$, $sl(1|1) \oplus F_{4}$, $sl(1|2) \oplus sp(6)$,
$sl(1|3) \oplus sl(3)$, $sl(1|4) \oplus sl(2)$, $osp(9|2)$ \\
\#2 : $E_{6}^{(2)}$, $sl(1|1) \oplus F_{4}$, $sl(1|2) \oplus so(7)$,
$sl(1|3) \oplus sl(3)$, $sl(1|4) \oplus sl(2)$, $osp(2|8)$ \\
\#3 : $F_{4}^{(1)}$, $osp(1|2) \oplus F_{4}$, $osp(1|4) \oplus sp(6)$,
$osp(1|6) \oplus sl(3)$, $osp(1|8) \oplus sl(2)$, $sl(1|9)^{(4)}$ \\
\#4 : $E_{6}^{(2)}$, $osp(1|2) \oplus F_{4}$, $osp(1|4) \oplus so(7)$,
$osp(1|6) \oplus sl(3)$, $osp(1|8) \oplus sl(2)$, $osp(1|8)^{(1)}$ \\
\#5 : $so(10)$, $sl(1|1) \oplus sl(5)$, $sl(1|2) \oplus sl(3) \oplus
sl(2)$, $sl(1|4) \oplus sl(2)$, $osp(8|2)$, $sl(1|5)$ \\
\#6 : $osp(8|2)$, $sl(1|1) \oplus sl(1|4)$, $sl(2) \oplus 2\, sl(1|2)$,
$sl(2|4)$ \\
\#7 : $sl(8)^{(2)}$, $sl(1|1) \oplus sp(8)$, $sl(1|2) \oplus sl(2) \oplus
sp(4)$, $sl(1|4) \oplus sl(2)$, $osp(8|2)$, $osp(2|8)$ \\
\#8 : $so(9)^{(1)}$, $sl(1|1) \oplus so(9)$, $sl(1|2) \oplus sl(2) \oplus
so(5)$, $sl(1|4) \oplus sl(2)$, $osp(8|2)$, $osp(9|2)$ \\
\#9 : $so(10)$, $osp(1|2) \oplus sl(5)$, $sl(2) \oplus sl(3) \oplus
osp(1|4)$, $sl(2) \oplus osp(1|8)$, $sl(1|8)^{(2)}$, $osp(1|10)$ \\
\#10 : $so(9)^{(1)}$, $osp(1|2) \oplus so(9)$, $osp(1|4) \oplus sl(2)
\oplus so(5)$, $osp(1|8) \oplus sl(2)$, $sl(1|8)^{(2)}$, $sl(1|9)^{(4)}$ \\
\#11 : $sl(8)^{(2)}$, $osp(1|2) \oplus sp(8)$, $osp(1|4) \oplus sl(2)
\oplus sp(4)$, $osp(1|8) \oplus sl(2)$, $sl(1|8)^{(2)}$, $osp(1|8)^{(1)}$
\\
\#12 : $sl(1|8)^{(2)}$, $osp(1|2) \oplus osp(1|8)$, $osp(1|4) \oplus sl(2)
\oplus osp(1|4)$, $osp(2|8)^{(2)}$ \\
\#13 : $sl(1|8)^{(2)}$, $sl(1|1) \oplus osp(1|8)$, $sl(1|2) \oplus sl(2)
\oplus osp(1|4)$, $sl(1|4) \oplus osp(1|2)$, $osp(8|2)$, $osp(3|8)$ \\
\#14 : $so(8)^{(1)}$, $sl(1|1) \oplus so(8)$, $sl(1|2) \oplus 3\, sl(2)$,
$osp(8|2)$ \\
\#15 : $so(8)^{(1)}$, $osp(1|2) \oplus so(8)$, $osp(1|4) \oplus 3\, sl(2)$,
$sl(1|8)^{(2)}$ \\

\end{document}